%% file: main.tex
\documentclass[sigconf, nonacm, screen]{acmart}

\usepackage{tikz}
\usepackage{amsfonts}
\usepackage{amsmath}
\usepackage{caption}
\usepackage{subcaption}
\usepackage{algorithm}
\usepackage[noend]{algpseudocode}
\usepackage{multirow}
\usepackage{newtxtext,newtxmath}
\usepackage{multicol}
\usepackage{booktabs}
\usepackage{graphicx}
\usepackage[draft]{minted}

\makeatletter
\def\PYG@reset{\let\PYG@it=\relax \let\PYG@bf=\relax%
    \let\PYG@ul=\relax \let\PYG@tc=\relax%
    \let\PYG@bc=\relax \let\PYG@ff=\relax}
\def\PYG@tok#1{\csname PYG@tok@#1\endcsname}
\def\PYG@toks#1+{\ifx\relax#1\empty\else%
    \PYG@tok{#1}\expandafter\PYG@toks\fi}
\def\PYG@do#1{\PYG@bc{\PYG@tc{\PYG@ul{%
    \PYG@it{\PYG@bf{\PYG@ff{#1}}}}}}}
\def\PYG#1#2{\PYG@reset\PYG@toks#1+\relax+\PYG@do{#2}}

\@namedef{PYG@tok@w}{\def\PYG@tc##1{\textcolor[rgb]{0.73,0.73,0.73}{##1}}}
\@namedef{PYG@tok@c}{\let\PYG@it=\textit\def\PYG@tc##1{\textcolor[rgb]{0.24,0.48,0.48}{##1}}}
\@namedef{PYG@tok@cp}{\def\PYG@tc##1{\textcolor[rgb]{0.61,0.40,0.00}{##1}}}
\@namedef{PYG@tok@k}{\let\PYG@bf=\textbf\def\PYG@tc##1{\textcolor[rgb]{0.00,0.50,0.00}{##1}}}
\@namedef{PYG@tok@kp}{\def\PYG@tc##1{\textcolor[rgb]{0.00,0.50,0.00}{##1}}}
\@namedef{PYG@tok@kt}{\def\PYG@tc##1{\textcolor[rgb]{0.69,0.00,0.25}{##1}}}
\@namedef{PYG@tok@o}{\def\PYG@tc##1{\textcolor[rgb]{0.40,0.40,0.40}{##1}}}
\@namedef{PYG@tok@ow}{\let\PYG@bf=\textbf\def\PYG@tc##1{\textcolor[rgb]{0.67,0.13,1.00}{##1}}}
\@namedef{PYG@tok@nb}{\def\PYG@tc##1{\textcolor[rgb]{0.00,0.50,0.00}{##1}}}
\@namedef{PYG@tok@nf}{\def\PYG@tc##1{\textcolor[rgb]{0.00,0.00,1.00}{##1}}}
\@namedef{PYG@tok@nc}{\let\PYG@bf=\textbf\def\PYG@tc##1{\textcolor[rgb]{0.00,0.00,1.00}{##1}}}
\@namedef{PYG@tok@nn}{\let\PYG@bf=\textbf\def\PYG@tc##1{\textcolor[rgb]{0.00,0.00,1.00}{##1}}}
\@namedef{PYG@tok@ne}{\let\PYG@bf=\textbf\def\PYG@tc##1{\textcolor[rgb]{0.80,0.25,0.22}{##1}}}
\@namedef{PYG@tok@nv}{\def\PYG@tc##1{\textcolor[rgb]{0.10,0.09,0.49}{##1}}}
\@namedef{PYG@tok@no}{\def\PYG@tc##1{\textcolor[rgb]{0.53,0.00,0.00}{##1}}}
\@namedef{PYG@tok@nl}{\def\PYG@tc##1{\textcolor[rgb]{0.46,0.46,0.00}{##1}}}
\@namedef{PYG@tok@ni}{\let\PYG@bf=\textbf\def\PYG@tc##1{\textcolor[rgb]{0.44,0.44,0.44}{##1}}}
\@namedef{PYG@tok@na}{\def\PYG@tc##1{\textcolor[rgb]{0.41,0.47,0.13}{##1}}}
\@namedef{PYG@tok@nt}{\let\PYG@bf=\textbf\def\PYG@tc##1{\textcolor[rgb]{0.00,0.50,0.00}{##1}}}
\@namedef{PYG@tok@nd}{\def\PYG@tc##1{\textcolor[rgb]{0.67,0.13,1.00}{##1}}}
\@namedef{PYG@tok@s}{\def\PYG@tc##1{\textcolor[rgb]{0.73,0.13,0.13}{##1}}}
\@namedef{PYG@tok@sd}{\let\PYG@it=\textit\def\PYG@tc##1{\textcolor[rgb]{0.73,0.13,0.13}{##1}}}
\@namedef{PYG@tok@si}{\let\PYG@bf=\textbf\def\PYG@tc##1{\textcolor[rgb]{0.64,0.35,0.47}{##1}}}
\@namedef{PYG@tok@se}{\let\PYG@bf=\textbf\def\PYG@tc##1{\textcolor[rgb]{0.67,0.36,0.12}{##1}}}
\@namedef{PYG@tok@sr}{\def\PYG@tc##1{\textcolor[rgb]{0.64,0.35,0.47}{##1}}}
\@namedef{PYG@tok@ss}{\def\PYG@tc##1{\textcolor[rgb]{0.10,0.09,0.49}{##1}}}
\@namedef{PYG@tok@sx}{\def\PYG@tc##1{\textcolor[rgb]{0.00,0.50,0.00}{##1}}}
\@namedef{PYG@tok@m}{\def\PYG@tc##1{\textcolor[rgb]{0.40,0.40,0.40}{##1}}}
\@namedef{PYG@tok@gh}{\let\PYG@bf=\textbf\def\PYG@tc##1{\textcolor[rgb]{0.00,0.00,0.50}{##1}}}
\@namedef{PYG@tok@gu}{\let\PYG@bf=\textbf\def\PYG@tc##1{\textcolor[rgb]{0.50,0.00,0.50}{##1}}}
\@namedef{PYG@tok@gd}{\def\PYG@tc##1{\textcolor[rgb]{0.63,0.00,0.00}{##1}}}
\@namedef{PYG@tok@gi}{\def\PYG@tc##1{\textcolor[rgb]{0.00,0.52,0.00}{##1}}}
\@namedef{PYG@tok@gr}{\def\PYG@tc##1{\textcolor[rgb]{0.89,0.00,0.00}{##1}}}
\@namedef{PYG@tok@ge}{\let\PYG@it=\textit}
\@namedef{PYG@tok@gs}{\let\PYG@bf=\textbf}
\@namedef{PYG@tok@gp}{\let\PYG@bf=\textbf\def\PYG@tc##1{\textcolor[rgb]{0.00,0.00,0.50}{##1}}}
\@namedef{PYG@tok@go}{\def\PYG@tc##1{\textcolor[rgb]{0.44,0.44,0.44}{##1}}}
\@namedef{PYG@tok@gt}{\def\PYG@tc##1{\textcolor[rgb]{0.00,0.27,0.87}{##1}}}
\@namedef{PYG@tok@err}{\def\PYG@bc##1{{\setlength{\fboxsep}{\string -\fboxrule}\fcolorbox[rgb]{1.00,0.00,0.00}{1,1,1}{\strut ##1}}}}
\@namedef{PYG@tok@kc}{\let\PYG@bf=\textbf\def\PYG@tc##1{\textcolor[rgb]{0.00,0.50,0.00}{##1}}}
\@namedef{PYG@tok@kd}{\let\PYG@bf=\textbf\def\PYG@tc##1{\textcolor[rgb]{0.00,0.50,0.00}{##1}}}
\@namedef{PYG@tok@kn}{\let\PYG@bf=\textbf\def\PYG@tc##1{\textcolor[rgb]{0.00,0.50,0.00}{##1}}}
\@namedef{PYG@tok@kr}{\let\PYG@bf=\textbf\def\PYG@tc##1{\textcolor[rgb]{0.00,0.50,0.00}{##1}}}
\@namedef{PYG@tok@bp}{\def\PYG@tc##1{\textcolor[rgb]{0.00,0.50,0.00}{##1}}}
\@namedef{PYG@tok@fm}{\def\PYG@tc##1{\textcolor[rgb]{0.00,0.00,1.00}{##1}}}
\@namedef{PYG@tok@vc}{\def\PYG@tc##1{\textcolor[rgb]{0.10,0.09,0.49}{##1}}}
\@namedef{PYG@tok@vg}{\def\PYG@tc##1{\textcolor[rgb]{0.10,0.09,0.49}{##1}}}
\@namedef{PYG@tok@vi}{\def\PYG@tc##1{\textcolor[rgb]{0.10,0.09,0.49}{##1}}}
\@namedef{PYG@tok@vm}{\def\PYG@tc##1{\textcolor[rgb]{0.10,0.09,0.49}{##1}}}
\@namedef{PYG@tok@sa}{\def\PYG@tc##1{\textcolor[rgb]{0.73,0.13,0.13}{##1}}}
\@namedef{PYG@tok@sb}{\def\PYG@tc##1{\textcolor[rgb]{0.73,0.13,0.13}{##1}}}
\@namedef{PYG@tok@sc}{\def\PYG@tc##1{\textcolor[rgb]{0.73,0.13,0.13}{##1}}}
\@namedef{PYG@tok@dl}{\def\PYG@tc##1{\textcolor[rgb]{0.73,0.13,0.13}{##1}}}
\@namedef{PYG@tok@s2}{\def\PYG@tc##1{\textcolor[rgb]{0.73,0.13,0.13}{##1}}}
\@namedef{PYG@tok@sh}{\def\PYG@tc##1{\textcolor[rgb]{0.73,0.13,0.13}{##1}}}
\@namedef{PYG@tok@s1}{\def\PYG@tc##1{\textcolor[rgb]{0.73,0.13,0.13}{##1}}}
\@namedef{PYG@tok@mb}{\def\PYG@tc##1{\textcolor[rgb]{0.40,0.40,0.40}{##1}}}
\@namedef{PYG@tok@mf}{\def\PYG@tc##1{\textcolor[rgb]{0.40,0.40,0.40}{##1}}}
\@namedef{PYG@tok@mh}{\def\PYG@tc##1{\textcolor[rgb]{0.40,0.40,0.40}{##1}}}
\@namedef{PYG@tok@mi}{\def\PYG@tc##1{\textcolor[rgb]{0.40,0.40,0.40}{##1}}}
\@namedef{PYG@tok@il}{\def\PYG@tc##1{\textcolor[rgb]{0.40,0.40,0.40}{##1}}}
\@namedef{PYG@tok@mo}{\def\PYG@tc##1{\textcolor[rgb]{0.40,0.40,0.40}{##1}}}
\@namedef{PYG@tok@ch}{\let\PYG@it=\textit\def\PYG@tc##1{\textcolor[rgb]{0.24,0.48,0.48}{##1}}}
\@namedef{PYG@tok@cm}{\let\PYG@it=\textit\def\PYG@tc##1{\textcolor[rgb]{0.24,0.48,0.48}{##1}}}
\@namedef{PYG@tok@cpf}{\let\PYG@it=\textit\def\PYG@tc##1{\textcolor[rgb]{0.24,0.48,0.48}{##1}}}
\@namedef{PYG@tok@c1}{\let\PYG@it=\textit\def\PYG@tc##1{\textcolor[rgb]{0.24,0.48,0.48}{##1}}}
\@namedef{PYG@tok@cs}{\let\PYG@it=\textit\def\PYG@tc##1{\textcolor[rgb]{0.24,0.48,0.48}{##1}}}

\def\PYGZus{\char`\_}

\def\PYGZsh{\char`\#}

\def\PYGZsq{\char`\'}

\makeatother

\usepackage{comment}
\newcommand{\COMMENT}[2][.5\linewidth]{%
  \leavevmode\hfill\makebox[#1][l]{//~#2}}
  
\newcommand{\sysname}[0]{GNNFlow} % or "GNNStream"

\usepackage{balance}

\newcommand\vldbdoi{XX.XX/XXX.XX}
\newcommand\vldbpages{XXX-XXX}
\newcommand\vldbvolume{17}
\newcommand\vldbissue{1}
\newcommand\vldbyear{2024}
\newcommand\vldbauthors{\authors}
\newcommand\vldbtitle{\shorttitle} 
\newcommand\vldbavailabilityurl{URL_TO_YOUR_ARTIFACTS}
\newcommand\vldbpagestyle{plain} 

\begin{document}
\title{\sysname: A Distributed Framework for Continuous Temporal GNN Learning on Dynamic Graphs}

%%
%% The "author" command and its associated commands are used to define the authors and their affiliations.
\author{Yuchen Zhong$^1$, Guangming Sheng$^1$, Tianzuo Qin$^1$, Minjie Wang$^2$, Quan Gan$^2$, Chuan Wu$^1$}
\affiliation{%
  \institution{$^1$The University of Hong Kong, $^2$AWS Shanghai AI Lab}
}
\email{{yczhong, gmsheng, tzqin, cwu}@cs.hku.hk, {minjiw, quagan}@amazon.com}

\begin{abstract}
Graph Neural Networks (GNNs) play a crucial role in various fields. However, most existing deep graph learning frameworks assume pre-stored static graphs and do not support training on graph streams. In contrast, many real-world graphs are dynamic and contain time domain information. We introduce GNNFlow, a distributed framework that enables efficient continuous temporal graph representation learning on dynamic graphs on multi-GPU machines. GNNFlow introduces an adaptive time-indexed block-based data structure that effectively balances memory usage with graph update and sampling operation efficiency. It features a hybrid GPU-CPU graph data placement for rapid GPU-based temporal neighborhood sampling and kernel optimizations for enhanced sampling processes. A dynamic GPU cache for node and edge features is developed to maximize cache hit rates through reuse and restoration strategies. GNNFlow supports distributed training across multiple machines with static scheduling to ensure load balance. We implement GNNFlow based on DGL and PyTorch. Our experimental results show that GNNFlow provides up to 21.1x faster continuous learning than existing systems.
\end{abstract}

\maketitle

\pagestyle{\vldbpagestyle}
\begingroup\small\noindent\raggedright\textbf{PVLDB Reference Format:}\\
\vldbauthors. \vldbtitle. PVLDB, \vldbvolume(\vldbissue): \vldbpages, \vldbyear.\\
\href{https://doi.org/\vldbdoi}{doi:\vldbdoi}
\endgroup
\begingroup
\renewcommand\thefootnote{}\footnote{\noindent
This work is licensed under the Creative Commons BY-NC-ND 4.0 International License. Visit \url{https://creativecommons.org/licenses/by-nc-nd/4.0/} to view a copy of this license. For any use beyond those covered by this license, obtain permission by emailing \href{mailto:info@vldb.org}{info@vldb.org}. Copyright is held by the owner/author(s). Publication rights licensed to the VLDB Endowment. \\
\raggedright Proceedings of the VLDB Endowment, Vol. \vldbvolume, No. \vldbissue\ %
ISSN 2150-8097. \\
\href{https://doi.org/\vldbdoi}{doi:\vldbdoi} \\
}\addtocounter{footnote}{-1}\endgroup

\ifdefempty{\vldbavailabilityurl}{}{
\vspace{.3cm}
\begingroup\small\noindent\raggedright\textbf{PVLDB Artifact Availability:}\\
The source code, data, and/or other artifacts have been made available at \url{\vldbavailabilityurl}.
\endgroup
}

\input{introduction}

\input{background_and_motivation}

\input{overview}

\input{system_design}

\input{implementation}
\input{evaluation}

\input{related_work}

\section{Conclusion}
We introduce GNNFlow, a distributed system for training temporal GNNs on CTDGs. We design a scalable time-indexed block-based data structure for dynamic graphs, cache lightweight graph metadata on GPU, and employ optimizations for temporal sampling. We also develop a dynamic, vectorized GPU-based dynamic cache with cache reuse and restoration. GNNFlow outperforms existing systems, achieving up to 21.1x speed-up in continuous learning compared to TGL and 1.46x higher throughput than distributed DGL.

%\clearpage

\bibliographystyle{ACM-Reference-Format}
\bibliography{citation.bib}

\end{document}

%% file: introduction.tex
\section{Introduction}

Graph Neural Networks (GNNs) have recently achieved remarkable success in graph representation learning, demonstrating excellent performance in various graph-related tasks such as recommendation~\cite{ying2018graph, wu2019session}, social network mining~\cite{fan2019graph} and molecule analysis~\cite{fout2017protein}. Several GNN training frameworks have been proposed for graph learning on pre-stored static
graphs using multiple GPUs or machines~\cite{euler, quiver, zhu2019aligraph, pgl, liu2021bgl, zhou2022tgl}. However, most real-world graphs are dynamic, with nodes and edges constantly emerging or disappearing over time. For instance, in social networks such as Facebook and Twitter, new users can join at any time, and users continuously interact with others by reacting to or commenting on their posts. In recommendation systems for e-commerce, users constantly interact with items by clicking or purchasing. 

Graph representation learning on dynamic graphs has gained significant attention in recent years. Specifically, \textit{temporal GNNs} (\S\ref{sec:preliminaries}) like TGN~\cite{rossi2021temporal} incorporate the temporal information from dynamic graphs and significantly outperform GNN models for static graphs on node classification and link prediction~\cite{trivedi2017know, trivedi2019dyrep, pareja2020evolvegcn, ma2020streaming, sankar2020dysat, Xu2020Inductive, rossi2021temporal}. Additionally, \textit{continuous GNN learning} (\S\ref{sec:preliminaries}) has been proposed on streaming graph data, to capture the changing patterns and properties of the graph over time~\cite{xu2020graphsail, wang2021graph, wang2020streaming, perini2022learning, ahrabian2021structure, ding2022causal}. By incorporating new data on the fly, continuous GNN learning allows for timely model updates, leading to more accurate and timely predictions. This adaptability is crucial in applications like recommendation systems, where it can swiftly track and respond to fast-paced changes in user preferences~\cite{platogl, concept-drift}. In this work, we focus on \textit{continuous temporal GNN learning on dynamic graphs}. 

However, existing deep graph learning frameworks like DGL~\cite{wang2019deep} and TGL~\cite{zhou2022tgl} have yet to support the workload efficiently. A significant issue is their reliance on static graph storage. For instance, DGL employs the Compressed sparse row (CSR) graph format, which enables efficient retrieval of a node's neighbors. This is crucial as GNN models aggregate the representations of a node’s inbound edges into its overall representation~\cite{kipf2017semisupervised}. However, adding new graph data in these frameworks requires a complete graph rebuild, which can incur significant time overhead (\S\ref{sec:limitation_of_existing_frameworks}). This overhead is exacerbated in real-world scenarios where graphs are often so large that they exceed a single machine's memory capacity. Consequently, existing frameworks such as DGL necessitate partitioning these graphs across multiple machines~\cite{zheng2020distdgl}. However, when adding new graph data, an expensive re-partitioning process is required, particularly when using popular edge-cut-minimizing graph partition algorithms like METIS~\cite{metis1998}. Therefore, managing graph updates calls for a dynamic, distributed graph storage system with an efficient data structure.

Moreover, calculating a node's representation using the representations of all its neighbors is computationally expensive. As a result, neighborhood sampling, in which a subset of neighbors is selected for computations, has become a common practice~\cite{hamilton2017inductive}. This situation becomes even more complex in temporal GNNs, where the temporal k-hop sampling operation is required (\S\ref{sec:preliminaries}). This operation recursively selects a subset of a node's neighbors based on specific timestamps across multiple ``hops'', forming a time-aware subgraph with the initial node at its root. Therefore, the dynamic distributed graph storage system should also incorporate support for efficient temporal k-hop sampling, thereby  facilitating continuous temporal GNN learning on dynamic graphs.

In addition, GPU acceleration is commonly employed for training GNNs, as it has been shown to enhance the efficiency of GNN training~\cite{wang2019deep, yang2022gnnlab, liu2021bgl}. However, due to the limited capacity of GPU memory, typically around 32 GB or less, when dealing with large-scale graphs that contain both topological and feature data (dense vectors attached to vertices or edges), most existing GNN systems store the graph data in the host memory of a machine and partition the graph when it exceeds the memory capacity of a single machine~\cite{zheng2020distdgl}. However, it incurs significant I/O overhead due to the multi-layered nature of GNNs that necessitate recursive neighbor sampling and cross-machine data transmission. This entails fetching neighbors and feature data from possibly non-local machine partitions, followed by transferring the resultant subgraph and feature data from host memory to GPU memory for GNN input. Neighborhood sampling and feature fetching have been identified as bottlenecks in training GNNs on static graphs due to the significant I/O overhead, which considerably slows down the overall training process~\cite{yang2022gnnlab, liu2021bgl}. The same holds true for GNN learning on dynamic graphs, where these operations can account for 43\%-73\% of the total training time (\S\ref{sec:limitation_of_existing_frameworks}).

Previous works on static graphs have proposed using GPUs to accelerate neighborhood sampling~\cite{jangda2021accelerating, yang2022gnnlab, yang2022wholegraph}. However, these studies exclusively focus on static graph storage. Graph formats designed for static graphs, such as CSR, are inherently well-suited for GPU operations. CSR employs three contiguous arrays, enabling efficient GPU memory access. However, the complex data structure of dynamic graphs presents unique challenges in optimizing this GPU-based sampling. Dynamic graph data structures usually use pointer-based data structures, such as adjacency lists. This results in highly scattered data storage and necessitates pointer chasing to find neighbors~\cite{besta2020practice}. The scattered data storage leads to poor memory access patterns that conflict with the contiguous memory access preferred by GPUs, thereby causing a degradation in performance~\cite{jangda2021accelerating}.

\begin{figure*}[t]
    \centering
    \includegraphics[width=0.95\textwidth]{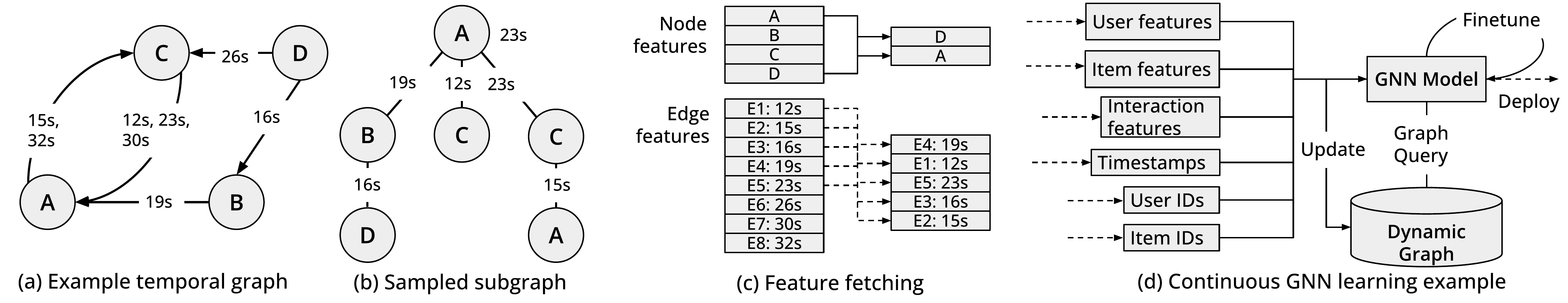}
    \vspace{-2mm}
    \caption{Figure (a) shows a temporal graph where edges are associated with timestamps. Figure (b) displays the result of temporal 2-hop sampling on node A at timestamp 23s, resulting in a subgraph. Figure (c) shows feature fetching for the subgraph in Figure (b). Figure (d) shows a concrete example of continuous GNN learning.}
    \label{fig:sampling_and_cont_learning_example}
\end{figure*}

Further, feature fetching of dense node and edge features  presents new challenges in continuous learning over dynamic graphs. Previous GNN systems proposed static GPU-based feature caches to reduce the transmission of node features, which requires careful cache initialization before training. For instance, PaGraph~\cite{lin2020pagraph} caches features of nodes with the highest node degrees, while GNNLab proposes a presampling method~\cite{yang2022gnnlab}, which samples the graph for several epochs and caches the features of the most visited nodes. These approaches work well in offline training scenarios involving tens of epochs using all data. However, in continuous learning, the cache needs to be re-initialized before each round, which brings two major issues: 1) the limited number of epochs trained in each round (typically 2-3 to avoid overfitting on new data~\cite{perini2022learning}) makes it challenging to amortize the cache initialization time (Figure~\ref{fig:presampling_time_ratio}); and 2) a substantial amount of features to be cached are already stored in the cache, making it unnecessary to perform costly cache initialization in each round (\S\ref{sec:limitation_of_existing_frameworks}). Moreover, previous works only focus on caching node features, but overlook the importance of caching edge features. Edge features are widely used in temporal GNN models to capture temporal dynamics~\cite{Xu2020Inductive, rossi2021temporal}. We discover that approximately 62.2\% to 99.3\% of the total feature communication is dedicated to transferring edge features for dense graphs (\S\ref{sec:limitation_of_existing_frameworks}). We further find that when training temporal GNNs on dynamic graphs, the access to edge features follows an exponential distribution with a notable degree of access to numerous edge features (\S\ref{sec:feature-fetching}). This makes previous static cache designs ineffective for edge features, highlighting the need for dynamic caching of features.

To address these challenges, we develop \sysname{}, a distributed deep graph learning framework for efficient continuous GNN training over dynamic graphs on multi-GPU machines. The main contributions of \sysname{} are summarized as follows. 

$\triangleright$ We propose an efficient time-indexed block-based data structure for storing dynamic graphs (\S\ref{sec:data-structure}). It uses adaptive block size to balance memory usage and the efficiency of graph updates and sampling operations.

$\triangleright$ We propose a hybrid GPU-CPU graph data placement for fast GPU-based temporal neighborhood sampling by storing frequently accessed lightweight graph metadata on the GPU and heavy graph edge data on the host shared memory (\S\ref{sec:neighborhood-sampling}). We also make kernel optimizations to improve the sampling process.

$\triangleright$ We develop a dynamic GPU cache for both node and edge features, improving the cache hit rate with cache reuse and restoration (\S\ref{sec:feature-fetching}). We also introduce a vectorized cache design for batch updates to boost efficiency.

$\triangleright$ We support distributed training by partitioning graphs (topology and features) on multiple machines (\S\ref{sec:distributed_training}). We propose static scheduling for distributed sampling to ensure load balance.

$\triangleright$ We implement \sysname{} using DGL~\cite{wang2019deep} and PyTorch~\cite{paszke2019pytorch}, and compare its performance with TGL~\cite{zhou2022tgl} and DGL. Note that while \sysname{} is designed for continuous temporal GNN learning on dynamic graphs, it also supports static GNN models like GCN~\cite{kipf2017semisupervised}, GraphSAGE~\cite{hamilton2017inductive}, and GAT~\cite{velickovic2018graph}, as well as training on static graphs. Our evaluations demonstrate that \sysname{} enables up to 21.1x speed-up compared to TGL in continuous learning of temporal GNN models (including TGN~\cite{rossi2021temporal}, TGAT~\cite{Xu2020Inductive}, and DySAT~\cite{sankar2020dysat}) on large real-world graphs with 8 GPUs. Additionally, we achieve up to 1.46x higher training throughput when training GraphSAGE~\cite{hamilton2017inductive} and GAT~\cite{velickovic2018graph} on a large partitioned static graph on 4 Amazon EC2 g4dn.metal instances, compared to distributed DGL. Notably, \sysname{} can efficiently train temporal GNNs on a dynamic graph with 5 billion edges on 8 g4dn.metal instances, while other systems cannot support this scale.

%% file: background_and_motivation.tex
\section{Background and Motivation}

\subsection{GNN Learning on Dynamic Graphs}
\label{sec:preliminaries}

\vspace{1mm}
\noindent\textbf{Dynamic graphs.}  There are two main models of dynamic graphs: {\em Discrete-time dynamic graphs} (DTDG) and {\em continuous-time dynamic graphs} (CTDG). DTDGs are sequences of static graph snapshots taken at regular intervals. CTDGs are represented by lists of timestamped events, such as edge and node additions or deletions. We focus on CTDGs for two main reasons.
First, CTDGs offer a more flexible and detailed representation of dynamic graphs compared to DTDGs, capturing graph evolution at the event level. Second, many recently proposed temporal GNN models are designed specifically for CTDGs, which outperform DTDG-based ones on tasks like node classification and link prediction~\cite{Xu2020Inductive, rossi2021temporal}. 

\vspace{1mm}
\noindent\textbf{Temporal graph neural networks.}  Traditional GNN models for static graphs like GCN~\cite{kipf2017semisupervised} and GAT~\cite{vaswani2017attention} employ a message-passing paradigm ~\cite{gilmer2017neural} to propagate node information to neighbors and compute node embeddings by aggregating neighborhood information.  % To improve computational efficiency, neighborhood sampling is commonly used \cite{hamilton2017inductive, chen2018fastgcn, huang2018adaptive, zeng2019graphsaint}, which samples a subset of a node's neighbors for embedding generation. Notable GNN models include GraphSAGE\cite{hamilton2017inductive}, which uses uniform neighbor sampling, and  GAT~\cite{vaswani2017attention}, which utilizes uniform sampling and attention mechanisms to determine which neighbors to focus on based on their relevance to the target node in embedding generation. 
Temporal GNNs are designed for temporal graphs, which adopt message passing and temporal neighborhood sampling for embedding generation. Models like TGN~\cite{rossi2021temporal} and TGAT~\cite{Xu2020Inductive} use temporal graph attention, with TGN also using a node memory module (a software module different from device memory). DySAT~\cite{sankar2020dysat}, designed for DTDGs, can also be extended to CTDGs~\cite{zhou2022tgl}. % Training of temporal GNNs is done in strict time order using mini-batches on multi-GPU machines for efficiency~\cite{zhou2022tgl}. Each mini-batch iteration consists of three phases: temporal neighborhood sampling, feature fetching and GNN computation.

\vspace{1mm}
\noindent\textbf{Temporal GNNs training.} 
The training (or target) nodes are divided into mini-batches and iteratively trained, with one mini-batch processed in one training iteration. Temporal GNN training
processes the dataset in strict time order~\cite{zhou2022tgl}. Each training
iteration can be divided into three phases: temporal k-hop sampling, feature fetching, and model training. 

Temporal k-hop sampling involves selecting a subset of the node $v$'s neighbors based on a specific timestamp $t$~\cite{Xu2020Inductive}. The edges connecting these neighbors must have timestamps that fall within a given time range of $[t - \delta, t]$. Temporal multi-hop sampling is a recursive process that samples each hop, resulting in a tree (subgraph) with node $v$ at its root. Figure~\ref{fig:sampling_and_cont_learning_example} (b) shows a toy example. Note that in the first hop, node C is sampled twice because there are two edges (12s, 23s) that meet the condition. TGN uses recent sampling to select a given number of neighbors closest to the given timestamp $t$, while TGAT uses uniform sampling, which randomly selects a subset of neighbors from all candidate neighbors. DySAT uses a specified time window for $\delta$. Static GNN models like GraphSAGE~\cite{hamilton2017inductive} and GAT~\cite{vaswani2017attention} can also adapt to temporal neighbor sampling, which is similar to the uniform sampling used in TGAT. After obtaining a sampled subgraph, we need to fetch the features of the sampled nodes/edges from disk/CPU memory, and copy them to GPUs for GNN computation. Figure~\ref{fig:sampling_and_cont_learning_example} (c) provides an example. Only the features of the outermost sampled nodes (i.e., nodes D and A) and all sampled edges in the subgraph are required. 

In distributed GNN training, a large graph is partitioned %and distributed 
among multiple machines, and %for parallel processing. 
each GPU/machine processes a subset of the training nodes using data-parallel training \cite{liu2021bgl}. %, with the same GNN model trained on each partition. 
During each phase of a training iteration, communication is necessary for data exchange among the machines~\cite{zheng2020distdgl}. %For neighborhood sampling, necessary 
Neighbor information  
of training nodes not in the local graph partition must be transferred for neighborhood sampling.
During feature fetching, machines %communicate with each other to 
retrieve node/edge features and node memories from others for %the sampled nodes and edges which 
those not in the local graph. In GNN computation, the machines synchronize gradients among each other for model parameter updates. %must communicate to aggregate information (i.e., gradient) to update the model parameters.

\vspace{1mm}
\noindent\textbf{Continuous GNN learning on dynamic graphs.}
\label{sec:continuous_learning_example}
Figure~\ref{fig:sampling_and_cont_learning_example} (d) illustrates a concrete example of continuous learning in the context of a real-time recommendation system scenario, such as short video recommendation~\cite{platogl}. In this scenario, new user/item interaction events are constantly generated, with both users and items having features, and the interactions themselves also having features (e.g., viewing duration, number of likes, comments, etc.). These generated events are batched (e.g., every hour) and simultaneously update the live dynamic graph and are used to finetune the GNN model on GPU machines. %To facilitate real-time processing of large amounts of data, finetuning is typically performed on GPU machines~\cite{platogl}. 
The finetuned GNN model can then be deployed immediately, allowing for continuous improvement and adaptation to changing user preferences and item attributes. While finetuning a GNN with new data can improve adaptation to changing patterns, it risks catastrophic forgetting of previously learned knowledge. Experience replay addresses this by incorporating both new and historical data when finetuning to mitigate catastrophic forgetting~\cite{perini2022learning}.

% The key hyper-parameter of the last continuous learning method is the ratio of the selected historical data within all the training data 
% %\cwu{all the training data in each training iteration?}
% , i.e., replay ratio $\gamma$ ($0\leq \gamma \leq 1)$. The third method can be considered as a special case of $\gamma = 0$. Our system  will support the third and fourth approaches with $0\leq \gamma \leq 1$.
\begin{table}[t]
\caption{Qualitative comparisons of different systems.}
\vspace{-3mm}
\begin{tabular}{lcccc}
\toprule
\multirow{2}{*}{\small{System}} & \small{Dist.} & \small{Temp. Neighb.} & \small{GPU} & \small{GPU Feature}  \\
 &  \small{CTDGs} & \small{Sampling} & \small{Sampling} & \small{Cache}  \\
\midrule
\small{DGL~\cite{wang2019deep}} & & & \checkmark &   \\
\small{TGL~\cite{zhou2022tgl}} & & \checkmark  & & \\
\small{GNNLab~\cite{yang2022gnnlab}} & & & \checkmark & \checkmark   \\
% \small{BGL~\cite{liu2021bgl}} & & & \checkmark & \checkmark   \\
\small{PlatoGL~\cite{platogl}} & \checkmark & & &  \\
\small{GNNFlow} & \checkmark & \checkmark & \checkmark & \checkmark \\
\bottomrule
\end{tabular}
\label{tbl:system_comparison}
\end{table}

\begin{table}[!t]
\caption{Overhead of building and partitioning large graphs.}
\vspace{-3mm}
\label{tbl:build_graph_time}
\centering
\begin{tabular}{llll}
\toprule
Framework & Load and Build & Partition & Training \\
\midrule
TGL~\cite{zhou2022tgl} & 0.5 hr & N/A & 2.5 hr \\
\midrule
DGL~\cite{zheng2020distdgl} & 0.7 hr & 2.7 hr & 5.2 hr \\ 
\bottomrule
\end{tabular}
\end{table}

\subsection{Limitations of Existing Systems \& Challenges}
\label{sec:limitation_of_existing_frameworks}
\label{sec:challenges}
Table~\ref{tbl:system_comparison} shows a qualitative comparison among existing systems. DGL~\cite{wang2019deep}, a widely used GNN framework, assumes static graphs and supports distributed GNN training over multiple machines~\cite{zheng2020distdgl}. TGL~\cite{zhou2022tgl} trains temporal GNNs on pre-stored static temporal graphs on a single multi-GPU machine. Its following work DistTGL~\cite{zhou2023disttgl} focuses on memory-based temporal GNNs but still assumes a static graph storage and does not partition graphs in distributed training. GNNLab~\cite{yang2022gnnlab} focuses on training static GNNs on static graphs with a GPU-based sampler and static feature cache. PlatoGL~\cite{platogl}, though it supports continuous learning over distributed CTDGs, does not support temporal GNNs. Furthermore, it is not open-source, and thus we are unable to perform quantitative comparisons.

\vspace{1mm}
\noindent\textbf{Limitation 1: Large overhead to build and partition large static graph snapshots.}  We examine the overhead of graph reconstruction in training TGN using TGL and GraphSAGE using DGL on AWS g4dn.metal instances using the MAG dataset~\cite{zhou2022tgl}. TGL's graph construction was rewritten in C++ with multi-threading for efficiency. TGL does not support distributed training on partitioned graphs. Thus, its partition time has not been demonstrated. As shown in Table~\ref{tbl:build_graph_time}, despite the improvements, graph construction and partitioning can take 30 minutes to several hours, making up 20\% to 65\% of training time in an epoch. This significant overhead limits the training frequency for both TGL and DGL on updated data.

\vspace{1mm}
\noindent\textbf{Challenge 1: Design an efficient CTDGs storage that efficiently supports temporal k-hop sampling.} Choosing the right data structure for CTDGs that supports graph updates and temporal neighborhood sampling without excess memory use is critical. While an adjacency list is a simple choice, querying neighbors requires traversing the whole lists, which is inefficient with long lists. Block-based adjacency lists can help by dividing neighbors into blocks, but the block size needs careful selection to balance efficiency and memory use. Previous dynamic graph processing systems like Tegra~\cite{iyer2021tegra} inefficiently handle fine-grained temporal queries, treating dynamic graphs as snapshot sequences. Temporal k-hop sampling in this approach would create millions of snapshots for large graphs, equal to the number of edges. In Aspen~\cite{dhulipala2019low}, a graph is represented as tree-of-trees: a purely-functional tree stores the set of vertices (vertex-tree), and each vertex stores the edges in its own C-tree (edge-tree). However, it does not support temporal k-hop sampling, and its tree structure is not optimized for GPU operations. To address this challenge, we propose a time-indexed block-based adjacency list with adaptive block size to address the challenge (\S\ref{sec:data-structure}). 

\begin{figure}[t]
    \centering
    \includegraphics[width=0.75\columnwidth]{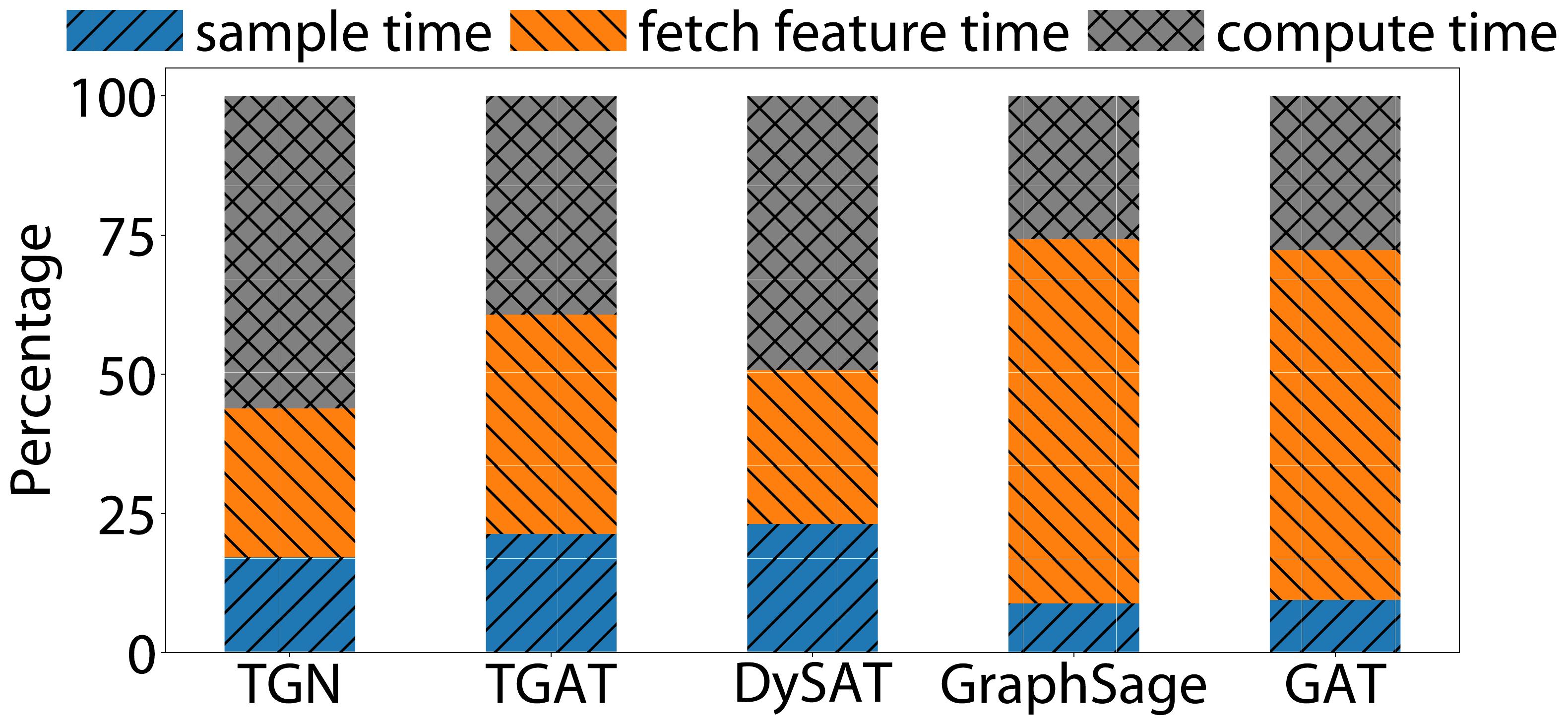}
    \caption{GNN training time breakdown for TGL and DGL.} 
    \label{fig:baseline-breakdown}
\end{figure}

\vspace{1mm}
\noindent\textbf{Limitation 2: Inefficient graph queries becoming the main bottleneck of GNN training on dynamic graphs.} Existing systems, like DGL and TGL, are not well-optimized for neighborhood sampling and feature fetching in dynamic graphs. They use CPUs for sampling and require copying features from CPU to GPU. Later releases of DGL utilize unified virtual addressing (UVA) for sampling. We run TGL to train three temporal GNNs (TGN, TGAT, DySAT) and DGL to train GraphSAGE and GAT with graph partition on the GDELT graph~\cite{zhou2022tgl}. Figure~\ref{fig:baseline-breakdown} shows that neighborhood sampling and feature fetching account for 43\% to 73\% of total training time, creating bottlenecks in model learning. Inefficient graph queries also limit retraining frequency. Some studies propose GPU-based sampling and feature cache~\cite{lin2020pagraph, yang2022gnnlab, quiver} for static graphs, but they do not address dynamic graphs.

\vspace{1mm}
\noindent\textbf{Challenge 2: Efficiently use GPUs to accelerate temporal neighborhood sampling on dynamic graph.} Due to the limited memory capacity of GPUs, previous works have proposed hosting the graph in pinned host memory and using unified virtual addressing (UVA) to access necessary graph data for large static graphs~\cite{quiver, sun2023legion}. However, for dynamic graphs, this method would incur substantial CPU-to-GPU data copying overheads during temporal neighborhood sampling. This is because dynamic graphs typically utilize a pointer-based data structure, leading to scattered graph data. Therefore, finding neighbors requires pointer-chasing, where each memory access inevitably results in a CPU-to-GPU data copy. Moreover, in a distributed multi-GPU setting, achieving load balancing among GPUs when performing sampling poses a challenge. To address this challenge, we propose an efficient GPU-based temporal neighborhood sampling design for distributed GNN training, which places frequently accessed graph metadata on GPUs, employs several kernel optimizations, and evenly distributes the sampling workload among workers (\S\ref{sec:neighborhood-sampling}).

\begin{table}[!t]
    \centering
    \caption{The percentage of edge feature transmission in the total feature transmission.}
    \vspace{-3mm}
    \begin{tabular}{ccccc}
        \toprule
         \small{Model} & \small{Reddit~\cite{kumar2019predicting}} & \small{GDELT~\cite{zhou2022tgl}} & \small{Netflix~\cite{bennett2007netflix}} \\ 
        \midrule 
        \small{TGN} & 82.2\% & 72.4\% &  26.9\% \\ 
        \small{TGAT} & 88.6\%  & 99.3\% &  35.9\%  \\ 
        \small{DySAT} & 62.2\% & 89.2\% & 28.0\% \\ 
        \bottomrule
    \end{tabular}
    \vspace{-3mm}
    \label{tab:edge_feat_percentage}
\end{table}

\vspace{1mm}
\noindent\textbf{Limitation 3: Ineffective feature caching in existing GPU-based designs for dynamic graph learning.} Previous works on GPU feature cache focus on static cache, requiring careful cache initialization to cache features of frequently accessed nodes. PaGraph~\cite{lin2020pagraph} uses a node-degree-based method. GNNLab ~\cite{yang2022gnnlab} proposes a presampling method, which samples the graph for several epochs and caches the features of the most visited nodes. However, The initialization time for the cache may account for up to 90\% of its feature fetching time (including cache initialization) (Figure~\ref{fig:presampling_time_ratio}). Furthermore, we have also found that nearly 92\% of node features and 22\% of edge features to cache are actually already on the cache (Table~\ref{tab:jaccard_similarty}). This indicates that it is unnecessary to perform the expensive cache initialization every time retraining is carried out. Moreover, most existing studies focus on static caching of node features, overlooking edge feature caching~\cite{rossi2021temporal, zhou2022tgl}. Table~\ref{tab:edge_feat_percentage} shows that, for dense Reddit and GDELT graphs, edge feature communication accounts for 66.2\% to 99.3\% of total feature communication (including TGN's node memory communication). For the sparse Netflix graph, this percentage is lower due to its large node count and high node feature dimension (768). Therefore, a new cache design is required to optimize GPU feature caching for continuous learning on dynamic graphs.

\vspace{1mm}
\noindent\textbf{Challenge 3: Optimize GPU feature cache for continuous learning.} 
We discover that node feature access patterns follow a power-law distribution, allowing us to cache only a small portion of frequently accessed nodes (Figure~\ref{fig:node-edge-access-pattern}). In contrast, edge feature access patterns follow an exponential distribution, resulting in a significant degree of access to many edge features, rendering the aforementioned approach ineffective. To address this challenge, we propose using a vectorized dynamic cache with cache reuse and restoration to address the challenge (\S\ref{sec:feature-fetching}).

%% file: overview.tex
\section{Overview}
\label{sec:overview}

\begin{figure}[t]
    \centering
    \includegraphics[width=0.90\columnwidth]{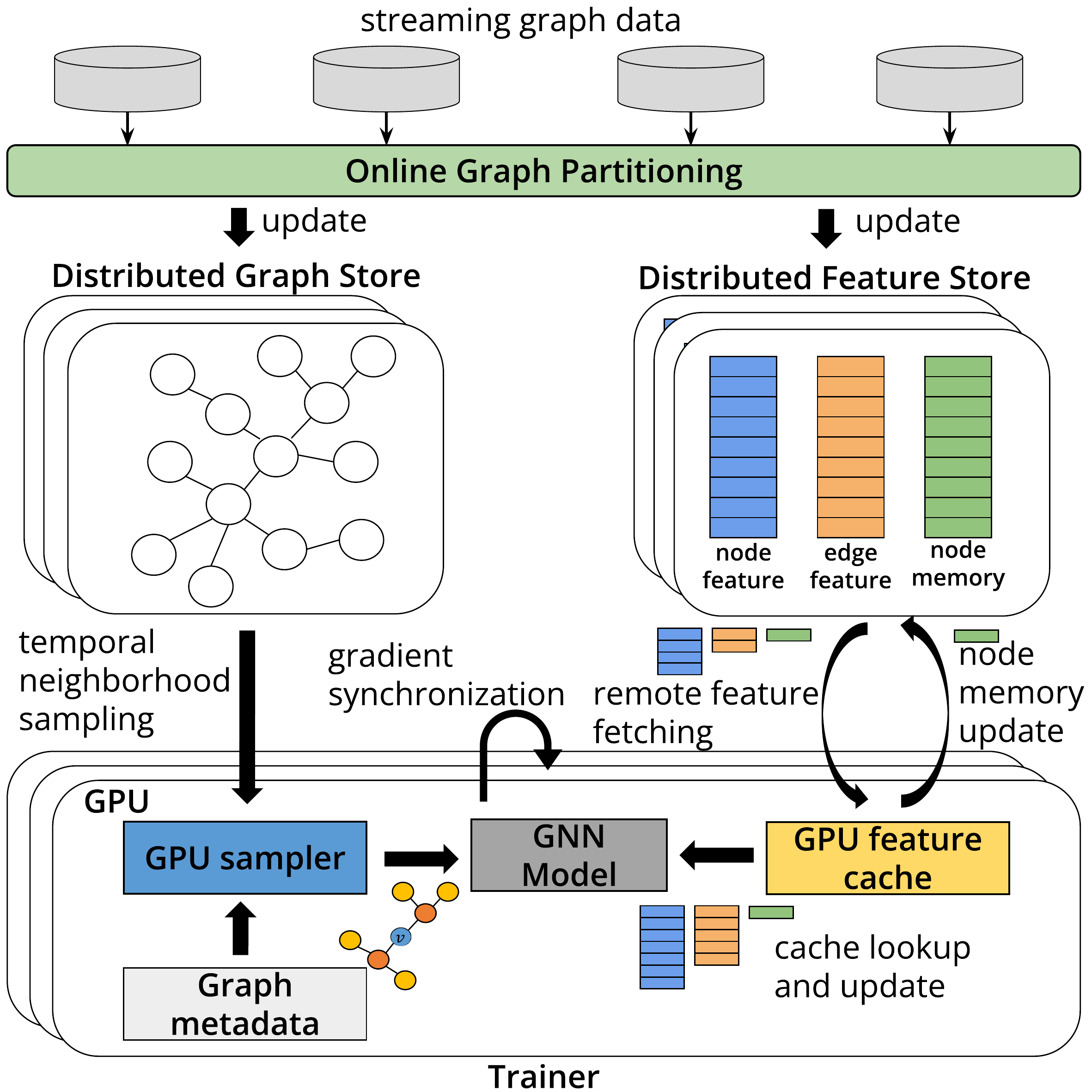}
    \caption{Architecture of \sysname{}} 
    \label{fig:architecture}
\end{figure}

We design \sysname{}, a distributed graph learning framework %for graph learning systems 
tailored for continuous temporal GNN training on dynamic graphs. It also supports static GNN models and offline training on static graphs. %(\S\ref{sec:preliminaries}). %and multiple online partitioning algorithms. 
%Following the common approach used by dynamic graph processing systems, we batch graph updates~\cite{ediger2012stinger}.
We consider a dynamic graph $\mathcal{G}(t)$ that constantly receives updates (i.e., node and edge insertion and deletion) over time. %$t$. %from streaming data, 
Our system obtains a GNN model $M(t)$ that learns up-to-date information from $\mathcal{G}(t)$.

%Figure~\ref{fig:workflow} shows our end-to-end solution to train on the streaming graph continuously. 
Figure~\ref{fig:architecture} illustrates the architecture of \sysname{}, comprising a distributed graph store, a distributed feature store (both in host memory), and a trainer (sampler and feature cache) on each GPU. The system processes streaming graph updates in incremental batches, following common practice in dynamic graph processing systems~\cite{ediger2012stinger}. New graph data (nodes, edges, features) received between $t$ and $t+1$ are batched into incremental batch $\mathcal{G}(t, t+1)$. Upon batch arrival, the system dispatches new data to machines using an online partitioning method (\ref{sec:partition}) and stores graph structure data and features in respective stores. Model retraining (aka finetuning) is triggered, where the current GNN model $M(t)$ is finetuned with new data $\mathcal{G}(t, t+1)$ and possibly historical data.

Iterative mini-batch training is adopted for model retraining upon each incremental batch, %in a distributed manner 
using GPUs on multiple machines. %on the current training set. 
In each training iteration, a trainer %(typically one on a GPU) 
queries the distributed graph store for temporal neighborhood sampling %that obtains sampled subgraphs according to a sampling algorithm %in the training method 
(\S\ref{sec:neighborhood-sampling}).
Then the trainer fetches node/edge features of the sampled subgraphs: it looks up %the features in 
its GPU feature cache and invokes the feature fetching service to retrieve features not in the cache. The feature cache is updated %according to the {\em cache replacement algorithm} when a dynamic cache is used 
dynamically (\S\ref{sec:feature-fetching}). Afterward, the trainers perform GNN computation, including forward computation, backward computation, and gradient synchronization. Finally, if node memories are used, the updated memory is written back to the distributed feature store. 
The dynamic graph storage is not updated during this retraining process; the incoming data received during this time are buffered as the next batch.

%% file: system_design.tex
\section{System Design}

\subsection{In-memory Dynamic Graph Storage}
%\vspace{-2mm}
\label{sec:data-structure}

\begin{figure}
    \centering
    \includegraphics[width=0.9\columnwidth]{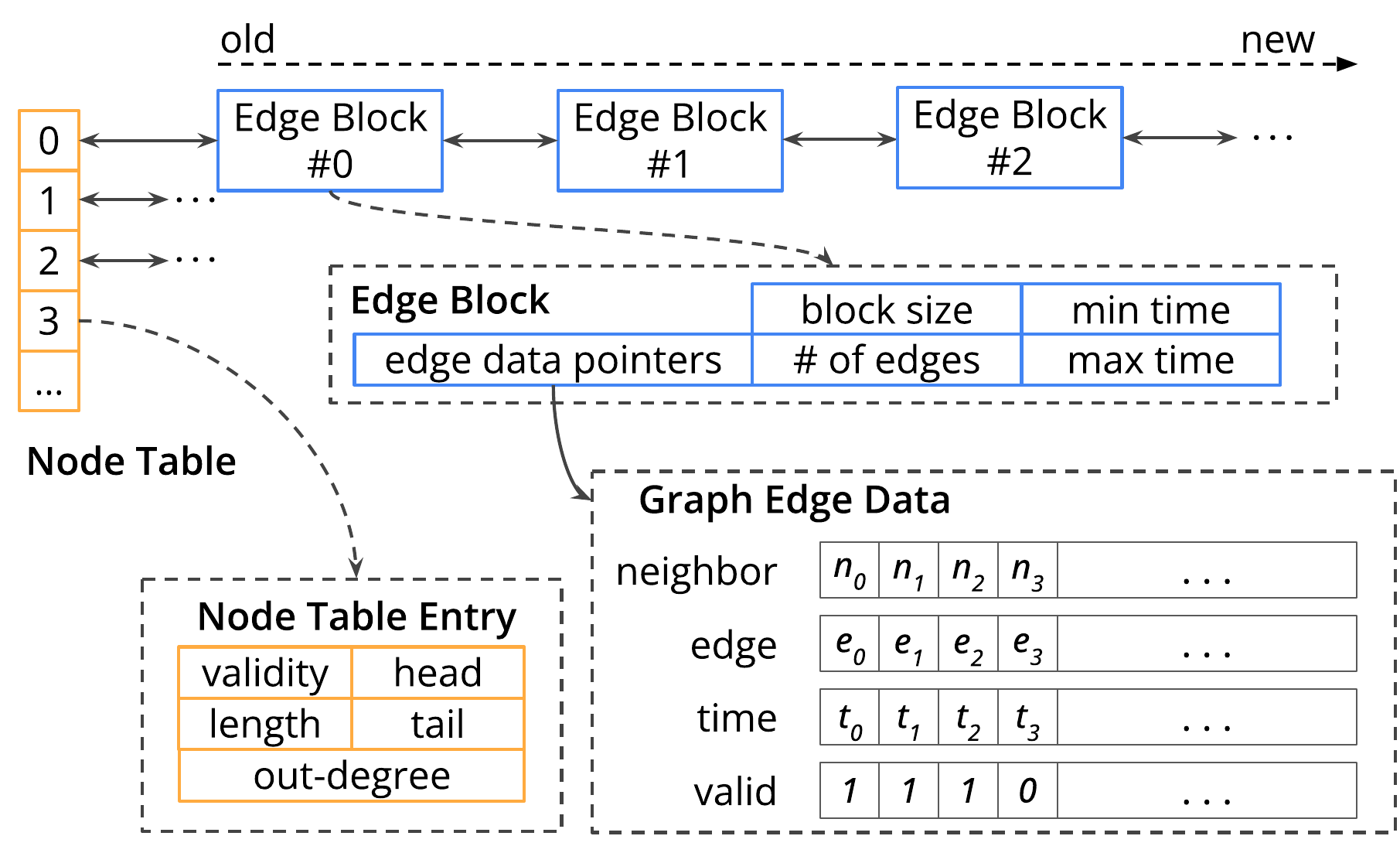}
    \caption{Block-based neighborhood storage}   
    \label{fig:block_based_adjacency_list}
\end{figure}

We propose block-based neighborhood storage, allowing efficient updates and queries in dynamic graph learning, as illustrated in Figure~\ref{fig:block_based_adjacency_list}. This data structure exploits temporal locality to enable fast insertion of new nodes/edges and efficient temporal neighborhood sampling (\S\ref{sec:neighborhood-sampling}). It consists of a \textit{node table} and a list of \textit{edge blocks} for each node. 

The node table is a contiguous array, with each entry corresponding to one node and a doubly-linked list of edge blocks that store all edges connected to that node.\footnote{Our system allows both undirected (Reddit~\cite{kumar2019predicting}, Netflix~\cite{bennett2007netflix})
and directed graphs (GDELT~\cite{zhou2022tgl}, MAG~\cite{zhou2022tgl}). A directed edge is stored with the source node while an undirected edge is stored in the edge blocks of both end nodes.} Each node entry contains pointers to the first and last edge blocks in the node's doubly-linked list, the number of edge blocks (list length), the total number of edges connected to the node (node degree in an undirected graph and out-degree in a directed graph) and a validity field indicating whether the node still exists. %(1)  or has been deleted (0). 
Adding a node is achieved by simply appending an entry to the table. Node deletion is indicated by setting the validity field of the node to 0.

The edge blocks are lightweight structures (72 bytes), %that store edge block metadata (e.g., block size) and pointers to the actual \textit{graph edge data} (e.g., neighbor ID, edge ID, timestamp). 
each storing \textit{graph metadata} about a group of edges, including 
the block size (maximal number of edges that can be contained in the block), the actual number of edges in the block, pointers to the \textit{graph edge data}, %(e.g., neighbor ID, edge ID, edge timestamp), 
and the earliest and latest timestamps of edges in the block. 
%The edge blocks do not store the actual graph edge data, but instead store indirections to the graph edge data. Therefore, we refer to it as \textit{graph metadata}.  
The graph edge data are lists of neighbor IDs, edge IDs, edge timestamps, and validity indicators. The same entry in the lists corresponds to one edge in the edge block: the respective neighbor ID is the ID of the neighbor to which the edge is connected, edge ID is the ID of the edge, the timestamp is when the edge is inserted, and the validity indicates if the edge is present or %has been 
deleted. The lists are organized in increasing order of edge timestamps. %It is worth noting that the graph metadata is typically several orders of magnitude smaller than the actual graph edge data (Table~\ref{tbl:adaptive-block-size}).  

Edge blocks at each node are chronologically ordered, with the oldest edges in the edge block at the list head and the newest edges in the edge block at the tail. %Within an edge block, edges are sorted chronologically. 
Chronologically ordering edge blocks and edges in each edge block brings two main advantages. %{\em First}, 
(1) It enables efficient temporal queries, such as finding all edges within a given time range, by sequentially scanning only a subset of the edge blocks rather than the entire list (\S\ref{sec:neighborhood-sampling}). Within each block, logarithmic-time neighborhood search is allowed over the sorted edges. %{\em Second}, 
(2) It enables efficient edge insertion: new edges can be appended to the tail of an existing edge block %(as new edges have larger timestamps) 
when the last block still has capacity, or added into a new block to be linked to the tail of the edge block list, without requiring costly re-sorting. %operations. Overall, this time-indexed approach allows good performance and scalability of the dynamic graph storage data structure.

For node/edge deletions, we %use tombstones to 
mark the corresponding validity fields %invalidated nodes and edges 
while preserving data layouts and pointers. Invalid nodes/edges are ignored during sampling.

%\vspace{1mm}
%\noindent\textbf{Offload old data.} 
In dynamic graph storage, continuous data addition may cause out-of-memory errors, while old data may no longer be relevant for training. We provide an API that can offload data older than a specified timestamp to a file. %In this way, we reduce memory pressure and ensure that only the most relevant data is kept in memory for training. %This approach not only addresses the problem of out-of-memory errors but also improves the efficiency of training by eliminating unnecessary memory usage.

%\vspace{1mm}
%\noindent\textbf
\vspace{1mm}
\noindent\textbf{Adaptive block sizing with threshold.} 
The edge block size in our dynamic graph data structure is a critical hyperparameter, balancing memory usage and graph operation efficiency. Small block sizes increase memory consumption and overhead due to more edge blocks per node and costly temporal neighborhood sampling as there are more blocks to traverse. Conversely, large block sizes lead to space wastage due to less densely filled edge blocks.

We adopt adaptive block sizing %with threshold 
for each node that adapts to the changing graph over time. %Due to the power-law nature of 
Many real-world graphs follow a power-law degree distribution, with most nodes having only a few neighbors and a small set of nodes densely connected~\cite{faloutsos1999power}. %Therefore, 
We set a threshold of node out-degree (decided using grid search in \S\ref{sec:ablation-study}):  %(out-degree in case of a directed graph) 
when the out-degree of a node is smaller than the threshold, the edge block size of the node is set to its out-degree; %when the node's out-degree exceeds the threshold, 
otherwise, the block size is set to the threshold. Therefore, the size of the allocated block for a node $v$ , denoted as $b_v$, is determined by $b_v=\min(deg(v), \tau)$, where $\tau$ is the threshold and $deg(v)$ is the node $v$'s current degree. For the majority of nodes whose neighbor numbers are small, setting an edge block size equivalent to the out-degree significantly reduces the length of the block list, and a new block's addition (with space reserved for future neighbors) does not incur excessive space overhead. % the capacity for storing edges can be directly doubled, without causing excessive memory usage. 
For nodes with many neighbors, adding a new block with a large block size may %that doubles the capacity of the linked list may %result in very large blocks, 
result in significant memory waste (when the new block is sparsely filled), and hence we limit its block size by the threshold. The upper bound of wasted space (i.e., not filled by edges) is less than half of the total number of edges in the graph. %which prevents the block size from exceeding a certain limit. 
By adaptively adjusting the block size based on each node's out-degree, %(subject to the threshold), 
we balance %the trade-off between 
memory consumption and graph update/query efficiency. % and can be applied to a wide range of real-world graphs with varying degrees of sparsity and density.
Note that the block size change applies only to new blocks to be appended, and the sizes of existing edge blocks remain unchanged.

% \noindent\textbf{Reallocate block.} For hub-vertex (i.e., vertex with many neighbors), the number of insertions and edges will be very large, which will lead to a very long edge block list, resulting in reduced query efficiency. Therefore, for these nodes, an alternative insertion strategy can be used, i.e., replacing the current edge block with a larger capacity edge block. This has the advantage of reducing the length of the linked list and increasing the efficiency of queries. The disadvantage is that it requires a copy of the original data when inserting, which causes a lot of overhead.
\subsection{GPU-based Graph Sampling}
\label{sec:neighborhood-sampling}

Algorithm~\ref{alg:temporal-neighborhood-sampling} shows the process of temporal neighborhood sampling on GPUs. Given a batch of target nodes and their time ranges, the algorithm traverses through each target node's linked list in parallel, skipping any edge blocks that do not coincide within the given time range. Within each edge block, if the minimum to maximum timestamps overlap with the specified time range, the algorithm employs a binary search on the candidate neighbors.  Subsequently, a predetermined number of these candidate neighbors are selected through a sampling method, such as recent sampling or uniform sampling (\S\ref{sec:preliminaries}).
For multi-hop neighborhood sampling, the process is repeated for each GNN layer, using the sampled nodes from one layer as the input to the next-layer sampling~\cite{zhou2022tgl}. Note that the system can also efficiently support other sampling methods. For example, we can support temporal random walk~\cite{nguyen2018continuous} by setting the neighborhood sample size of each layer to one~\cite{pandey2020c}.

\begin{algorithm}[t]
\caption{Temporal Neighborhood Sampling}
\label{alg:temporal-neighborhood-sampling}
\begin{algorithmic}[1]
    \State \textbf{Input}: block-based graph storage %$G$ 
    (node table $V$), target nodes $\mathbf{n}$, start timestamps $\mathbf{t_s}$, end timestamps $\mathbf{t_e}$, neighborhood sample size $f$
    \State \textbf{Output}: temporal neighbors for each $n \in \mathbf{n}$
    \State \textbf{parallel for} $n \in \mathbf{n}$, $t_s \in \mathbf{t_s}$, $t_e \in \mathbf{t_e}$ \textbf{do}
    \State \hspace*{\algorithmicindent} $\text{candidates} = []$
    \State \hspace*{\algorithmicindent} $lst = V[n]$ 
    \State \hspace*{\algorithmicindent} $b = lst.tail$  \COMMENT{the newest block}
    \State \hspace*{\algorithmicindent} \textbf{while} $b \neq null$ \textbf{do}
    \State \hspace*{\algorithmicindent}\hspace*{\algorithmicindent} \textbf{if} $t_e < b.t_{\text{min}}$:\quad $b = b.prev$ \quad \textbf{continue}
    \State \hspace*{\algorithmicindent}\hspace*{\algorithmicindent} \textbf{if} $t_s > b.t_{\text{max}}$:\quad \textbf{break}
    \State \hspace*{\algorithmicindent}\hspace*{\algorithmicindent} find neighbors within time window $[t_s, t_e)$ 
 using binary search in $b$ and add them to candidates
    \State \hspace*{\algorithmicindent}\hspace*{\algorithmicindent} $b = b.prev$ 
    \State \hspace*{\algorithmicindent}\textbf{end while}
    \State \hspace*{\algorithmicindent}sample $f$ neighbors from candidates 
    \State \textbf{end for}
\end{algorithmic}
\end{algorithm}

\vspace{1mm}
\noindent\textbf{Hybrid GPU-CPU graph data placement.} %memory layout
%Some previous works use GPUs to accelerate neighborhood sampling, placing the graph entirely on a GPU~\cite{yang2022gnnlab, jangda2021accelerating}. 
Due to the limited memory capacity of GPUs, the graph data are commonly stored in host memory. Substantial data copying overheads from CPU to GPU are incurred during GNN training, especially %poses a significant challenge for GNN applications, particularly when dealing with 
on dynamic graphs, which requires frequent access to graph data. %Dynamic graphs have many indirections, which can result in more accesses, such as 
For instance, during temporal neighborhood sampling (Algorithm~\ref{alg:temporal-neighborhood-sampling}), several edge blocks are accessed, followed by retrieving the graph edge data referenced by these edge blocks. Previous works have proposed hosting the graph in pinned host memory and using unified virtual addressing (UVA) to access necessary graph data~\cite{quiver}, which still requires frequent CPU-to-GPU copies. %and can result in performance degradation due to the large number of indirections involved in accessing the data. % may need some data here to show the impact of graph metadata on overall performance
%Despite the significant impact of graph metadata on overall performance, previous works have not adequately addressed the problem of managing graph metadata storage in a way that minimizes indirections and maximizes GPU memory usage. Therefore, 
%Some previous works use GPUs to accelerate neighborhood sampling, placing the graph 

We note that graph metadata (i.e., node table and edge blocks) is accessed more frequently than graph edge data: the former is first accessed before the latter are retrieved, and many edge data are skipped during neighborhood sampling (e.g., because they are not in the specified time range) while their metadata is accessed. We %propose a novel approach that
leverage the fact that graph metadata is usually several orders of magnitude smaller in size than the graph edge data (Table~\ref{tbl:adaptive-block-size}) and store graph metadata in global memory on GPUs to reduce frequent CPU-to-GPU copies, while placing the large graph edge data in host memory (Graph Store in Figure~\ref{fig:architecture}). In this way, we can significantly %improve the performance of 
accelerate GPU-based graph queries %such as neighborhood sampling 
(\S\ref{sec:neighborhood-sampling}). %particularly for dynamic graphs that require frequent accesses to graph metadata.

%Within a single multi-GPU server, each GPU is assigned to one worker process. 
On a multi-GPU machine, the graph edge data is shared among trainers through page-locked host memory mapped into the CUDA address space of each GPU, allowing direct access by any GPU across the PCIe bus. The trainer of rank 0 is responsible for updating the graph storage over time, %(node/edge insertion/deletion), 
while all trainers can query the graph data. %using GPU sampling (\S\ref{sec:neighborhood-sampling}).
By mapping page-locked host memory into CUDA, the edge data appears in local GPU memory for high-bandwidth, low-latency concurrent read accesses by multiple GPUs, while atomic updates from rank 0 are instantly propagated. %This enables fast data sharing and synchronization among GPUs on a multi-GPU machine. 

\vspace{1mm}
\noindent\textbf{Kernel optimizations.}
%Efficient sampling of neighbors of a target node is a critical operation in graph processing on GPUs. However, 
%Previous GPU-based graph sampling designs often suffer from warp divergence and inefficient memory access patterns
%with suboptimal performance.
We initiate a GPU thread to sample one neighbor of a given target node following previous works~\cite{yang2022wholegraph, jangda2021accelerating}. We sample neighbors of the same target node by adjacent threads, which helps to reduce warp divergence by minimizing the differences in the operations performed by each thread. We also ensure sampled results written to global GPU memory are consecutive to support coalesced access. To reduce GPU global memory access during sampling, each thread caches its currently accessed lightweight edge block (72 bytes) in its registers. For uniform sampling, we record candidate neighbors' positions in edge data lists before selection in GPU's shared memory to reduce GPU global memory access. With recent sampling, we directly select the latest candidates.

\subsection{Feature Cache on GPUs}
\label{sec:feature-fetching}
\begin{figure}[t]
    \centering
    \includegraphics[width=0.9\columnwidth]{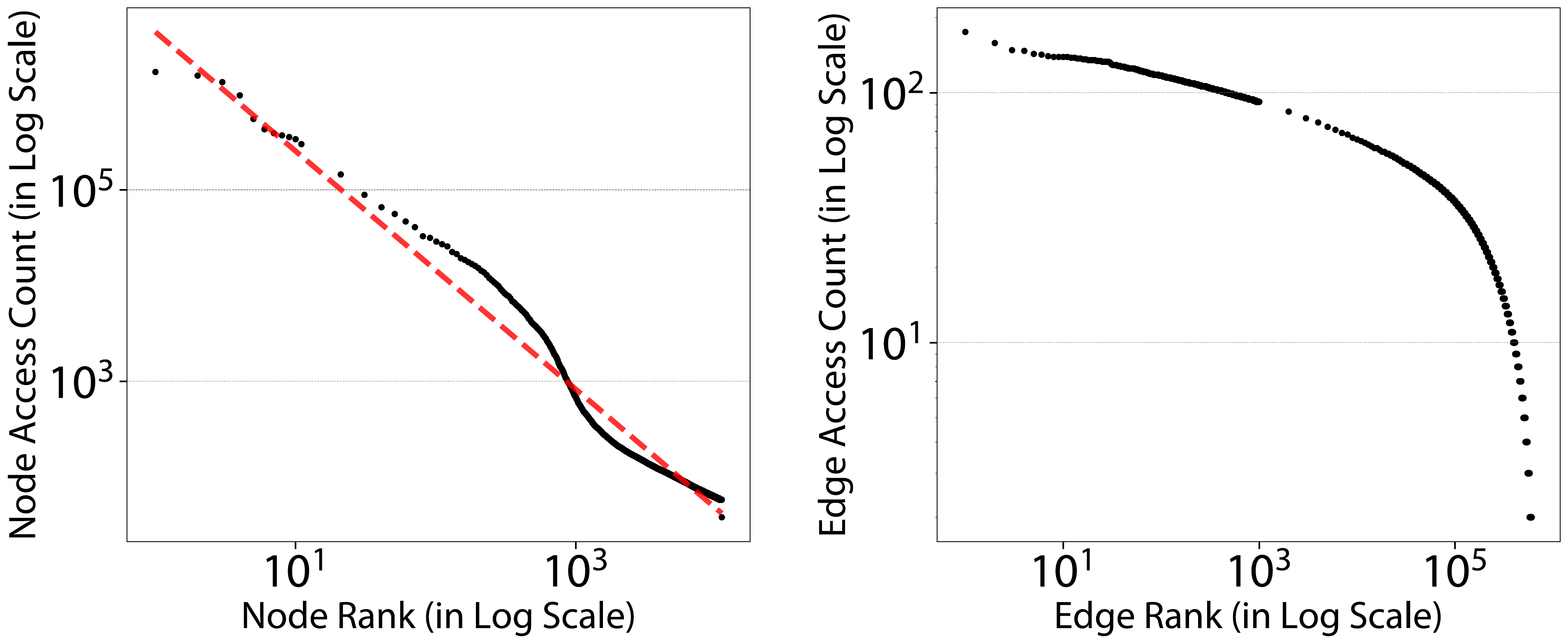}
    \caption{Distributions of node and edge accesses when training TGAT on Reddit. Red dashed line    denotes linear fitting.}
     \label{fig:node-edge-access-pattern}
\end{figure}

\begin{table}[t]
    \centering
    \caption{Jaccard similarity of the set of all sampled nodes/edges in adjacent retraining rounds of TGAT model.}
    \vspace{-3mm}
    \begin{tabular}{ccccc}
        \toprule
         \small{Dataset} & \small{Reddit~\cite{kumar2019predicting}} & \small{GDELT~\cite{zhou2022tgl}} & \small{Netflix~\cite{bennett2007netflix}} \\ 
        \midrule 
        \small{Node} & 99.5\% & 95.0\% &  87.5\% \\ 
        \small{Edge} & 91.9\%  & 27.6\% &  29.3\%  \\ 
        \bottomrule
    \end{tabular}
    \label{tab:jaccard_similarty}
\end{table}

\vspace{1mm}
\noindent\textbf{Observation.} 
We discover a notable amount of duplication in the nodes and edges sampled between adjacent continuous learning rounds. Table~\ref{tab:jaccard_similarty} displays the Jaccard similarity~\cite{jaccard1912distribution} of the sets of sampled nodes and edges for each dataset between adjacent continuous learning rounds. We observed a high similarity of 99.5\% in the sets of nodes, whereas the similarity in the sets of edges was lower, possibly due to the higher number of edges compared to nodes. These results indicate that there is an opportunity for data reuse. 

%For efficient caching of node/edge features on GPU, 
Moreover, we analyze node and edge access patterns when training GNNs. %different GNN models on representative graph datasets. 
Figure~\ref{fig:node-edge-access-pattern} shows the distributions of node/edge access counts when training TGAT % with pre-sampling %\cwu{clarify what the pre-sampling does, and what sampling alg. is used} 
on Reddit~\cite{rossi2021temporal}. We observe that the node access follows a power-law distribution, with a few nodes frequently retrieved and the majority of nodes accessed much less. This %observation 
can be attributed to the common structure of real-world graphs, where a small number of  ``hub'' nodes %often referred to as ``hubs'', 
are more connected and thus more likely to be accessed during training. %due to their higher impact on the overall graph structure. 
% %This observation 
It also helps to explain the success of the static node feature cache adopted in PaGraph~\cite{lin2020pagraph} (caching nodes with the highest out-degrees) and GNNLab~\cite{yang2022gnnlab} (caching a small subset of frequently accessed nodes). % achieving a high cache hit rate). %However, this observation does not hold for edges, as 
However, the edge access follows an exponential distribution, with most edges accessed to some extent. This pattern makes static cache unsuitable as it cannot effectively capture the wide range of edge accesses. %observed in the edge distribution, resulting in a lower cache hit rate and reduced efficiency.
We observe similar node/edge access patterns when training other temporal GNNs (e.g., TGN) on other datasets (including Wikipedia~\cite{rossi2021temporal}, MOOC~\cite{kumar2019predicting}, LastFM~\cite{kumar2019predicting} and GDELT). When training GNN models for static graphs (e.g., GraphSAGE), the edge access obeys a skewed distribution. 

Based on these observations, we advocate utilizing GPU-based dynamic feature caching (such as LRU~\cite{o1993lru}, LFU~\cite{matani20211}, FIFO) for continuous learning, and fully exploiting data reuse through inter-round cache reuse and intra-round cache restoration.

\vspace{1mm}
\noindent\textbf{Cache reuse and restoration.}
First, we exploit the inter-round feature fetching duplication (Table~\ref{tab:jaccard_similarty}) by recording the cache contents (e.g., to disk) at the end of each round and loading the cache into GPU memory at the beginning of the next round. This reuse avoids high cache miss rates caused by cold cache startups and avoids long cache initialization times caused by static cache methods.
Second, a retraining round typically requires multiple epochs to converge, and each epoch traverses the training samples in strictly temporal order (\S\ref{sec:preliminaries}). However, at the beginning of the second epoch, the dynamic cache has already been replaced with newer node/edge features, which can be considered ``polluted'', resulting in a large number of cache misses. To counter this, we save the cache contents (for example, to CPU memory) at the start of each round, and restore the cache at the beginning of every epoch. This technique improves the cache hit rate at the start of each new epoch.

\vspace{1mm}
\noindent\textbf{Vectorized cache.}
In each training iteration, feature fetching often requires accessing features of thousands of nodes and edges, with no particular order among them. Typical dynamic cache implementation is inherently inefficient, not able to handle batched accesses in parallel. For example, LRU cache usually assumes a single access at a time, updating cache entries one by one~\cite{o1993lru}. The standard doubly-linked list structure and locking mechanism required for thread safety in LRU cache may severely hinder efficiency~\cite{fan2013memc3}.

We design a highly efficient vectorized cache that enables parallel access and update of cache entries in batches. For example, in our vectorized LRU Cache, we store a ``recent score'' for each cache entry in a vector. The scores represent how recently each entry was last accessed, with 0 meaning most recently. Initially, all recent scores are 0. Upon each update, all recent scores are decremented by 1. An access resets the corresponding entry’s recent score to 0. For replacement, the entries with the lowest recent scores are evicted using an efficient vectorized top-k operation.
For vectorized LFU cache, we adopt a similar approach by changing the recent score to an access count. The vectorized FIFO cache is implemented using a ring buffer, which maintains a pointer to the most recently cached node or edge features. During each batch update, the pointer only needs to be moved by the number of entries that are to be replaced in the ring buffer. 

In addition, sometimes a large number of items need to be replaced in a dynamic cache (e.g., LRU), resulting in most or even all cache items being replaced. This could potentially lead to cache thrashing, where items are frequently replaced. We limit each cache update to at most a fraction $\lambda$ of its capacity ($0 < \lambda <= 1$). This strategy can help maintain the stability of the cache, avoid frequent replacement of cache items, and thus maintain a higher cache hit rate. In subsequent experiments, we set $\lambda = 0.2$ by default.

\subsection{Distributed Training on Partitioned Graphs}
\label{sec:distributed_training}

% \noindent\textbf{Distributed feature store.} 

\vspace{1mm}
\noindent\textbf{Distributed storage.}
\label{sec:partition} When graph data exceeds a single machine's RAM capacity, partitioning is needed. We use the edge-cut model~\cite{malewicz2010pregel, low2014graphlab}, storing each node and its associated edges and features on one machine. Edge-cut is favored over vertex-cut~\cite{dbh, petroni2015hdrf}, as it reduces cross-machine communication while vertex-cut distributes node neighbors to multiple machines and causes high latency for sampling due to multi-machine access. Existing edge-cut partitioning methods, such as the widely used METIS~\cite{metis1998}, are primarily designed for static graphs and not suitable for dynamic graphs due to high re-computation cost (\S\ref{sec:limitation_of_existing_frameworks}). Some online partitioning methods (e.g., Fennel~\cite{Streaming:fennel14wsdm}) aim at node balance but not edge balance among the partitions. Achieving edge balance  %with respect to the edges 
is important, %as node balance, 
to balance computation %(graph sampling) 
and communication %(feature fetching) 
workloads in distributed GNN training. %and uneven workloads can lead to slower training %times, 
%and decreased resource efficiency. %and increased communication overhead. 
We utilize a simple yet effective online hash partitioning method that assigns each node to a partition (machine) %based on a hash function. Specifically, the partition that a node is assigned to is 
of index ${hash}({n}) \% P$, where $n$ is the node ID and $P$ is the number of machines. We use an identity hash (i.e., $hash(x) = x$) due to %since it provides a straightforward mapping without any additional 
computation simplicity %low computation overhead. This approach leads to a more edge-balanced partition, 
and that it evenly distributes nodes and their associated edges across the machines~\cite{gandhi2021p3}. Since node IDs and their connectivity are typically %evenly distributed or 
randomly distributed across the graph, %\cwu{not clear why the identity hash can evenly distributed edges across machines since it only works on node ID}, 
this leads to more edge-balanced partitions~\cite{barabasi1999emergence}. %based on their IDs, which helps to avoid clustering and improve load balancing~\cite{gandhi2021p3}. 
%resulting in faster and more evenly distributed workloads. The balance of graph partitioning is essential for load balancing during training, as it affects both the computation and communication aspects of the process. By adopting a hash partitioning technique, we can achieve a more balanced partition with respect to the edges, leading to faster and more efficient training.

\begin{figure}
    \centering
    \includegraphics[width=0.9\columnwidth]{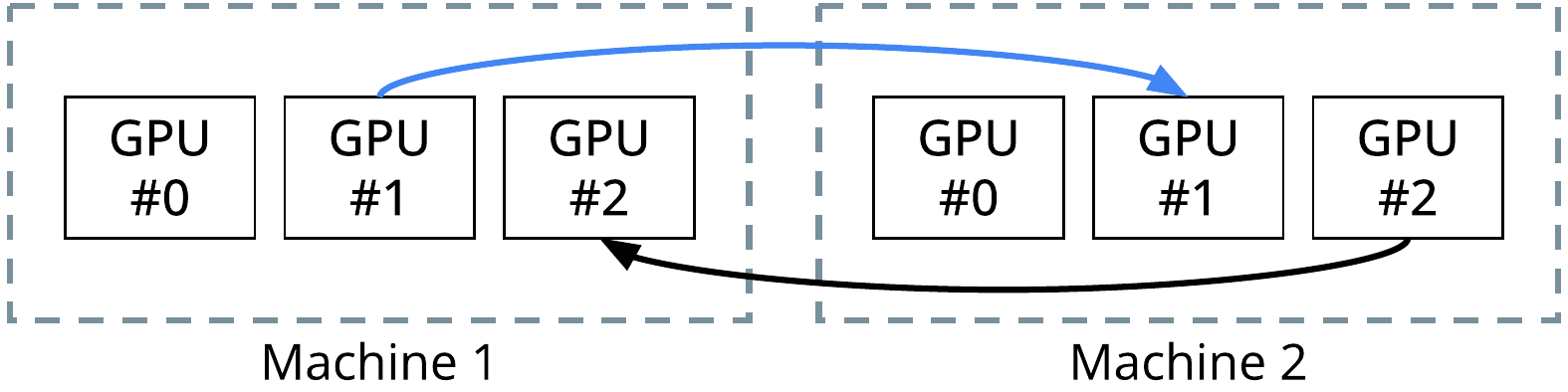}
    \caption{Static scheduling of distributed sampling } 
     \label{fig:static-scheduling}
\end{figure}

\vspace{1mm}
\noindent\textbf{Distributed sampling.}
To ensure balanced workloads among trainers, we may need to assign some trainers to nodes outside their local graph partition. In these cases, we use remote neighborhood sampling on machines that store the relevant nodes. When a trainer needs to perform multi-hop sampling beyond the local machine, remote sampling is needed. To manage these requests, we propose a static scheduling approach (shown in Figure~\ref{fig:static-scheduling}) that assigns GPUs on a machine to handle remote sampling requests from trainers. When a trainer needs to sample a remote target node, they send an RPC request to the machine that stores the node, targeting the GPU with the same rank as the trainer's local GPU. We empirically validate the effectiveness of this approach by measuring the average coefficient of variance (CV) of sampling times for temporal and static GNN models (TGN, TGAT, DySAT, GraphSAGE, and GAT) on the GDELT graph using four AWS g4dn.metal instances. The CVs are very low (less than 0.06), indicating a good load balance.

\vspace{1mm}
\noindent\textbf{Distributed feature store.} Inside each machine, node/edge features and node memories are stored in shared host memory for all trainers to share. We store node features and memories using a key-value map (Python dictionary) mapping node IDs to feature vectors. New edges have larger IDs, and edge features are stored in ascending order. This enables efficient edge feature queries using PyTorch's optimized binary search function \texttt{searchsorted}.
% motivation

%% file: implementation.tex
\section{Implementation}

We implemented \sysname{} in 8,400 lines of code (LoC) in C++ and Python. The dynamic graph storage data structure was implemented in 2,000 LoC of C++. GNN models, GPU feature cache, and distributed module (including online graph partitioning, distributed sampling, and remote feature fetching) were implemented in 5,000 LoC of Python. We also implemented CUDA kernels for temporal neighborhood sampling %(\S\ref{sec:neighborhood-sampling}) 
using 400 LoC of CUDA. We implement various static and temporal GNN models  on DGL 0.9.1 with PyTorch 1.13 as the underlying deep learning framework. 
We rely on PyTorch's RPC module~\cite{li13pytorch} with the TensorPipe backend~\cite{tensorpipe} to implement the distributed module, and use NCCL and Gloo for AllReduce operations of model gradient synchronization.

\vspace{1mm}
\noindent\textbf{User interface.} %We provide Python APIs for manipulating dynamic graphs and invoking temporal neighborhood sampler. 
We provide easy-to-use Python APIs for developers. Figure~\ref{fig:code-example} shows how developers can create and use dynamic graphs, temporal neighborhood samplers, and feature caches for training. 

\begin{figure}[t]

\begin{Verbatim}[commandchars=\\\{\}, fontsize=\footnotesize]
\PYG{k+kn}{import} \PYG{n+nn}{gnnflow} \PYG{k}{as} \PYG{n+nn}{gf}

\PYG{k}{def} \PYG{n+nf}{train\PYGZus{}step}\PYG{p}{(}\PYG{n}{model}\PYG{p}{,} \PYG{n}{edges}\PYG{p}{,} \PYG{n}{dgraph}\PYG{p}{,} \PYG{n}{sampler}\PYG{p}{,} \PYG{n}{cache}\PYG{p}{):}
  \PYG{c+c1}{\PYGZsh{} neighborhood sampling}
  \PYG{n}{mfgs} \PYG{o}{=} \PYG{n}{sampler}\PYG{o}{.}\PYG{n}{sample}\PYG{p}{(}\PYG{n}{edges}\PYG{p}{)}
  \PYG{c+c1}{\PYGZsh{} fetch node/edge features with cache}
  \PYG{n}{cache}\PYG{o}{.}\PYG{n}{fetch\PYGZus{}features}\PYG{p}{(}\PYG{n}{mfgs}\PYG{p}{)}
  \PYG{c+c1}{\PYGZsh{} GNN training}
  \PYG{n}{output} \PYG{o}{=} \PYG{n}{model}\PYG{p}{(}\PYG{n}{mfgs}\PYG{p}{)}
  \PYG{o}{...}

\PYG{n}{dgraph} \PYG{o}{=} \PYG{n}{gf}\PYG{o}{.}\PYG{n}{DynamicGraph}\PYG{p}{()}
\PYG{n}{model} \PYG{o}{=} \PYG{n}{gf}\PYG{o}{.}\PYG{n}{models}\PYG{o}{.}\PYG{n}{TGN}\PYG{p}{()}
\PYG{n}{sampler} \PYG{o}{=} \PYG{n}{gf}\PYG{o}{.}\PYG{n}{TemporalKhopSampler}\PYG{p}{(}\PYG{n}{dgraph}\PYG{p}{,} \PYG{n}{fanouts}\PYG{o}{=}\PYG{p}{[}\PYG{l+m+mi}{10}\PYG{p}{,} \PYG{l+m+mi}{10}\PYG{p}{])}
\PYG{n}{cache} \PYG{o}{=} \PYG{n}{gf}\PYG{o}{.}\PYG{n}{Cache}\PYG{p}{(}\PYG{n}{node}\PYG{o}{=}\PYG{l+s+s1}{\PYGZsq{}LRU\PYGZsq{}}\PYG{p}{,} \PYG{n}{node\PYGZus{}cache\PYGZus{}ratio}\PYG{o}{=}\PYG{l+m+mf}{0.01}\PYG{p}{,}
                 \PYG{n}{edge}\PYG{o}{=}\PYG{l+s+s1}{\PYGZsq{}LRU\PYGZsq{}}\PYG{p}{,} \PYG{n}{edge\PYGZus{}cache\PYGZus{}ratio}\PYG{o}{=}\PYG{l+m+mf}{0.01}\PYG{p}{)}

\PYG{c+c1}{\PYGZsh{} update the graph}
\PYG{n}{new\PYGZus{}edges} \PYG{o}{=} \PYG{n}{gf}\PYG{o}{.}\PYG{n}{get\PYGZus{}ingestion\PYGZus{}batch\PYGZus{}edges}\PYG{p}{()}
\PYG{n}{dgraph}\PYG{o}{.}\PYG{n}{add\PYGZus{}edges}\PYG{p}{(}\PYG{n}{new\PYGZus{}edges}\PYG{p}{)}
\PYG{o}{...}
\PYG{c+c1}{\PYGZsh{} a typical training loop}
\PYG{k}{for} \PYG{n}{batch\PYGZus{}edges} \PYG{o+ow}{in} \PYG{n}{data\PYGZus{}loader}\PYG{p}{:}
  \PYG{n}{train\PYGZus{}step}\PYG{p}{(}\PYG{n}{model}\PYG{p}{,} \PYG{n}{batch\PYGZus{}edges}\PYG{p}{,} \PYG{n}{dgraph}\PYG{p}{,} \PYG{n}{sampler}\PYG{p}{,} \PYG{n}{cache}\PYG{p}{)}
  
\end{Verbatim}
\caption{An example to demonstrate \sysname{}'s API}
\label{fig:code-example}
\end{figure}

\vspace{1mm}
\noindent\textbf{Dynamic graphs.} We allocate shared host memory for graph edge data storage. %allowing multiple trainers to share the same graph. 
Each trainer registers the shared memory for use %by CUDA 
with \texttt{cudaHostRegister}. For adding edges in a distributed setting, there is a dispatcher that first partitions the edges according to the hash partitioning method (\S\ref{sec:partition}), and then uses PyTorch's asynchronous RPC to call the operation of adding edges on the local partition of each trainer.

 % Only one of the workers (i.e., local rank 0) modifies the graph edge data with graph updates, while all workers can simultaneously query the graph using GPU-based samplers (\S\ref{sec:neighborhood-sampling}).

\vspace{1mm}
\noindent\textbf{GPU-based sampling.} We pre-allocate CPU pinned memory buffers and GPU buffers for input target nodes and output sampled neighbors. For each sampling layer, we invoke custom CUDA kernels, use \texttt{thrust::remove\_if} to remove invalid neighbors, and copy results to the CPU. In static scheduling for distributed sampling (\S\ref{sec:neighborhood-sampling}), each GPU may handle multiple requests from other machines. Thus, we use a task queue. The distributed sampler adds a sampling task to the queue, which returns a handle. The sampler polls the handle, and if the task is complete, it returns the result.

%% file: evaluation.tex
\section{Evaluation}

\vspace{1mm}
\noindent\textbf{Testbed.} We use Amazon EC2 g4dn.metal instances for all experiments. Each instance has 8 NVIDIA 16GB T4 GPUs, 96 vCPU cores, 384 GB memory, and 100Gbps network interconnectivity.

\vspace{1mm}
\noindent\textbf{Datasets.} We study four real-world graphs and one synthetic graph detailed in Table~\ref{tab:datasets}. The Reddit dataset~\cite{kumar2019predicting} is a bipartite graph mapping interactions between users and subreddits on Reddit. The Netflix dataset~\cite{bennett2007netflix} is a bipartite network connecting users and movies, with edge weights indicating rating scores from the 2006 Netflix Prize competition. The GDELT dataset~\cite{zhou2022tgl} is a dynamic graph documenting global events from news articles, with nodes for actors and temporal edges for events. The MAG dataset~\cite{zhou2022tgl} is a citation network with temporal edges and node features. The LDBC Social Network Benchmark dataset~\cite{angles2020ldbc} is a large synthetic graph which is a list of events of members joining forums.

    \begin{table}[t]
        \centering
        \caption{Graph datasets. $|d_v|$ and $|d_e|$ are dimensions of node features and edge features, respectively. $t_{\text{span}}$ is the total time span of the temporal dataset, and $t_{\text{interval}}$ indicates the %time interval 
        minimum time granularity recorded in the dataset.} 
        \vspace{-3mm}
        \bgroup
        \def\arraystretch{1.2}%  1 is the default, change whatever you need
        \begin{tabular}{lllllll}
            \toprule
            \small{\textbf{Dataset}} & \small{$|\mathbf{V}|$} & \small{$|\mathbf{E}|$} & \small{$\mathbf{d_v}$} & \small{$\mathbf{d_e}$} & \small{$\mathbf{t_{\text{span}}}$} & \small{$\mathbf{t_{\text{interval}}}$} \\
        \midrule
        \small{Reddit~\cite{kumar2019predicting}} & \small{11K} & \small{672K} & \small{128} & \small{172} & \small{30 days} & \small{few secs.} \\
        \small{Netflix~\cite{bennett2007netflix}} & \small{978K} & \small{100M} & \small{768} & \small{128} & \small{7.2 years} & \small{1 day} \\ 
        \small{GDELT~\cite{zhou2022tgl}} & \small{17K} & \small{191M} & \small{413} & \small{186} & \small{5 years} & \small{15 mins.}\\ 
        \small{MAG~\cite{zhou2022tgl}} & \small{122M} & \small{1.3B} & \small{768} & - & \small{10 years} & \small{1 month} \\
        \small{LDBC~\cite{angles2020ldbc}} & \small{113M} & \small{5.1B} & - & - & \small{3 years} & \small{few secs.} \\ 
        \bottomrule
    \end{tabular}
    \egroup
    \label{tab:datasets}
\end{table}

\vspace{1mm}
\noindent\textbf{GNNs.} We %evaluate \sysname{} using 
train three temporal GNN models (TGN, TGAT, and DySAT) and two static GNN models (GraphSAGE and GAT), using default model settings in their respective papers. %Specifically, 
%The temporal GNNs use different sampling methods to incorporate temporal information, %into their models, with 10 neighbors sampled per layer. 
The GNNs sample two layers of neighbors with 10 neighbors per layer, except for TGN, which samples one layer, and GraphSAGE, which samples 15 neighbors for the first layer.

\vspace{1mm}
\noindent\textbf{Baselines.} We compare the performance of \sysname{} with TGL~\cite{zhou2022tgl} (not support multi-machine training on partition graphs) and DGL~\cite{zheng2020distdgl} (not support temporal GNNs). Note that we rewrite TGL’s graph construction code in C++ with multi-threading (128 threads) since the original Python implementation is very slow. We tune the baselines to their best configurations. As PlatoGL~\cite{platogl} is not open source and its implementation details are not disclosed in the paper, we are unable to conduct an end-to-end comparison of the system. Nonetheless, we compare some of its design points by implementing them in \sysname{} during the ablation study (\S\ref{sec:ablation-study}).  %TGL is a unified framework for training temporal GNNs, but does not support distributed training with graph partitioning on multiple machines. %limiting its scalability for large-scale graphs. 
%DGL is a deep learning library for static graphs that supports distributed training, but does not support training temporal GNNs. Both TGL and DGL use CPU sampling and do not employ caching mechanisms. 
 % and implement a TGL w/ cache version as a state-of-the-art baseline for a fair comparison. 
We also compare with various GPU-based sampling and caching baselines when evaluating individual components in \sysname{}.

\vspace{1mm}
\noindent\textbf{Training.} We perform temporal link prediction as the downstream task (i.e., predicting unseen links that will be formed in the future) in all experiments. % GDELT is trained on a single machine and distributed settings, and MAG and the synthetic graph are trained over multiple machines. 
For TGN, TGAT, and DySAT, we use a default per-GPU batch size of 4000, 600, and 600 edges, respectively. %in all experiments. %We use default batch size for all of our experiments. 
Following TGL, we use random chunk scheduling for TGN to learn inter-batch dependency in large batches. % mitigating the issue that a large batch size can lead to outdated memory and less accurate predictions~\cite{zhou2022tgl}. 
For GraphSAGE and GAT, the default per-GPU batch size is 1200. 
% When learning GraphSAGE and GAT on DGL, we use a larger per-GPU batch size of 2400 since there is no GPU sampling nor GPU cache and the GPU memory is sufficient.
% (Do we need to say these baselines can't modify the graphs here?)
%During the end-to-end training, 
By default, we use vectorized LRU cache (\S\ref{sec:feature-fetching}) in \sysname{}, 
and cache 3\% node features and 3\textperthousand\quad edge features (due to the higher number of edges) in each GPU's memory for GDELT and Netflix, and 1\% node features for MAG.

\subsection{Continuous GNN Learning}
\label{sec:continuous_exp}
\begin{figure}[t]
    \centering
    \begin{subfigure}[t]{\columnwidth}
    \centering
    \includegraphics[width=0.8\columnwidth]{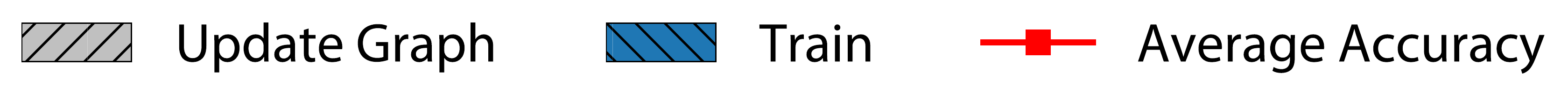}
    % \vspace{-5mm}
    \end{subfigure}
    \hfill
    \begin{subfigure}[t]{0.23\textwidth}
    \centering
    \includegraphics[width=\columnwidth]{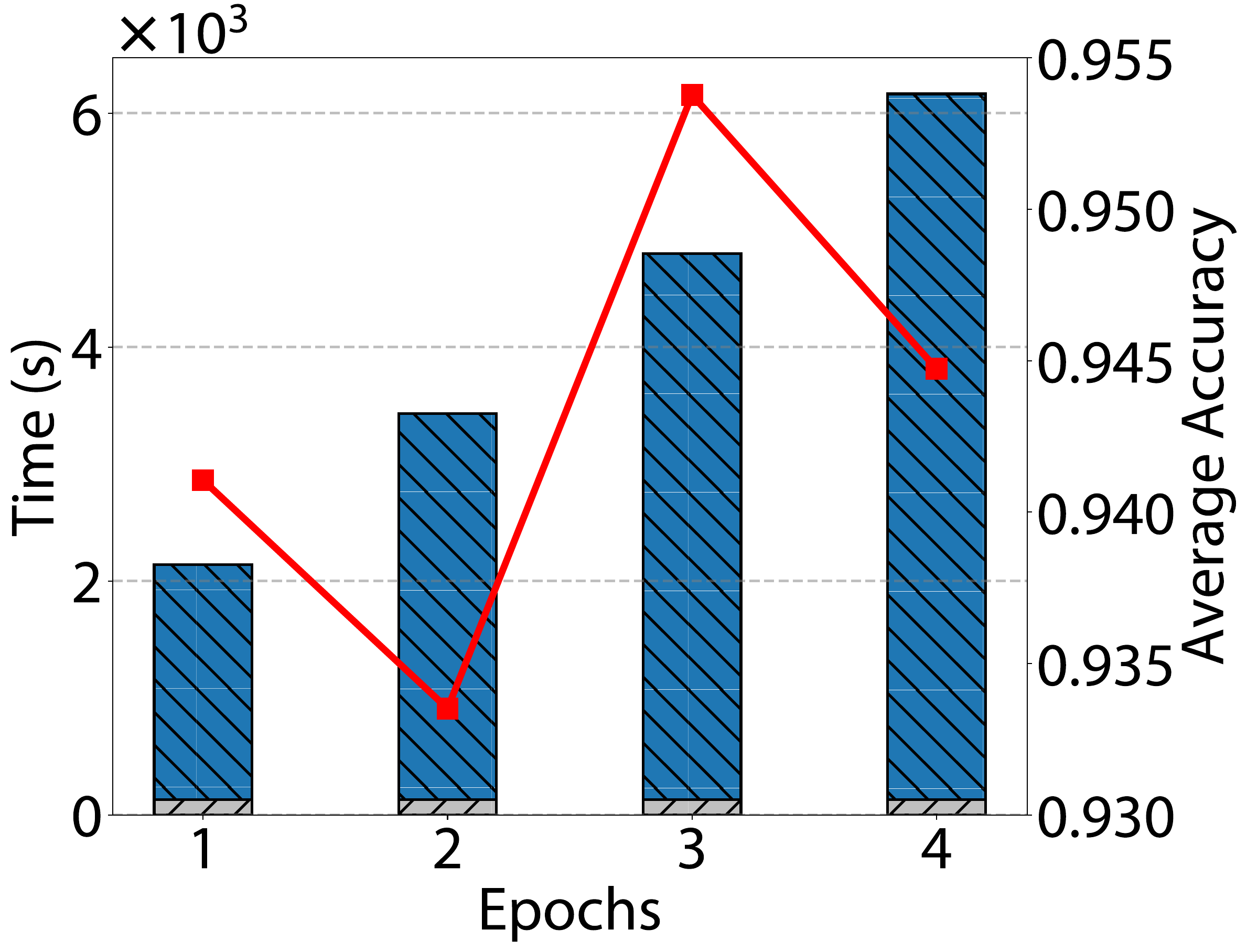}
     \caption{TGN}
    \end{subfigure}
    \hfill
    % \hspace{-4mm}
    \begin{subfigure}[t]{0.23\textwidth}
    \centering
    \includegraphics[width=\columnwidth]{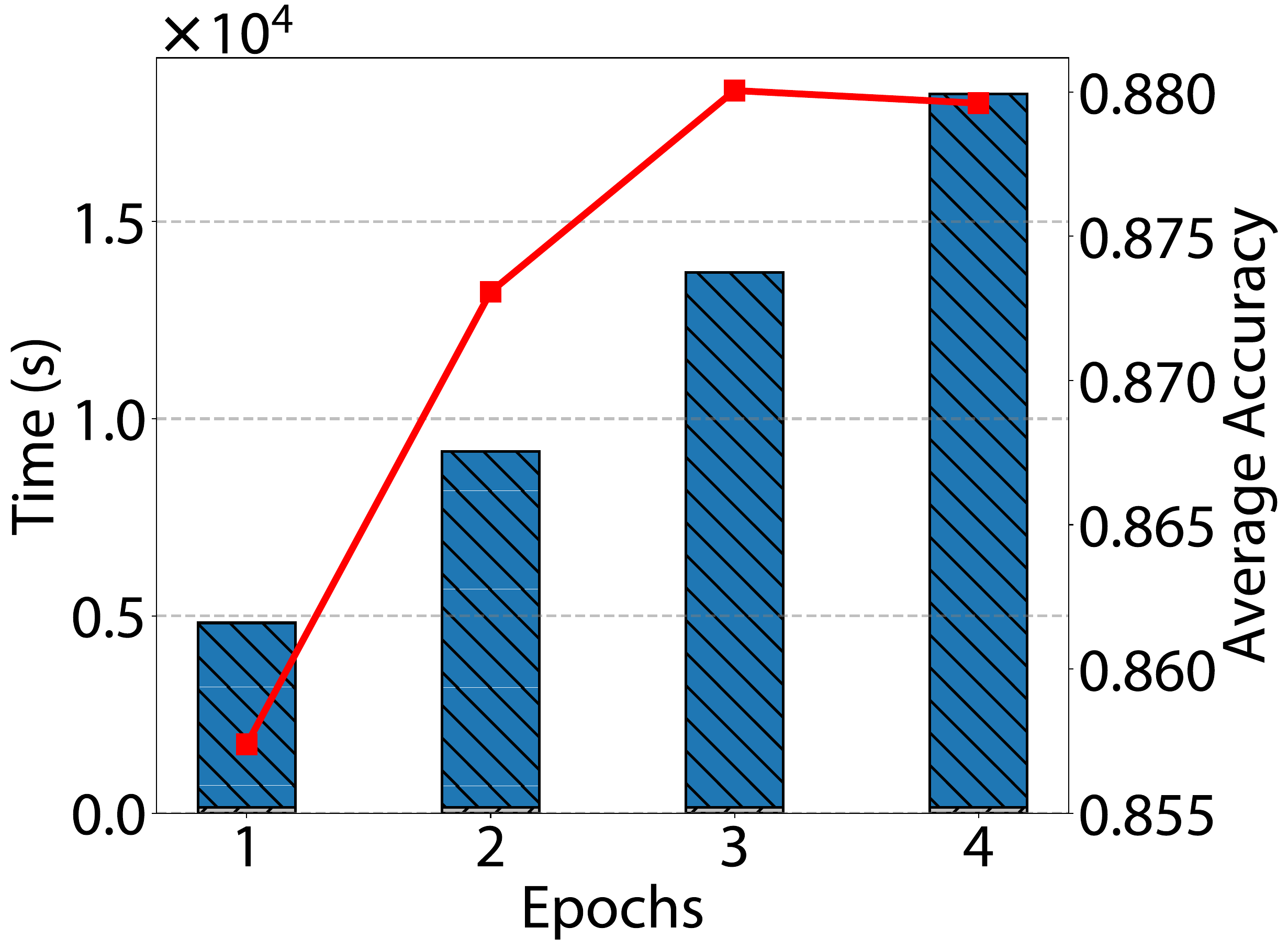}
     \caption{TGAT}
    \end{subfigure}
    \caption{Impact of the number of finetuning epochs.} 
    \label{fig:continuous_learning_epoch}
\end{figure}

\vspace{1mm}
\noindent\textbf{Methodology.}
We first evaluate \sysname{}'s continuous training on dynamic graphs, comparing with TGL (with our improvement of multi-thread graph building) on 8 GPUs. We use the first 30\% (in chronological order) of a dataset to train an initial model and the remaining 70\% for continuous streaming updates, which are divided into incremental batches. On each new batch, we first compute the accuracy (i.e., average precision) of the current model on the new data, and then use the new data for model finetuning. We finetune the GNN model periodically (\S\ref{sec:overview}), finetuning the model for three epochs on new data upon each incremental batch (with an experience replay ratio of 0), in this experiment. Three epochs are the sweet point we found for balancing accuracy and training cost (Figure~\ref{fig:continuous_learning_epoch}).  %every $t$ time, depending on the time unit of the dataset). 
We set the time interval $t$ for retraining to 1 day. %120 dayss and 42 days for learning over GDELT and Netflix, respectively. 

\begin{figure}[t]
    \centering
    \begin{subfigure}[t]{\columnwidth}
        \centering
        \includegraphics[scale=0.15]{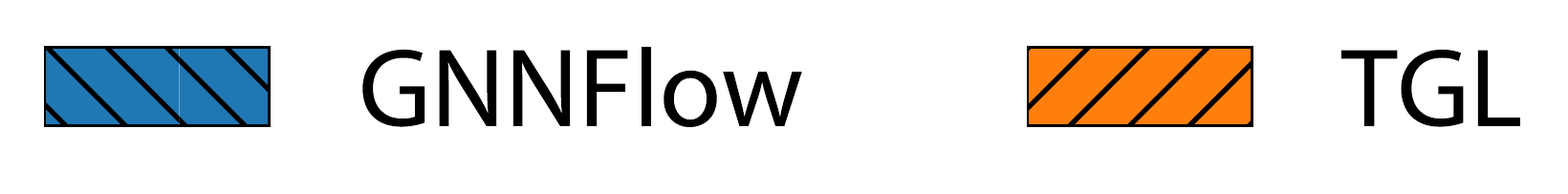}
        % \vspace{-5mm}
    \end{subfigure}
    \hfill
    \begin{subfigure}[t]{0.23\textwidth}
        \centering
        \includegraphics[width=\columnwidth]{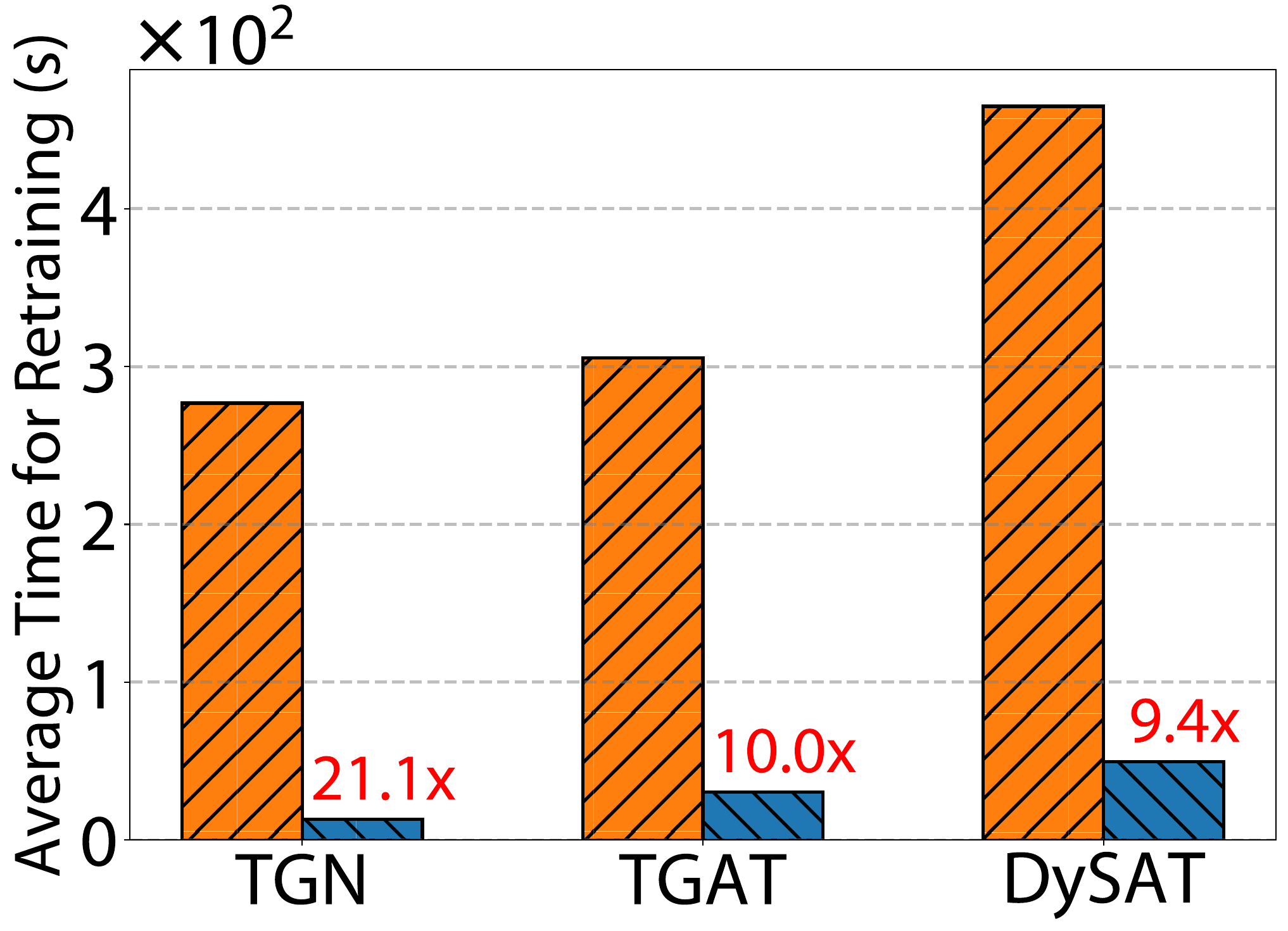}
        \caption{GDELT} 
        % \vspace{-5mm}
    \end{subfigure}
    \hfill
        \begin{subfigure}[t]{0.23\textwidth}
        \centering
        \includegraphics[width=\columnwidth]{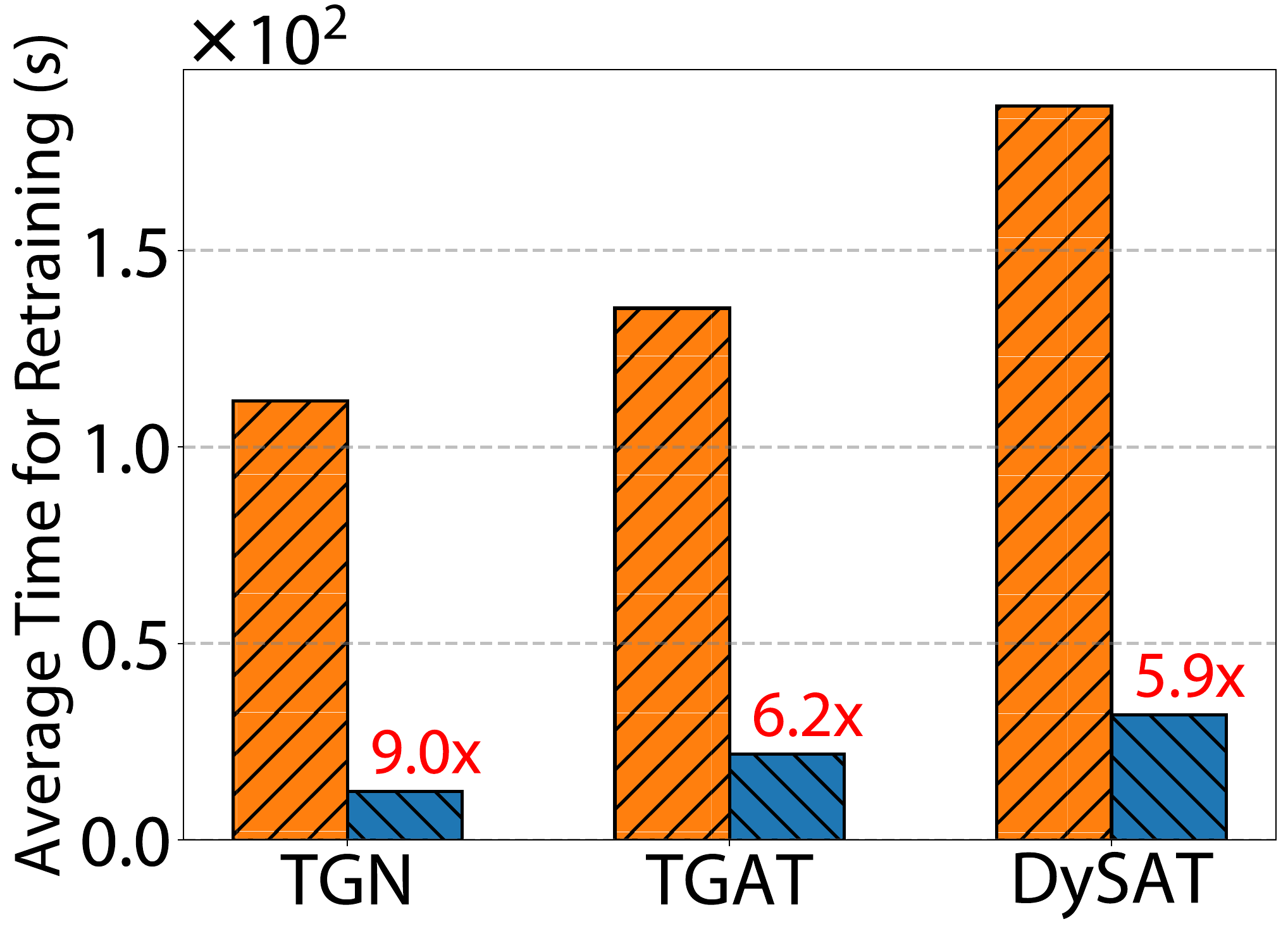}
        \caption{Netflix} 
        % \vspace{-5mm}
    \end{subfigure}
   \caption{The average retraining time of continuous GNN learning per day for \sysname{} and TGL.}
     \label{fig:continuous_per_day_cost}
\end{figure}

\vspace{1mm}
\noindent\textbf{Overall performance.} We measure the average total time of continuous learning 
upon each incremental batch (i.e., graph building/update time and model finetuning time).  Figure~\ref{fig:continuous_per_day_cost} demonstrates that \sysname{} achieves 9.4x to 21.1x faster retraining than TGL on the GDELT graph and 5.9x to 9.0x speed-up on the Netflix graph. %when TGL and \sysname{} are retrained the same number of times. 
The main time consumption of TGL lies in graph reconstruction, taking on average 170.8s and 94.3s for GDELT and Netflix, respectively, accounting for up to 36.7\% and 61.3\% of the total time upon one day's events. In contrast, GNNFlow only requires an average of 0.12s and 0.56s to update GDELT and Netflix graphs, with an average insertion of 96K and 130K edges. Therefore, GNNFlow can complete graph updates and training in half a minute, achieving real-time learning; whereas TGL, due to the need for graph reconstruction, requires at least several minutes for three epochs of finetuning.

\begin{figure}[t]
    \centering
    \begin{subfigure}[t]{\columnwidth}
    \centering
    \includegraphics[scale=0.16]{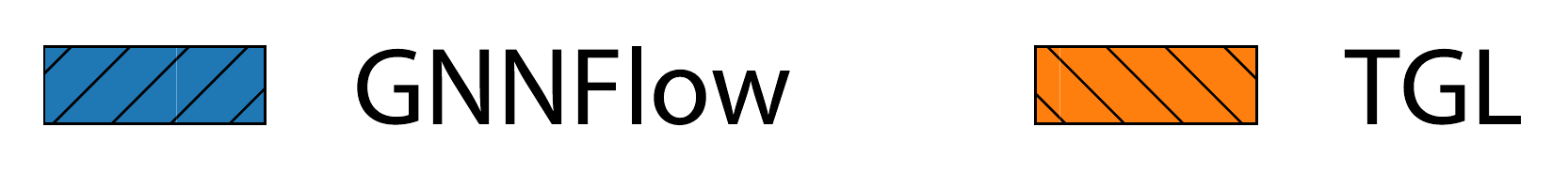}
    % \vspace{-5mm}
    \end{subfigure}
    \hfill
    \begin{subfigure}[t]{0.23\textwidth}
    \centering
    \includegraphics[width=\columnwidth]{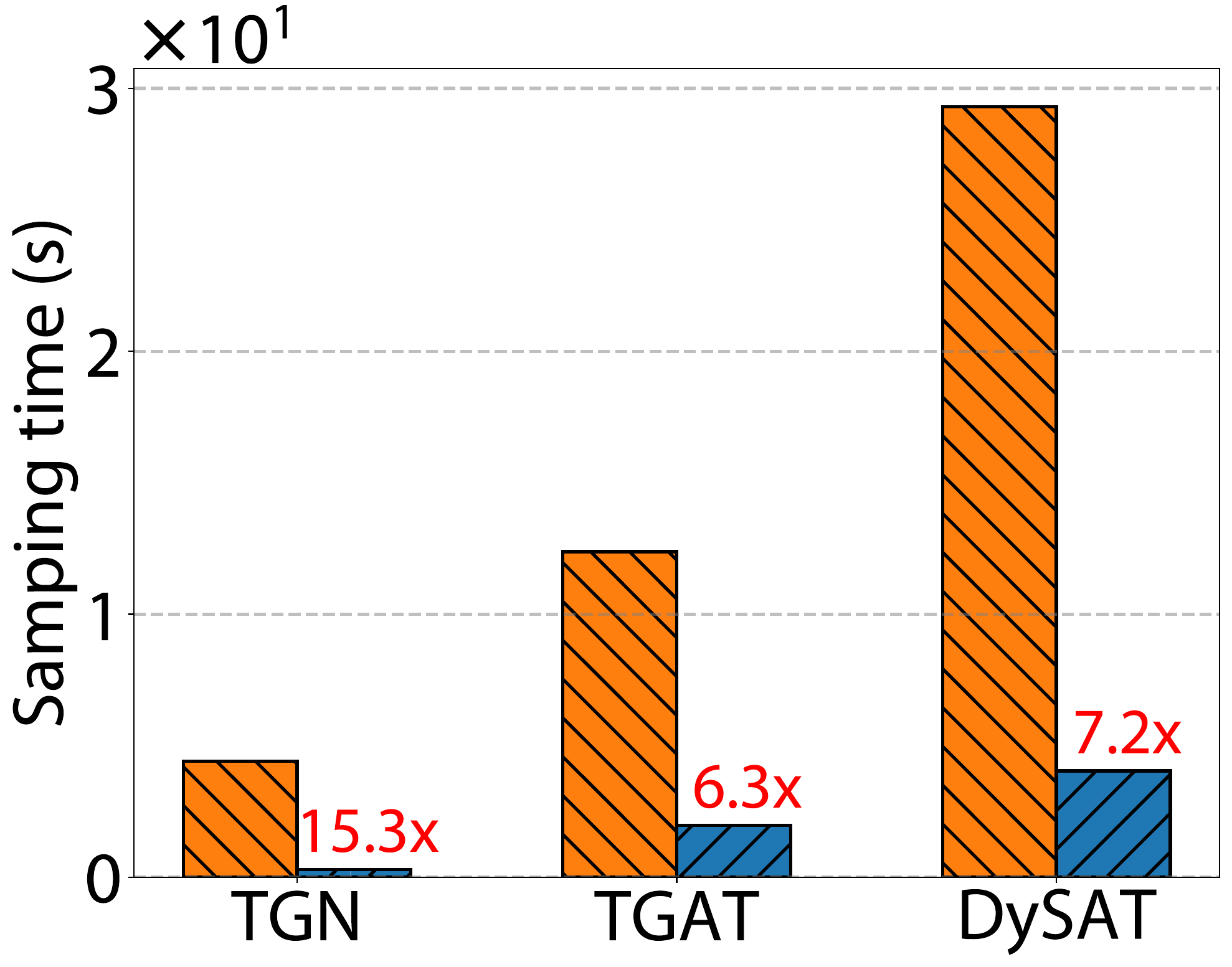}
     \caption{Sampling speed-up}
     \label{fig:sample_time}
    \end{subfigure}
    \hfill
    % \hspace{-4mm}
    \begin{subfigure}[t]{0.23\textwidth}
    \centering
    \includegraphics[width=\columnwidth]{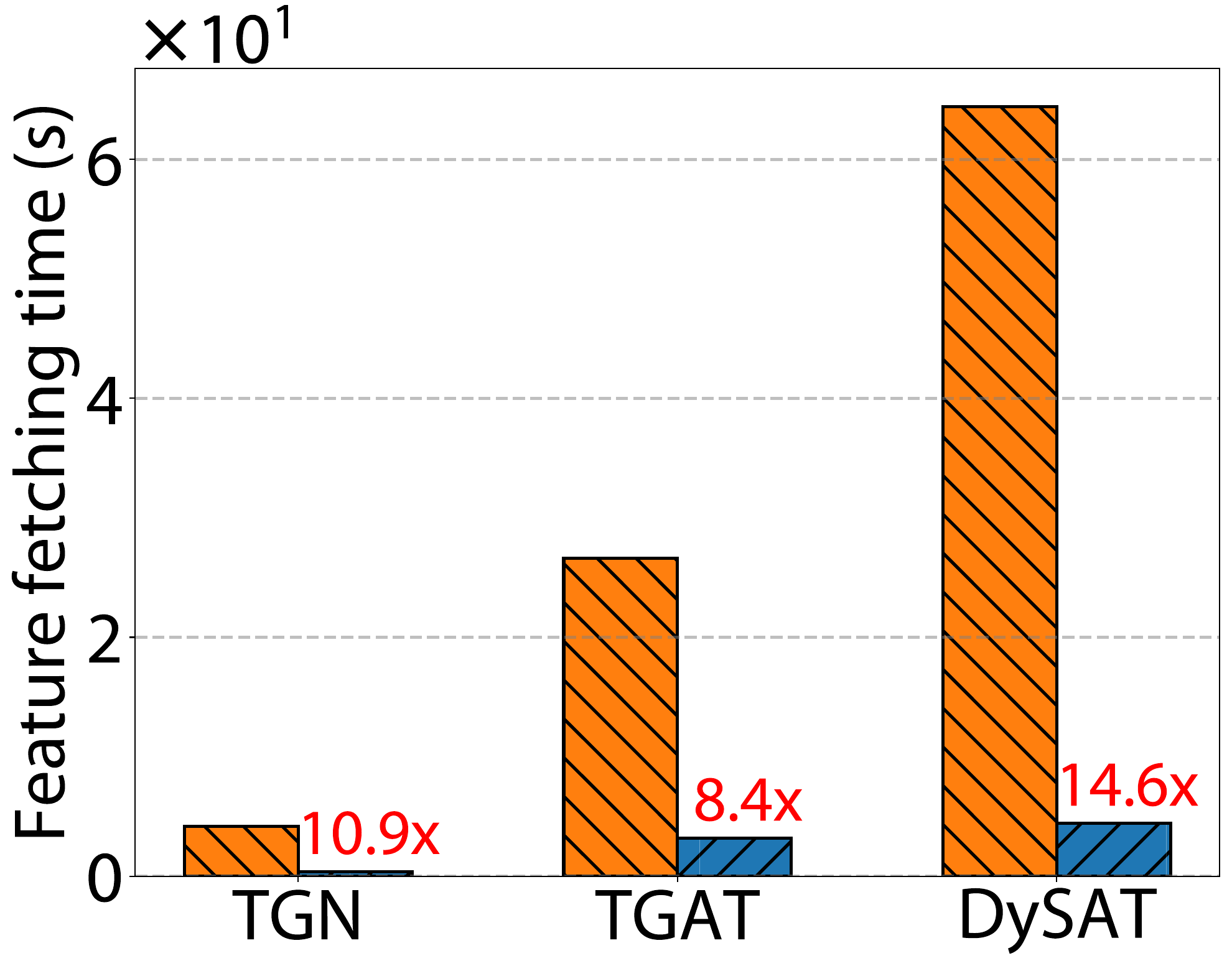}
     \caption{Feature fetching speed-up}
     \label{fig:feature_time}
    \end{subfigure}
    \caption{Averaged sampling and feature fetching time in continuous training on GDELT.} 
    \label{fig:time_speedup}
\end{figure}

\vspace{1mm}
\noindent\textbf{Performance breakdown.} %\yzhong{to show performance of graph update, sampling, feature fetching}
% Figure~\ref{fig:continuous_breakdown} presents a breakdown of the continuous training time, in terms of the average graph building time and average training time in each training epoch.
 %That's because, whenever new edges are inserted, TGL has to rebuild the whole graph and \sysname{} only need to update the dynamic graph in milliseconds.   
Figure~\ref{fig:time_speedup} further compares the average time taken by temporal neighborhood sampling and feature fetching during the model finetuning phase on GDELT. Our GPU-based temporal sampling (\S\ref{sec:neighborhood-sampling}) achieves $6.3\times$ - $15.3\times$ speed-ups on the three temporal GNN models. %Larger sampling speed-ups are achieved on TGAT and DySAT because they adopt two-layer sampling with larger computational efforts. %The effectiveness of GPU temporal sampling increases when the number of sampling layers increases with a heavier sampling workload. %\cwu{clarify which models sample more layers}. %As for snapshot-based temporal sampling \cwu{snapshot-based temporal sampling has never been mentioned before}, \sysname{} can outperform TGL by $2.73\times$. %During the end-to-end training, we use LRU cache with caching 3\% edge features and all node features on the GPU memory to accelerate the feature fetching. 
%Figure~\ref{fig:feature_time} shows that 
In addition, our GPU-based vectorized LRU feature cache (\S\ref{sec:feature-fetching}) decreases the feature fetching time by up to $14.6\times$. These results validate the benefit of our GPU optimization for sampling and feature fetching.

\vspace{1mm}
\noindent\textbf{Impact of finetuning epoch number.} 
We investigate the impact of the number of finetuning epochs upon each incremental batch, on model accuracy and finetuning time. We increase the finetuning epochs from 1 to 4 when continuously training TGN and TGAT on GDELT with \sysname{}. We record the %average accuracy (i.e., the 
average of all test accuracies on all batches and the overall time taken by updating the graph and finetuning.    %in continuous GNN learning using \sysname{} 
% \cwu{clarify how you do the continuous GNN learning with GNNFlow that produces the results here}.
%In Figure~\ref{fig:continuous_learning_epoch}, the average accuracy is obtained on all streaming batches %\cwu{should the accuracy be obtained on streaming batches that have arrive so far instead of all?}, 
%and the graph update and training times are sumed over all trainging times. %\cwu{per epoch?}. 
In Figure~\ref{fig:continuous_learning_epoch}, increasing finetuning epochs %for continuous GNN learning 
results in diminishing returns on accuracy but substantial increases in finetuning time. %For TGN, accuracy improves from 0.93 to 0.95 when increasing epochs from 2 to 3 but does not increase further with 4 epochs, while training time grows from 2008 to 4666 seconds. For TGAT, accuracy rises from 0.86 to 0.88 across the same epoch range, but training time escalates from 4676 to 18088 seconds. 
%\cwu{explain why accuracy drops when increasing epochs from 3 to 4 and from 1 to 2 for TGN finetuning - potentially due to overfitting on the limited new data in each streaming batch?} 
The accuracy drops when increasing the epoch number from 3 to 4 for TGN and TGAT, %and from 1 to 2 for TGN finetuning - 
potentially due to overfitting on the limited new data in each incremental batch.
The results suggest that 2-3 finetuning epochs per incremental batch may strike a good balance in continuous GNN learning. %in practice, achieving solid model accuracy with feasible training costs. Using more epochs leads to marginal accuracy benefits but unrealistic time requirements, potentially due to overfitting on the limited new data in each streaming batch.

\begin{figure}[t]
    \centering
    \begin{subfigure}[t]{0.235\textwidth}
    \centering
    \includegraphics[width=\columnwidth]{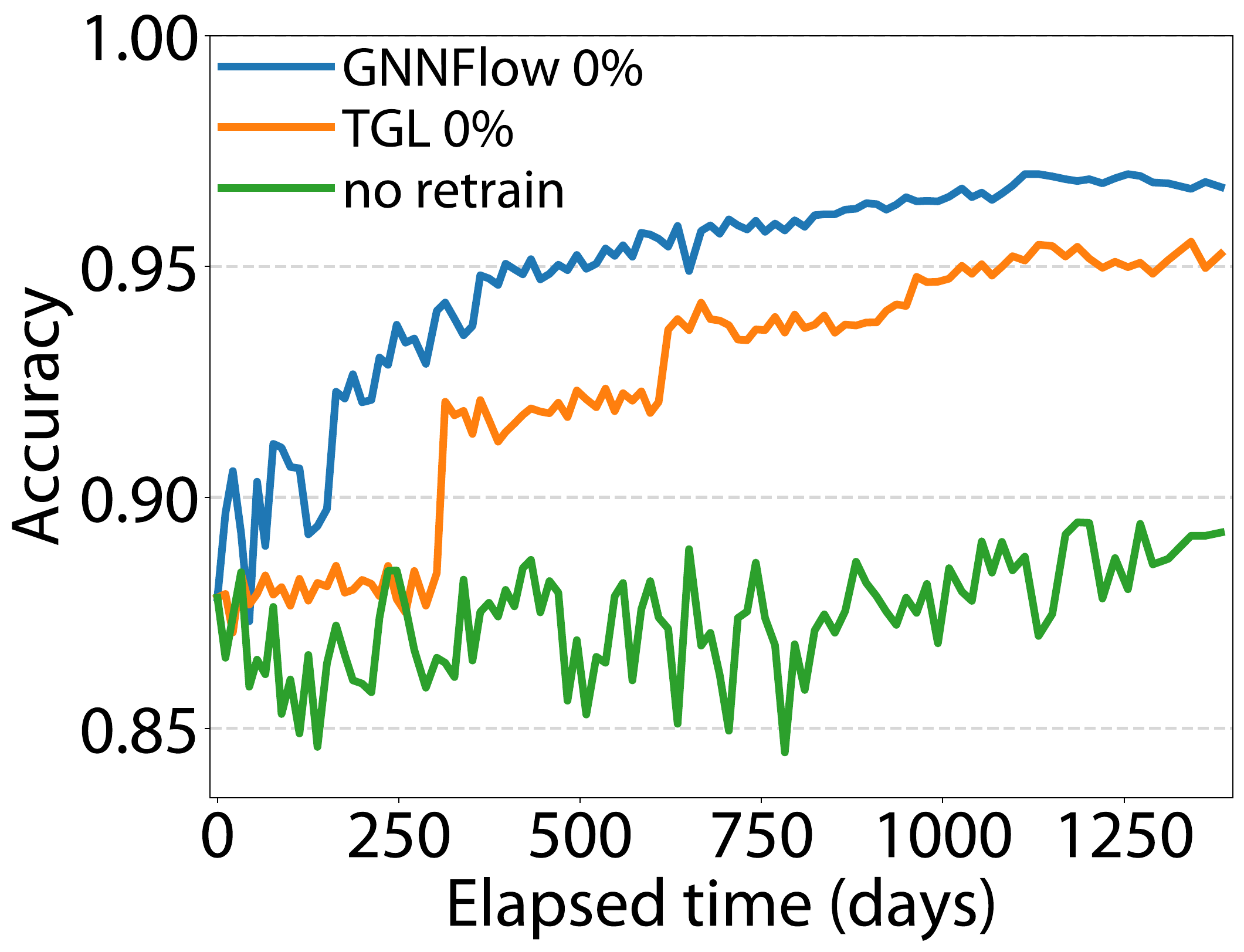}
     \caption{TGN}
     \label{fig:online_accu_tgn}
    \end{subfigure}
    \hfill
    \begin{subfigure}[t]{0.235\textwidth}
    \centering
    \includegraphics[width=\columnwidth]{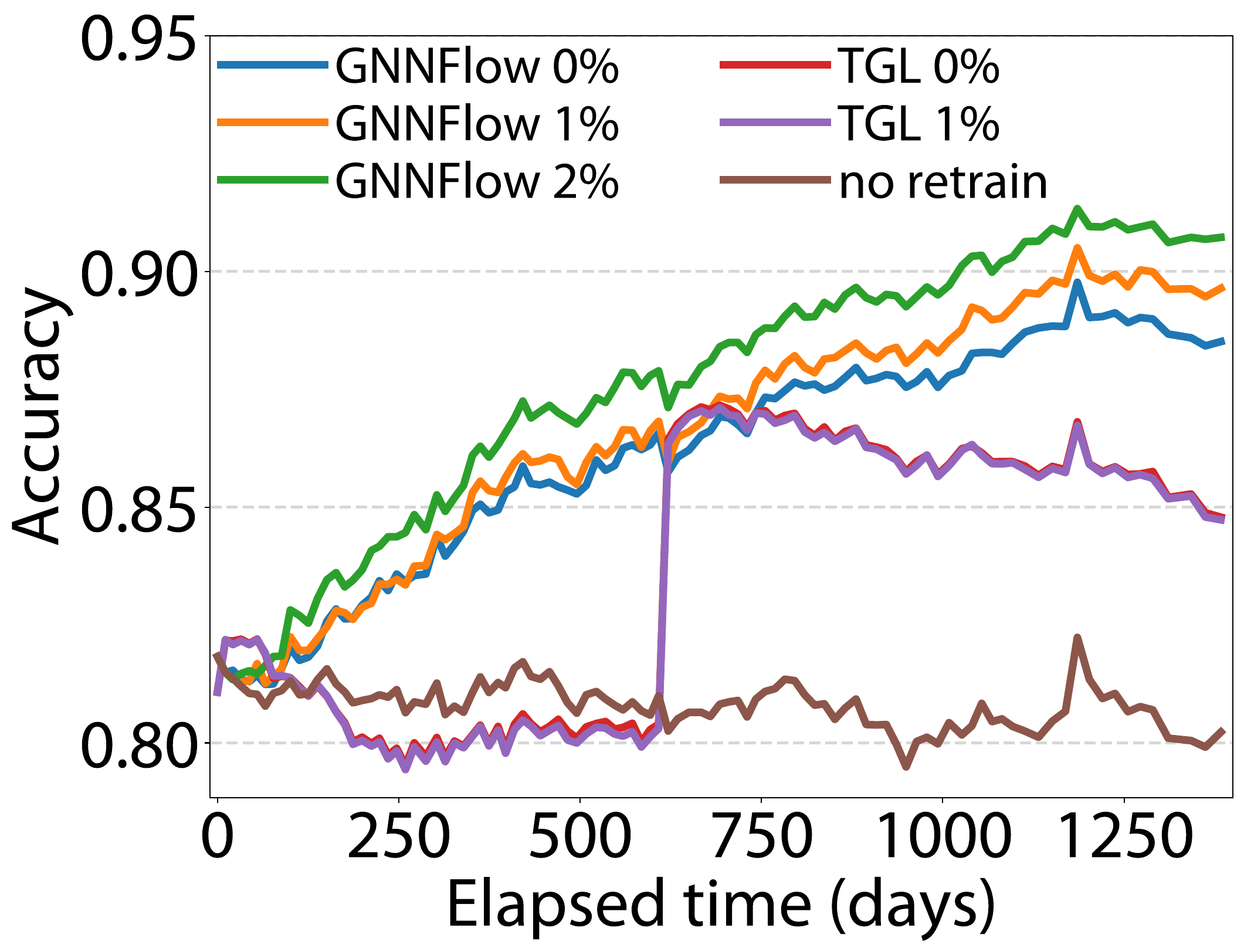}
     \caption{TGAT}
     \label{fig:online_accu_tgat}
    \end{subfigure}
    \caption{Model accuracy during continuous training. The elapsed time %of the x-axis
    refers to the time in the GDELT dataset.}
    \label{fig:online_accu}
\end{figure}

\vspace{1mm}
\noindent\textbf{Accuracy.}
We further study how the retraining frequency and the experience replay ratio influence model accuracy. %In order to see the fine-grained test accuracy, 
%Based on previous results, 
We observe that the total time needed for \sysname{} to finetune TGN (TGAT) for one epoch upon each incremental batch over the course of 100 incremental batches is roughly the same as that for TGL to retrain the model for one epoch once every 25 (TGN) or 50 (TGAT) incremental batches. %each time and retrain the model 4 (2) times to complete 
%for the 100 batches. %TGN (TGAT) continuously is roughly the same as that for TGL to train 25 (50) batches each time and retrain the model 4 (2) times to complete the 100 batches.
%perform continuous learning on each streaming batches (100 in total) with \sysname{} and record the time usage $t_2$. With higher graph reconstruction latency and slower training speed, TGL is unable to update the graph and retrain the model using the same frequency as \sysname{} in $t_2$. We find that TGL can achieve comparable time (still larger) with $t_2$ when using the training interval of 25 batches and 50 batches for TGN and TGAT respectively.
%Whenever a new batch stream in, we first update the graph and validate the average precision (i.e. AP) of the whole batch. The time for validation is excluded and the build graph time will only be counted when the system performs training on this batch. 
Figure~\ref{fig:online_accu} shows the accuracy tested on each incoming incremental batch on the current model, when \sysname{} is retrained on each batch and TGL is retrained per 25 and 50 batches for TGN and TGAT, respectively. %It should be noted that the x-axis elapsed time corresponds to the time within the GDELT dataset. 
The percentage in the legend indicates the experience replay ratio. The `no retrain' case gives the performance of the model trained using the first 30\% of data, tested on each incremental batch. 
Being able to retrain the model more frequently within the same retraining time budget (implying the same amounts of resources), the accuracy results of TGN and TGAT trained with \sysname{} outperform those of TGL on all batches, with a gap of up to 7.2\% and 9.0\%, respectively. %The AP of \sysname{} increases gradually with the streaming batches while the ap of TGL experience a leap in each training because the model get retrained with all batches. However, the performance of the models in TGL may stay unchanged or even get worse before next retraining. Besides, 
Figure~\ref{fig:online_accu_tgat} shows that a higher replay ratio results in better model accuracy because it avoids forgetting important information and thus performs well. The sudden jump in accuracy for TGL is because TGL underwent retraining at that time. % on both old and new data.   %\cwu{why experience replay helps if the model is tested only on new incoming data?} %but it may cause inevitable training overhead. 
%\sysname{} supports frequent dynamic graph updates and model retraining, achieving better model accuracy. %, which is considered a powerful platform for continous learning.

% Figure~\ref{fig:online_accu}
% figure. explain how we conduct the experiment

\subsection{Individual Components}
\label{sec:ablation-study}

\begin{figure}[t]
    \centering
    \begin{subfigure}[t]{\columnwidth}
    \centering
    \includegraphics[scale=0.12]{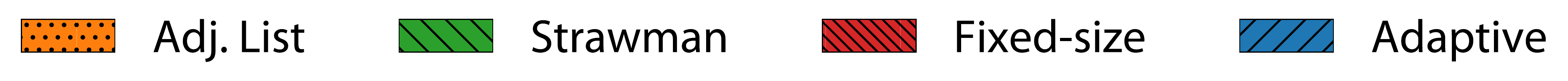}
    \end{subfigure}
    \hfill
    \begin{subfigure}[t]{0.23\textwidth}
    \centering
    \includegraphics[width=0.9\columnwidth]{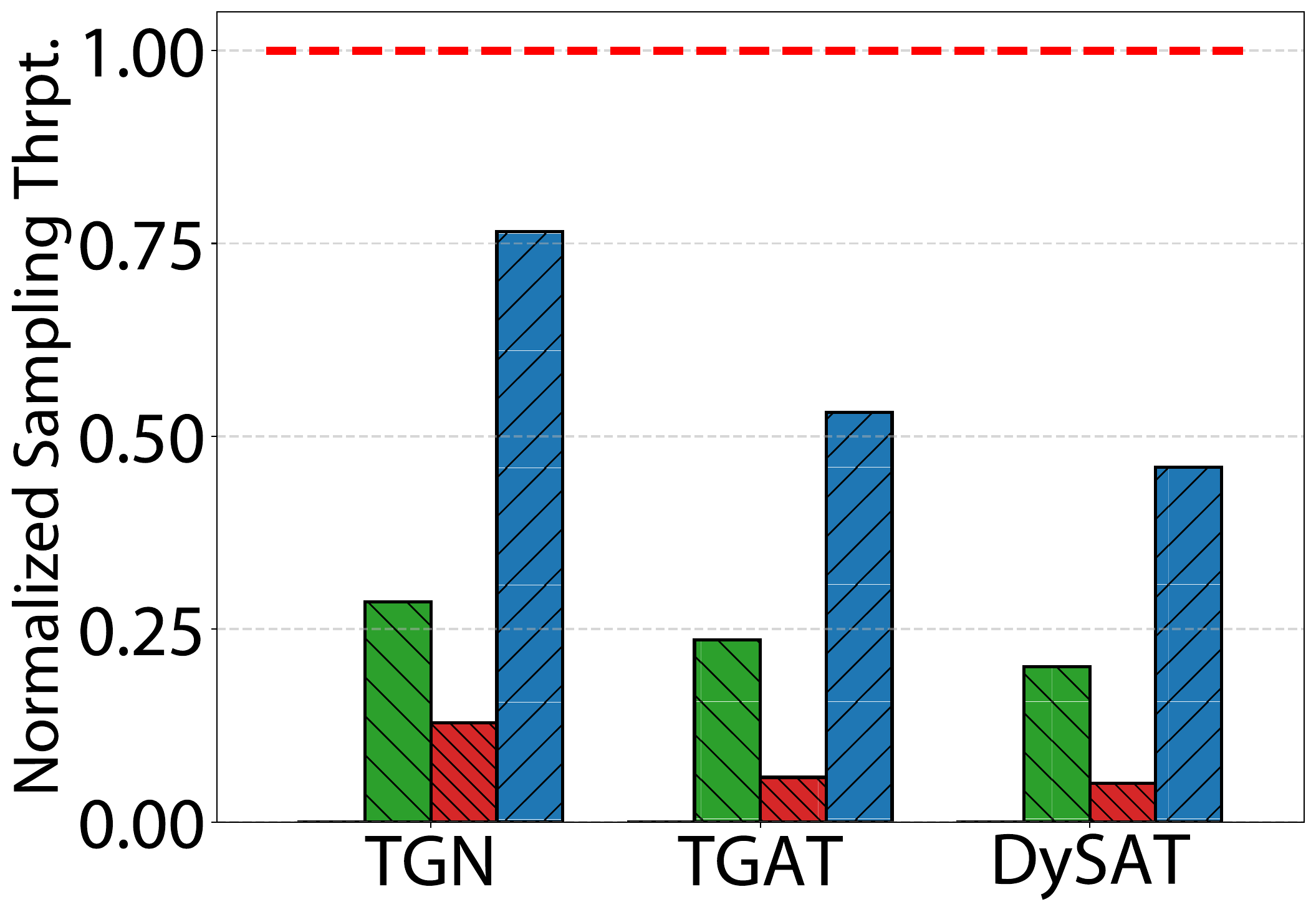}
     \caption{GDELT}
    \end{subfigure}
    \hfill
    \begin{subfigure}[t]{0.23\textwidth}
    \centering
    \includegraphics[width=0.9\columnwidth]{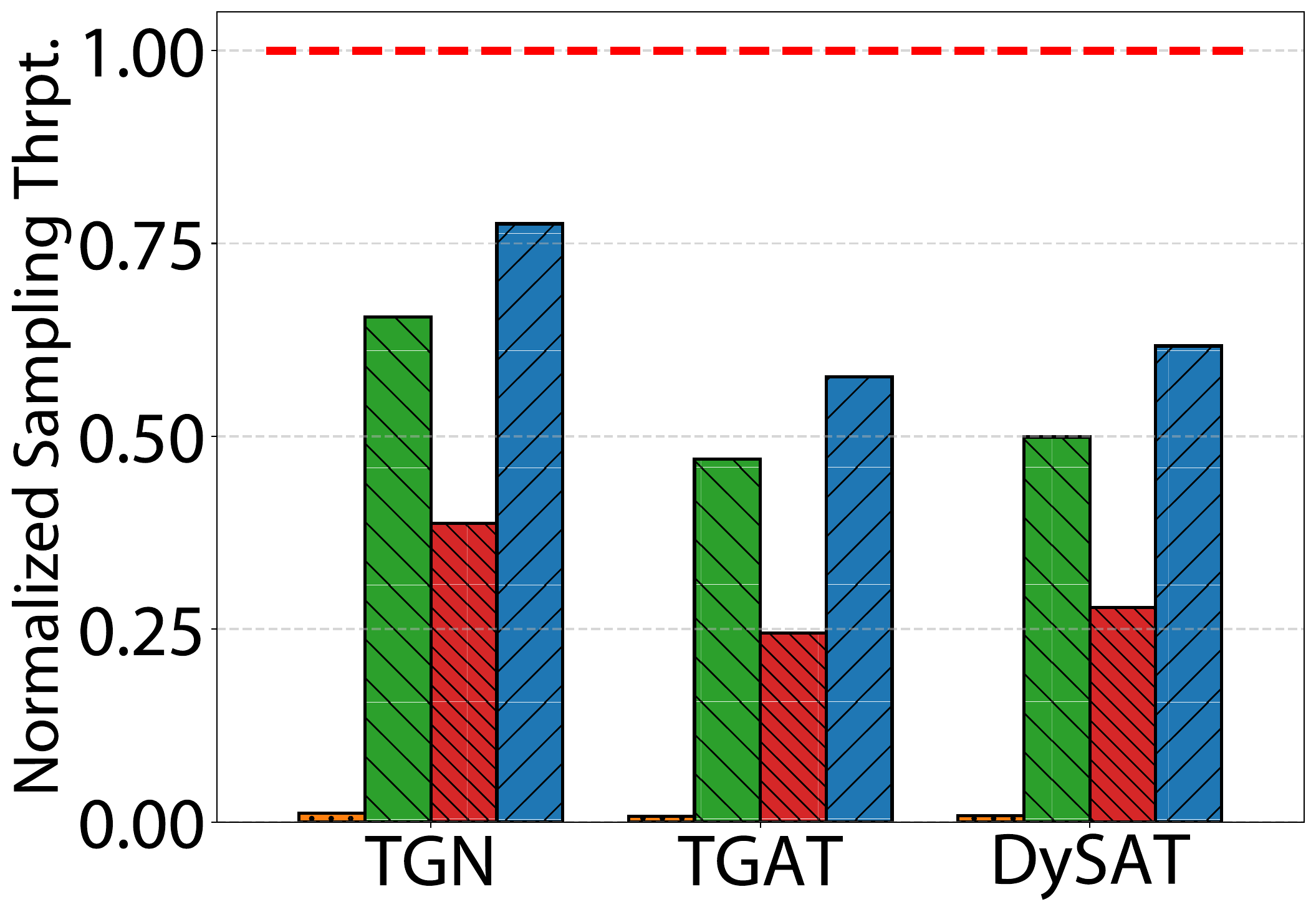}
     \caption{Netflix}
    \end{subfigure}
    \caption{Normalized sampling throughput (w.r.t. static system) for different block sizing methods. %on the GDELT and Netflix graphs.
    }
    \label{fig:adaptive_block_size_ablation}
\end{figure}

\vspace{1mm}
\noindent\textbf{Adaptive block sizing with threshold.} We compare the sampling performance of our adaptive block sizing (\S\ref{sec:data-structure}) against several baselines: (1) adjacency list; (2) a strawman approach, which sets a new block's size to the number of new edges to add in each incremental batch; (3) the fixed-size approach used in PlatoGL~\cite{platogl}, assigning each block a fixed size. We also give the results for a static system, where the complete graph is constructed all at once, resulting in an adjacency array.  We use grid search to find the appropriate size/threshold for the fixed-size approach and our method, leading to a graph edge data memory overhead of around 5\% (as compared to graph edge data memory usage of the static system). For GDELT and Netflix, the fixed sizes are 1024 and 48, respectively, and the thresholds in our method are 8192 and 48, respectively. In this experiment, we build each graph from scratch on one machine by injecting the graph dataset in batches with a batch size of 100,000 edges. 

We observe from Table \ref{tbl:adaptive-block-size} that using an adjacency list can result in very long linked lists, and requires storing a large amount of graph metadata (i.e., pointers point to next or previous adjacent neighbor, etc.). Compared to the strawman approach, our adaptive block sizing reduces the final linked list length (averaged among nodes) by about 36.7x with only less than 5\% more memory usage for graph edge data and reduced memory for graph metadata (due to reduced length of the linked list). With the fixed-size method, %also reduces the average linked list length, but 
nodes with many neighbors may have very long linked lists. Our adaptive method uses larger block sizes for these nodes, which reduces their list length. %As a result, as shown in 
Figure \ref{fig:adaptive_block_size_ablation} shows that the performance of sampling using adjacency lists is extremely poor (almost 0) as nodes with many neighbors become stragglers. The fixed-size method also performs poorly compared to the strawman method for the same reason. Our method significantly improves sampling performance.
The lower sampling throughput as compared to that in the static system reflects the overhead of using a dynamic graph structure.
% \end{comment}

\begin{table}[t]
\centering
\caption{%Impact of adaptive block size (with threshold) on 
Linked list length and memory usage with different block sizing methods %on the GDELT graph and the Netflix graph.
}
\vspace{-3mm}
\label{tbl:adaptive-block-size}
\bgroup
\def\arraystretch{1}%  1 is the default, change whatever you need
\begin{tabular}{llllll}
\toprule
\multirow{2}{*}{\small{\textbf{Dataset}}}  & \multirow{2}{*}{\small{\textbf{Method}}} & \multicolumn{2}{c}{\small{\textbf{Linked List Len.}}}  & \multicolumn{2}{c}{\small{\textbf{Graph Data Size (MB)}}} \\
\cline{3-6}
 & & \small{\textbf{Avg.}} & \small{\textbf{Max.}} & \small{\textbf{Edge Data}} & \small{\textbf{Metadata}} \\
\midrule

% \hline
\multirow{4}{*}{\scriptsize{GDELT}} & \small{Adj. List} & \small{2866.91} & \small{4.8M} & \small{3649.06} & \small{2919.25} \\
 & \small{Strawman} & \small{233.97} & \small{1913} & \small{3649.01} & \small{238.22} \\
% \hline
& \small{Fixed-size} & \small{12.74} & \small{18680} & \small{3843.77} & \small{12.97} \\
% \hline
& \small{Adaptive} &  \small{6.38} & \small{1896} & \small{3818.22} & \small{6.49} \\
% \hline
& \small{Static} & \small{0.95} & \small{1}  & \small{3648.69} & \small{0.96} \\
\midrule
\multirow{4}{*}{\scriptsize{Netflix}} & \small{Adj. List} & \small{101.27} & \small{58K} & \small{3847.28} & \small{3077.83} \\

 & \small{Strawman} & \small{32.63} & \small{984} & \small{3845.94} & \small{991.69} \\
% \hline
& \small{Fixed-size} & \small{9.68} & \small{4854} & \small{4064.20} & \small{294.33} \\
% \hline
& \small{Adaptive} &  \small{9.04} & \small{974} & \small{4020.28} & \small{274.82} \\
% \hline
& \small{Static} & \small{1.00} & \small{1}  & \small{3833.22} & \small{30.39} \\

\bottomrule
\end{tabular}
\egroup
\end{table}

\begin{figure}[t]
    \centering
    \begin{subfigure}[t]{\columnwidth}
    \centering
    \includegraphics[scale=0.15]{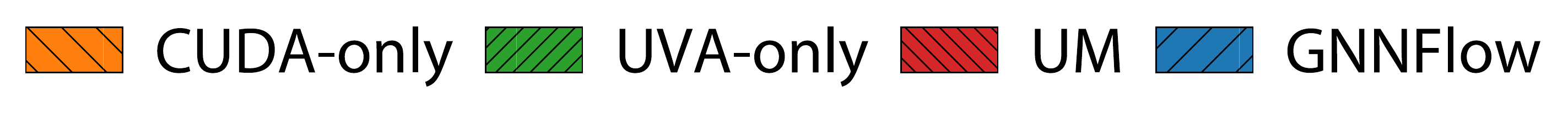}
    \end{subfigure}
    \hfill
    \begin{subfigure}[t]{0.23\textwidth}
    \centering
    \includegraphics[width=0.9\columnwidth]{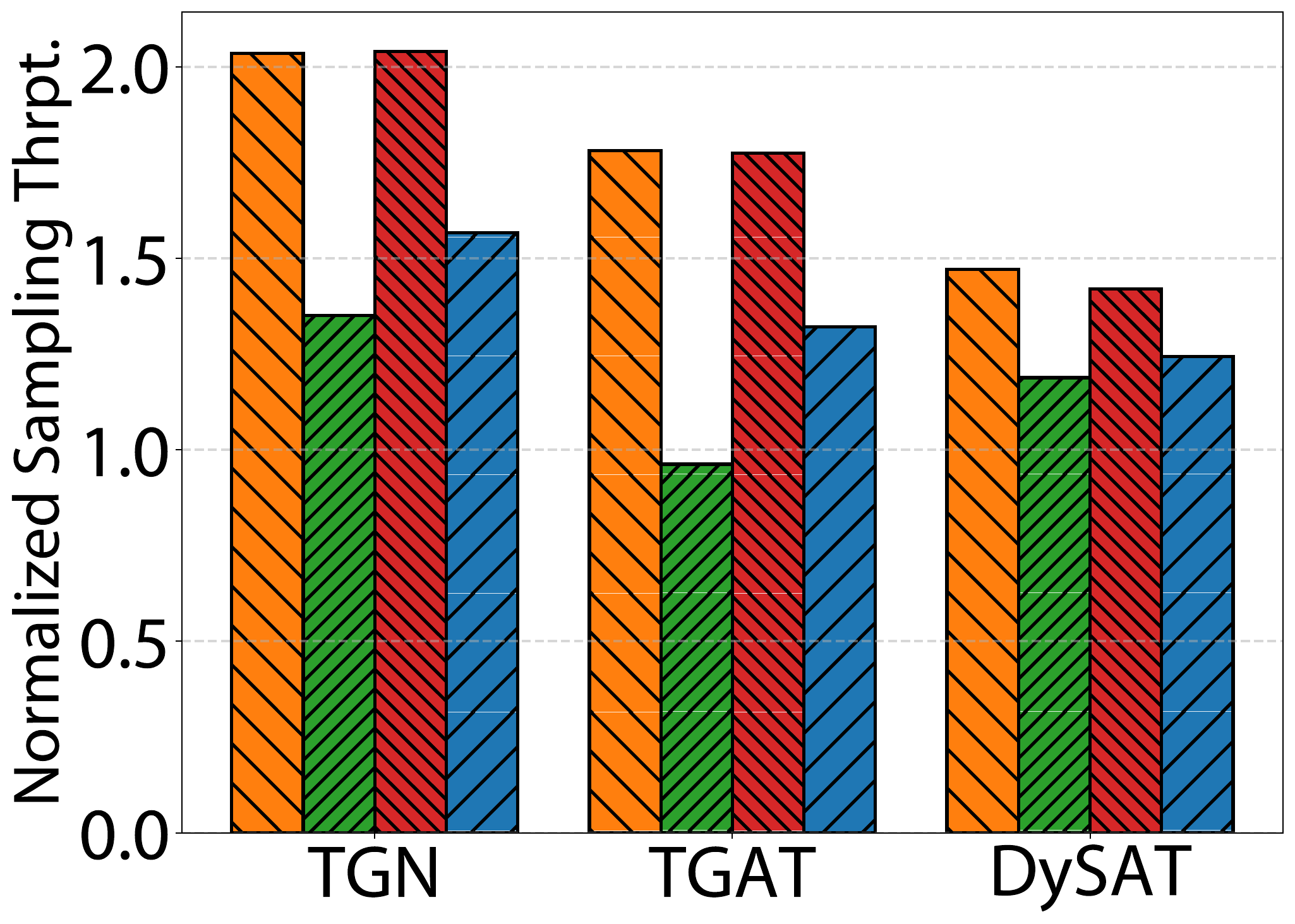}
     \caption{Reddit}
    \end{subfigure}
    \hfill
    \begin{subfigure}[t]{0.23\textwidth}
    \centering
    \includegraphics[width=0.9\columnwidth]{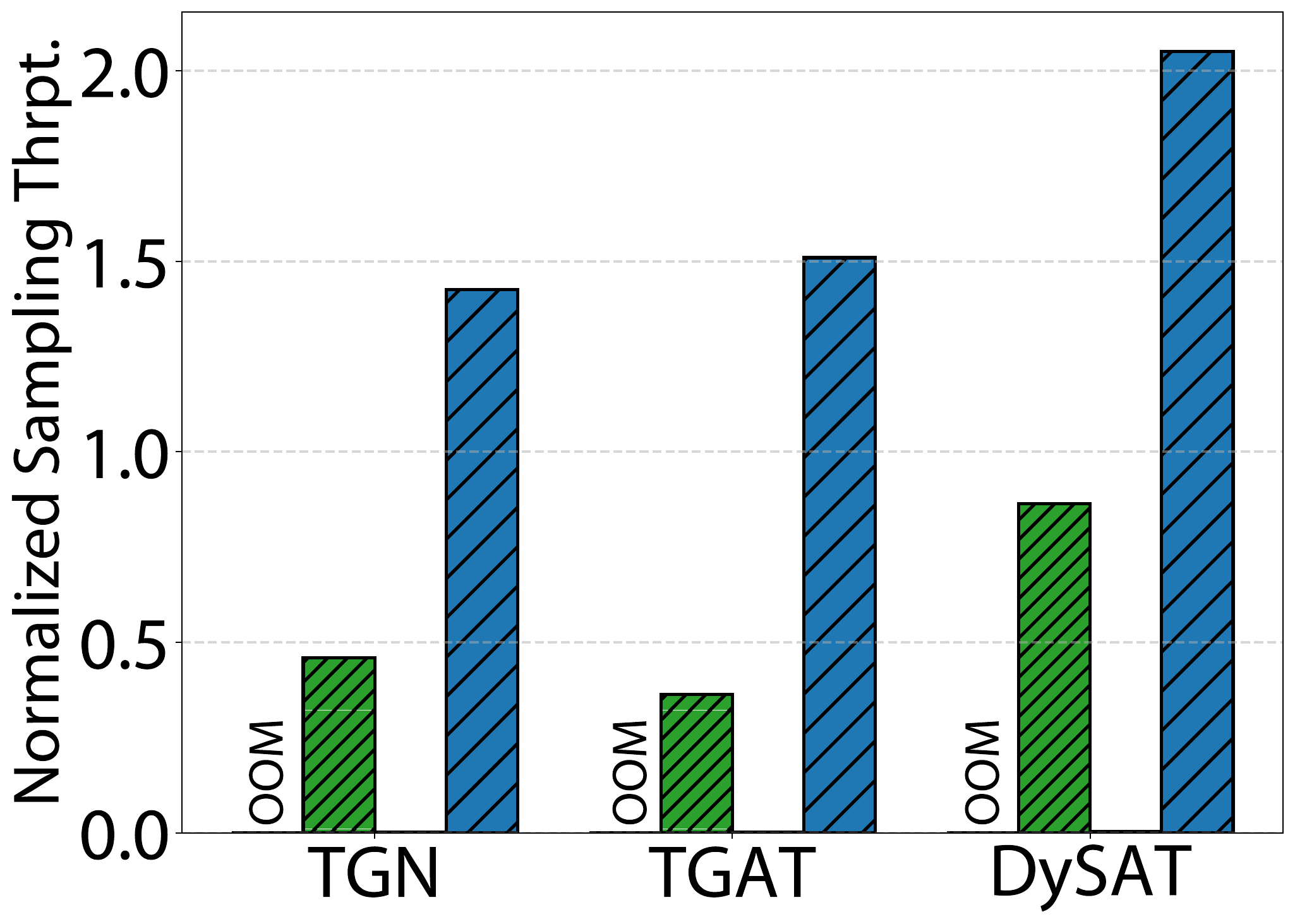}
     \caption{Netflix}
    \end{subfigure}
    \caption{Normalized sampling throughput (w.r.t. CPU sampling) for different sampling methods. %on the Reddit and Netflix graphs.
    }
    \label{fig:gpu_sampling_ablation}
\end{figure}

\vspace{1mm}
\noindent\textbf{GPU-based sampling.}
 We compare our GPU-based sampling with the following: (1) CPU sampling (used by TGL~\cite{zhou2022tgl}), with our highly optimized CPU sampler implementation for dynamic graphs; (2) CUDA-only, where all graph structure data are stored on GPU; (3) UVA-only (used by Quiver~\cite{quiver} and DGL with UVA sampling feature~\cite{wang2019deep}), where all graph structure data are placed on host pinned memory; and (4) unified memory (UM), which uses software interrupt and page migration with a one-page granularity for automatic migration of all graph structure data between CPU and GPU memory. %\cwu{which data is put on CPU and which on GPU?}. All data is initially on CPU. Then some may copied to GPU, and some may copied back to CPU, managed by UM itself.
We experiment on a single GPU and limit the maximum amount of GPU memory used for graph structure data to 4 GB because %real training requires 
enough GPU memory should be reserved for GPU-based feature cache and GNN training. 

As shown in Figure~\ref{fig:gpu_sampling_ablation}, CUDA-only and UM methods achieve the best performance on Reddit because the graph is small enough to fit entirely in GPU memory. %resulting in excellent performance. 
Our method performs better than the UVA-only method as we place metadata on GPU, reducing costly copying from CPU. On the large GDELT graph, the CUDA-only method exhausts the available memory. The UVA-only method is even slower than CPU sampling because it requires frequent access to the graph metadata located on CPU. %incurring high latency. 
The UM method exhibits extremely poor performance %(nearly zero throughput) 
due to thrashing, which is a situation that arises when GPU memory is limited and data transfers between the CPU and GPU occur frequently~\cite{yu2020quantitative}. %In contrast, our method achieves the best performance.

% \noindent\textbf{Effect of kernel optimizations.} 

\begin{figure}[t]
    \centering
    \begin{subfigure}[t]{0.47\columnwidth}
    \centering
    \includegraphics[width=\columnwidth]{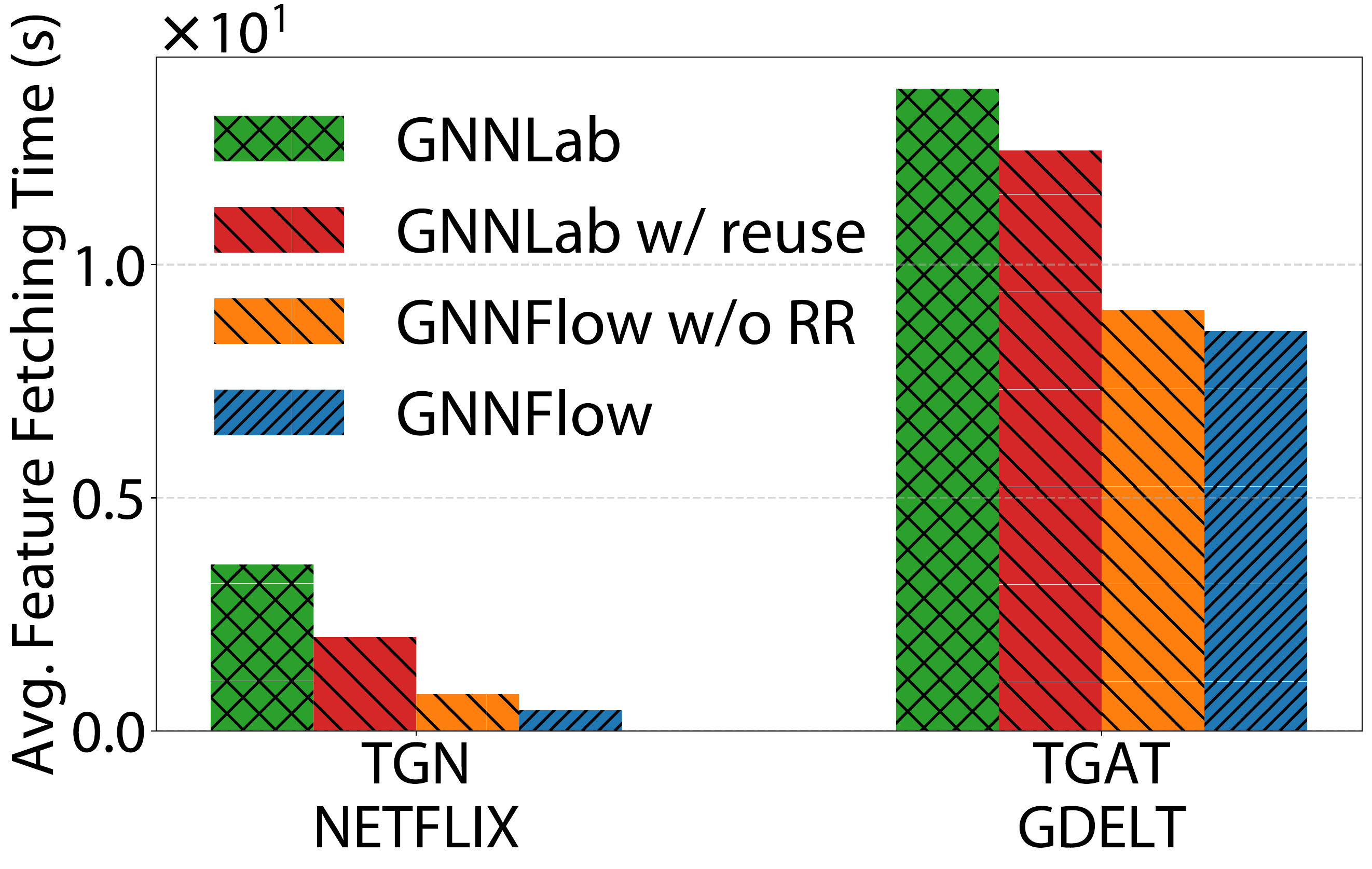}
     \caption{Effect on different models}
     \label{fig:cache-different-model}
    \end{subfigure}
    \hfill
    \begin{subfigure}[t]{0.47\columnwidth}
    \centering
    \includegraphics[width=\columnwidth]{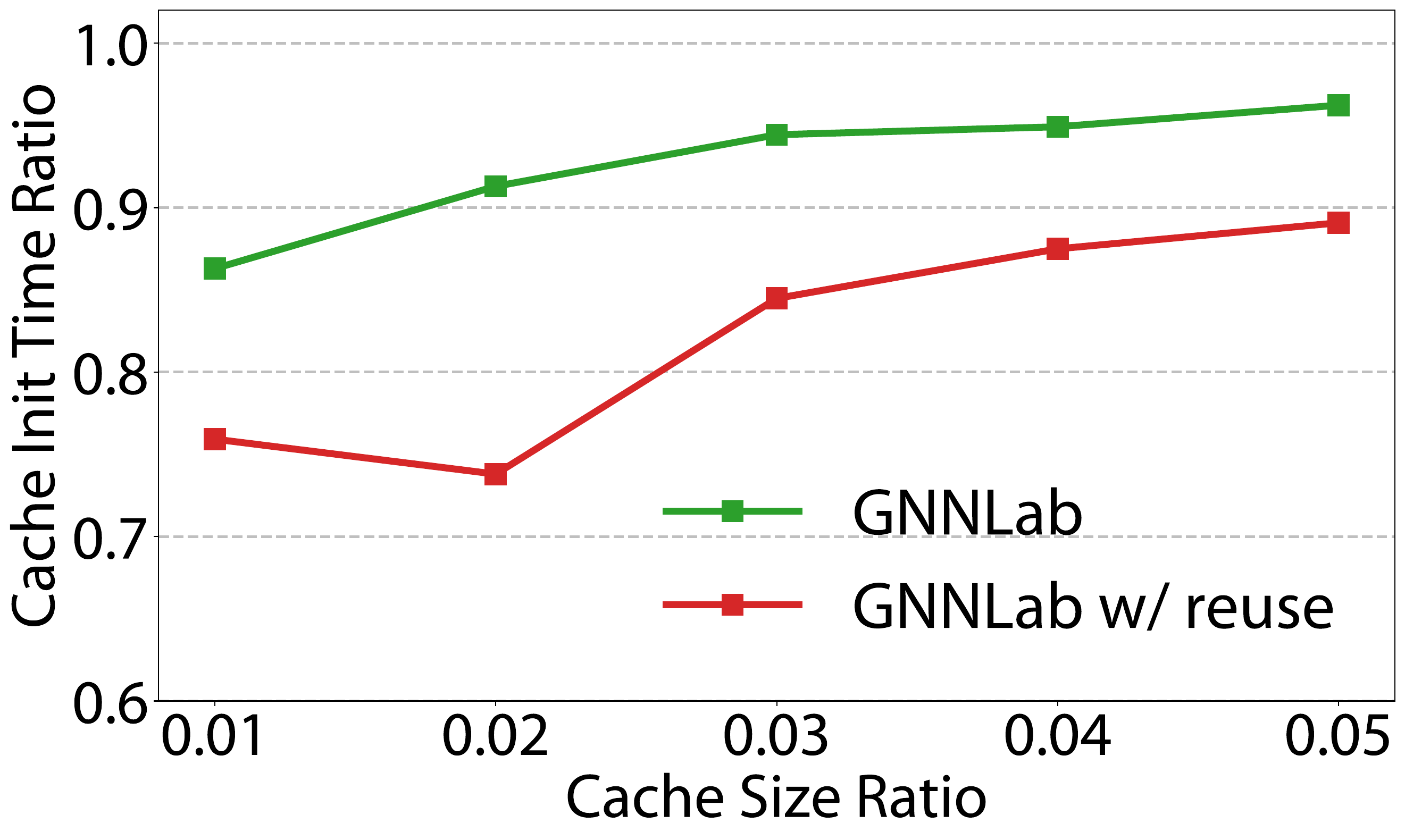}
     \caption{Cache init. time ratio}
     \label{fig:presampling_time_ratio}
    \end{subfigure}
    \hfill
    \begin{subfigure}[t]{0.47\columnwidth}
    \centering
    \includegraphics[width=\columnwidth]{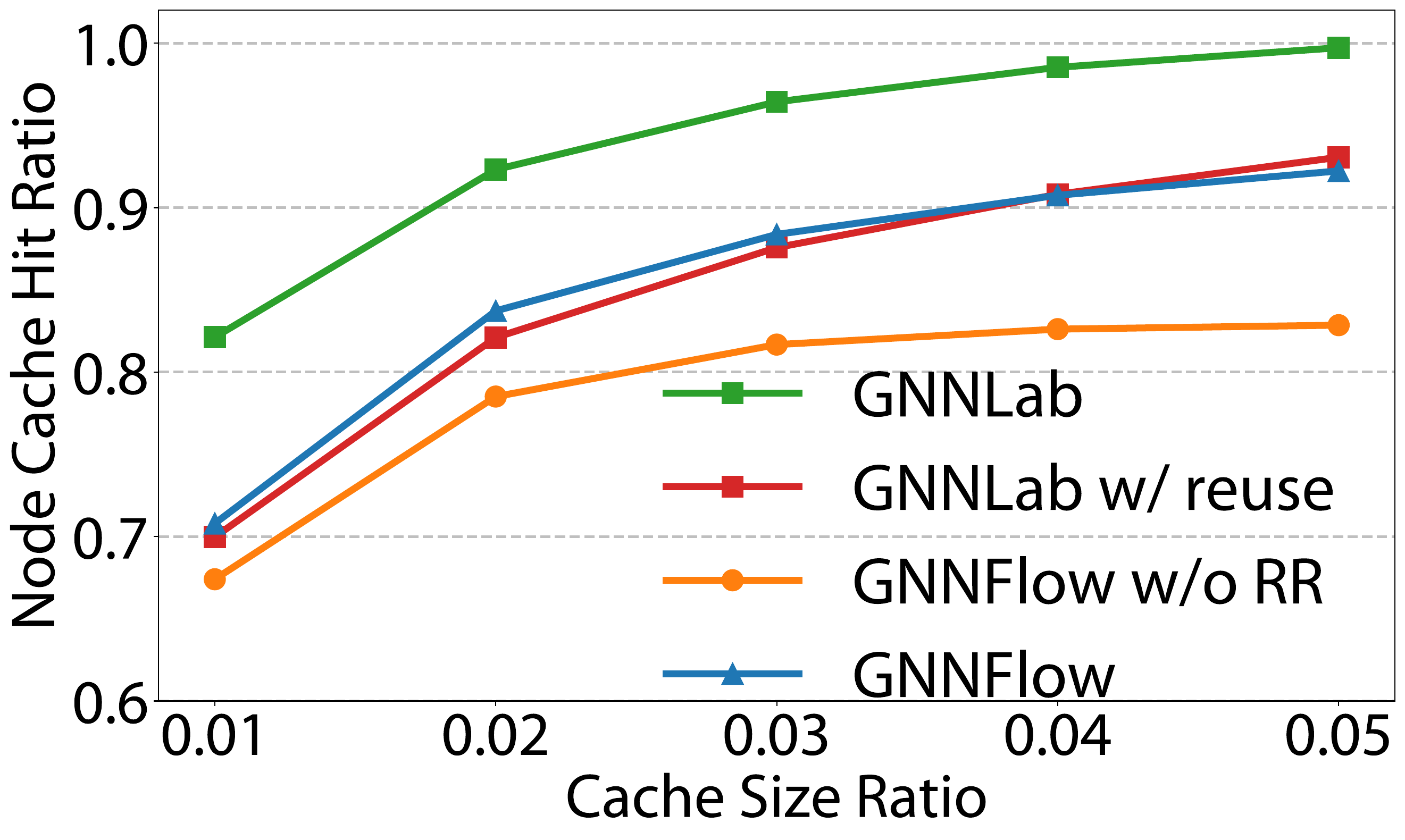}
     \caption{Effect of node feature}
     \label{fig:cache_node_feat}
    \end{subfigure}
    \hfill
    \begin{subfigure}[t]{0.47\columnwidth}
    \centering
    \includegraphics[width=\columnwidth]{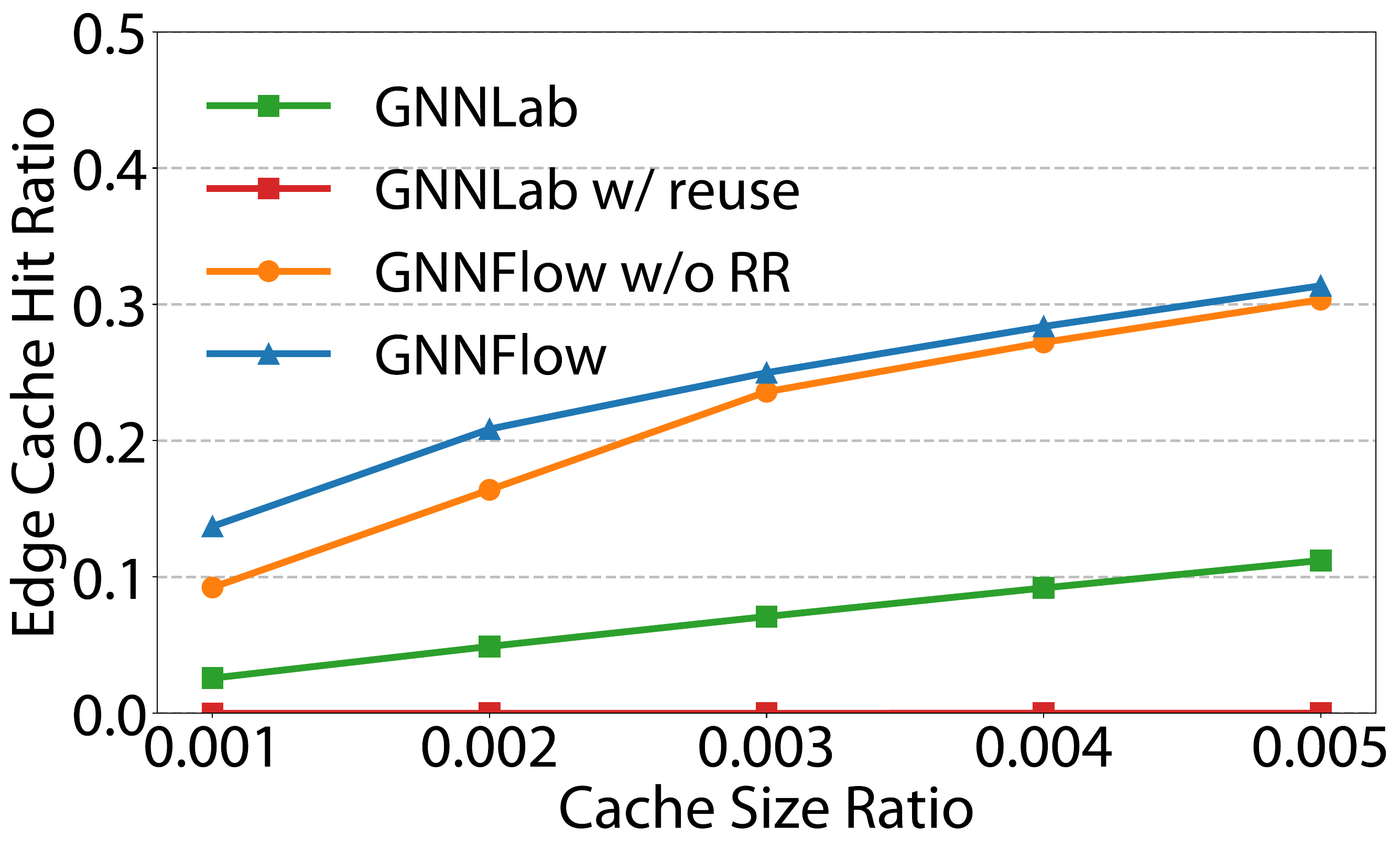}
     \caption{Effect of edge feature}
     \label{fig:cache_edge_feat}
    \end{subfigure}
    \caption{Performance of dynamic feature cache on diff. models and diff. cache sizes.}
    \label{fig:cache-abalation}
\end{figure}

\vspace{1mm}
\noindent\textbf{GPU dynamic feature cache.} 
We compare three baselines in the same \S\ref{sec:continuous_exp} scenario: (1) the static cache proposed by GNNLab~\cite{yang2022gnnlab}, which requires presampling with a new incremental batch for two epochs (as suggested in their paper) to re-initialize the cache in each retraining round; (2) the GNNLab static cache with cache reuse, which re-initializes the cache every two retraining rounds to allow for cache reuse between rounds; and (3) our approach without the cache reuse and restoration (noted as RR) optimizations (\S\ref{sec:feature-fetching}). We use LRU cache for all our methods, as it performs better than other dynamic cache methods on average. Figure~\ref{fig:cache-different-model} shows the average feature fetching time per retraining round (including cache initialization time) for continuous learning on two workloads. Our approach significantly reduces the feature fetching time by 8.1x and 1.6x compared to GNNLab in two workloads, respectively. Figure~\ref{fig:presampling_time_ratio}, Figure~\ref{fig:cache_node_feat} and Figure~\ref{fig:cache_edge_feat} show the effect of cache initialization time and cache hit rate for node and features when training TGN on the Netflix graph, respectively. Figure~\ref{fig:presampling_time_ratio} shows the proportion of time spent on cache initialization in GNNLab, which accounts for nearly 90\% of the total feature fetching time (including cache initialization), and slightly decreases to around 80\% when using cache reuse. On the other hand, our proposed vectorized dynamic cache eliminates the need for expensive cache re-initialization in every retraining round, and the cache reuse and restoration optimizations (\S\ref{sec:feature-fetching}) in GNNFlow can utilize data duplication between adjacent rounds to significantly increase the cache hit rate, as shown in Figure~\ref{fig:cache_node_feat} and Figure~\ref{fig:cache_edge_feat}, thereby reducing the feature fetching time. 

In addition, Figure~\ref{fig:cache_edge_feat} illustrates the ineffectiveness of GNNLab's static cache strategy for edge features. The cache hit rate is nearly zero without cache re-initialization in each round (GNNLab cache w/ reuse). The edges cached in previous rounds are seldom used in subsequent training. Even with cache re-initialization in each round (GNNLab), its cache hit ratio remains lower than our dynamic cache. This is because the edge access pattern is more dispersed than nodes, rendering the static cache unsuitable (S\ref{sec:feature-fetching}).

\subsection{Scalability}

\begin{figure}[t]
    \centering
    \begin{subfigure}[t]{0.49\columnwidth}
        \centering
        \includegraphics[width=\columnwidth]{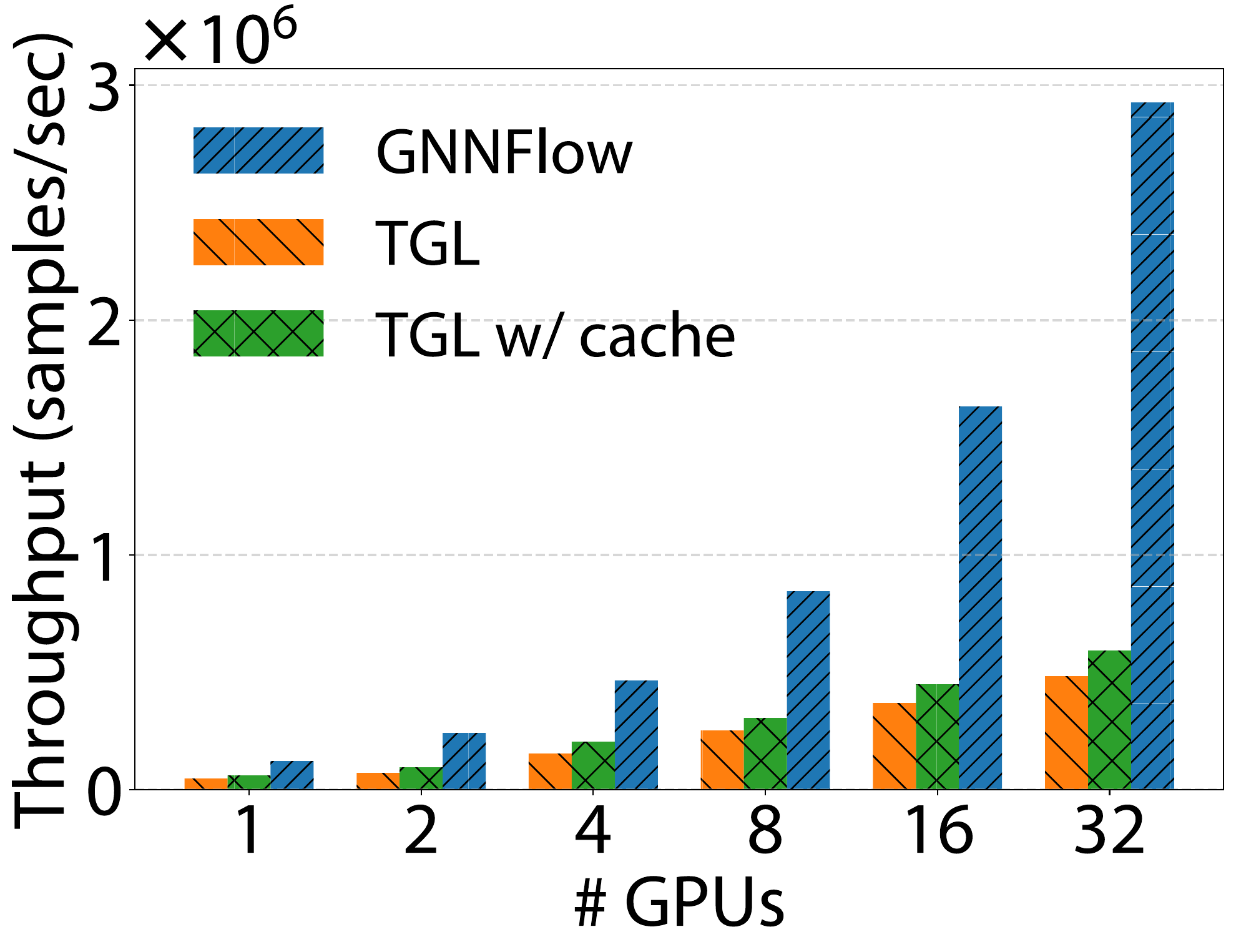}
         \caption{TGN}
    \end{subfigure}
    \hfill
    \begin{subfigure}[t]{0.49\columnwidth}
        \centering
        \includegraphics[width=\columnwidth]{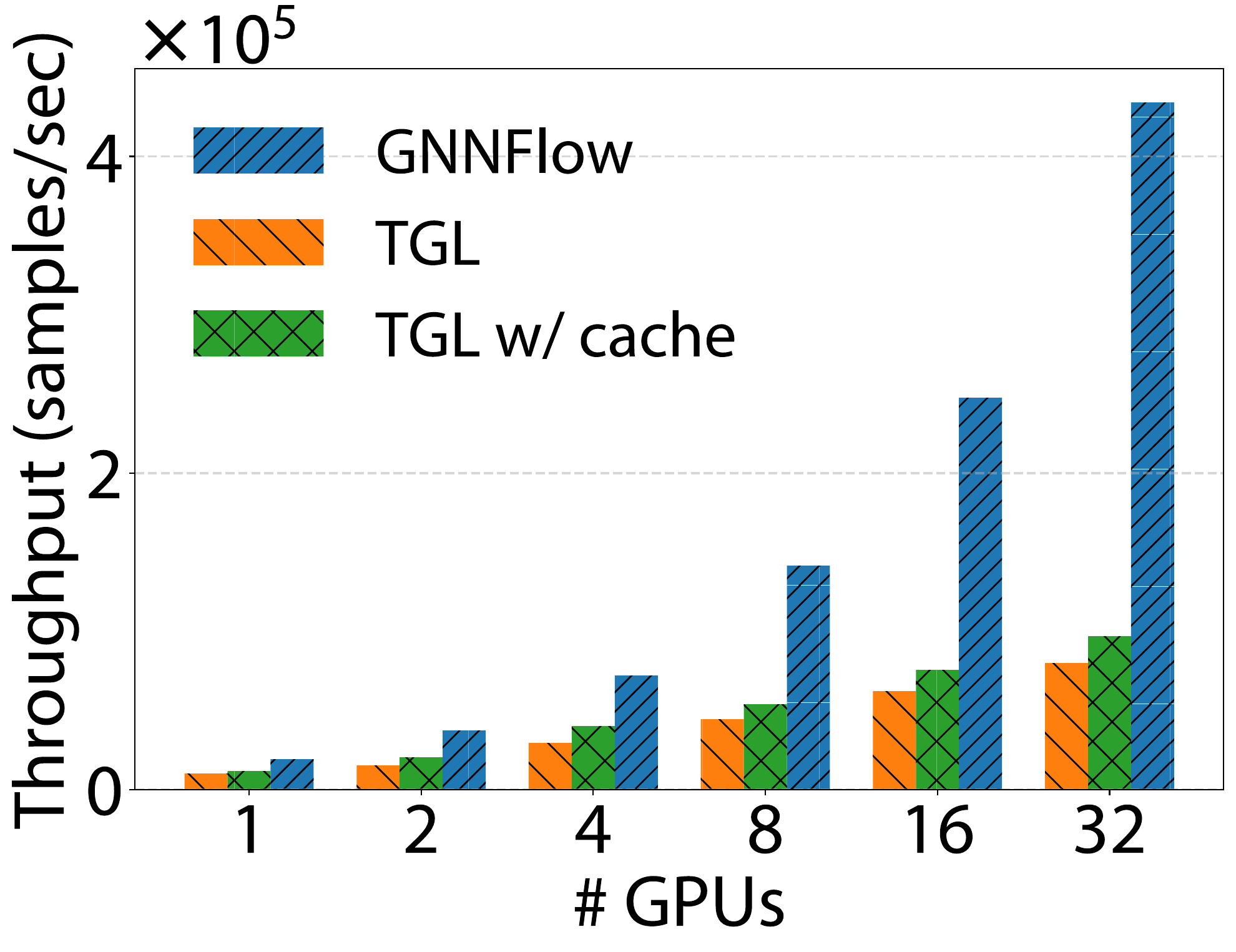}
         \caption{TGAT}
    \end{subfigure}
    \hfill
    \caption{Training throughput with different numbers of GPUs on the GDELT graph.}
    \label{fig:scalability}
\end{figure}

% \noindent\textbf{Scalability.}
We further evaluate the training throughput of \sysname{}, TGL and TGL with cache (the same LRU cache and cache ratios as \sysname{}) when training TGN and TGAT on GDELT in multi-GPU multi-machine scenarios by injecting 100 incremental batches consecutively. Since the graph (with features) can fit on a single machine, we do not partition it across multiple machines. We utilize 1 to 32 GPUs on four g4dn.metal instances. Figure~\ref{fig:scalability} shows the training throughput of model retraining on the last incremental batch. \sysname{} always %outperforms TGL and TGL w/ cache for both models, 
achieves the best training throughput %improvement of $1.9\times$ to $6.1\times$ compared to TGL.
while attaining near-linear scalability within a single machine and 71.9\% and 76.2\% of the ideal linear-scaling performance on 32 GPUs for TGN and TGAT, respectively.

\begin{table}[t]
    \centering
    \caption{Multi-machine training of GraphSAGE and GAT on partitioned MAG graph using 32 GPUs} 
    \vspace{-3mm}
    \bgroup
    \def\arraystretch{0.9}%  1 is the default, change whatever you need
    \begin{tabular}{ccccc}
        \toprule
        \multirow{2}{*}{\small{\textbf{System}}} & \small{\textbf{Graph Building}} & \multirow{2}{*}{\small{\textbf{Model}}} & \small{\textbf{Throughput}} \\ 
        & \small{\textbf{Time (s)}} & & \small{\textbf{(samples/s)}}  \\
        \midrule 
        \multirow{2}{*}{\small{DGL}} & \multirow{2}{*}{\small{7778}} & \small{GraphSAGE} & \small{124086.2}  \\ 
        &  & \small{GAT} & \small{146870.8}  \\ 
        \midrule
        \multirow{2}{*}{\small{GNNFlow}} & \multirow{2}{*}{\small{5246}} & \small{GraphSAGE} & \small{181608.9} \\ 
        & & \small{GAT} & \small{193720.9}   \\
        \bottomrule
    \end{tabular}
    \egroup
    \label{tab:throughput_dgl}
\end{table}

\subsection{Multi-machine Training on Partitioned Graphs}

%Since TGL does not support multi-machine training of partitioned graph, 
We compare \sysname{}'s performance of distributed training %over partitioned graphs 
with DGL on MAG, which is unable to fit into one g4dn.metal instance. CPU sampling is used since the DGL's UVA sampling feature is not available for distributed training. DGL employs METIS partitioning, %which is designed for static graphs, 
whereas \sysname{} utilizes online hash partitioning (\S\ref{sec:partition}) with an incremental batch size of 10 million edges. Table~\ref{tab:throughput_dgl} shows that \sysname{} achieves up to 1.46x training speed-up as compared to DGL, and a shorter time for graph construction. For both models, \sysname{} achieves similar or even higher final model accuracies.  %thanks to a simple yet effective hash-based online partition method. 

% \noindent\textbf{Training on the large synthetic graph.} 

We also train TGN on the LDBC graph %with 5.1 billion edges 
on eight g4dn.metal instances with incremental batches of 100 million edges each. %per incremental batch to the distributed dynamic graph. %It takes \sysname{} 3.3 hours to build the graph. CWu: this number is not very instrumental as how the edges are injected may not be continous
Its graph edge data size is 123 GB and graph metadata size is 1.65 GB, with our dynamic graph storage. \sysname{}'s training achieves a throughput of 556 thousand samples/second. %Since the graph can be partitioned on multiple servers, 
This further validates that \sysname{} enables efficient GNN training over large graphs partitioned across more servers.

%% file: related_work.tex
\section{Related Work}
%\subsection
\vspace{1mm}
\noindent\textbf{Other GNN training systems.}
PyG~\cite{fey2019fast} integrates with PyTorch~\cite{paszke2019pytorch} to provide a message-passing API for GNN training. AliGraph~\cite{zhu2019aligraph} and AGL~\cite{DBLP:journals/pvldb/ZhangHL0HSGWZ020} are scalable multi-machine GNN training systems; they do not exploit GPUs but use CPUs on multiple machines for GNN training. Euler~\cite{euler} is integrated with TensorFlow~\cite{abadi2016tensorflow} for GNN training, and uses CPU for graph sampling without feature caching. PaGraph~\cite{lin2020pagraph} and GNNLab~\cite{yang2022gnnlab} proposes using static GPU feature cache to reduce data transmission. BGL~\cite{liu2021bgl} is a distributed GNN training system which adopts dynamic node caching using FIFO policy. Legion~\cite{sun2023legion} also proposes a topology cache on GPUs to speed up neighbor sampling. PiPAD~\cite{wangpipad2023} is a dynamic GNN training system tailored for DTDGs, but still assumes a static graph storage. BLAD~\cite{fu2023blad} considers DTDGs and focus on speeding up distributed training for DTDG-based GNN models. All these systems focus on GNN training over static graphs or DTDGs, while we focus on continuous temporal GNN learning on CTDGs.

\vspace{1mm}
\noindent\textbf{Dynamic graph processing systems.}
Dynamic graph processing systems (that do not use GNN but traditional graph processing models) also adopt dynamic graph storage. 
Early works like STINGER~\cite{ediger2012stinger} adopt a block adjacency list~\cite{ediger2012stinger, macko2015llama, feng2015distinger} or an adjacency array \cite{green2016custinger} as the graph data structure, which stores the IDs of neighbors of each vertex contiguously. 
FaimGraph~\cite{winter2018faimgraph} and Hornet~\cite{busato2018hornet} use a block adjacency list as the graph data structure. 
GraphOne~\cite{kumar2019graphone} adopts an edge list recording the latest updates and a block adjacency list storing archived data. 
In Aspen~\cite{dhulipala2019low}, a graph is represented as a tree of trees, storing the set of vertices (vertex tree) where each vertex's edges are stored in a C-tree (edge tree). 
Tegra~\cite{iyer2021tegra} adopts an adaptive radix tree (ADT) for dynamic storage of graph snapshots.
While previous dynamic graph processing systems utilize variations of adjacency lists or alternative structures, they lack efficient support for temporal k-hop sampling, a critical need for temporal GNNs (\S\ref{sec:preliminaries}). Our proposed time-index dynamic graph structure storage (\S\ref{sec:data-structure}) is tailor-made for temporal k-hop sampling. Additionally, our data structure is both GPU-friendly (\S\ref{sec:neighborhood-sampling}), leveraging the computational power of modern GPUs for improving sampling efficiency, and designed for distributed settings (\S\ref{sec:distributed_training}), enabling scalability across multiple compute nodes.

%% file: main.bbl

\begin{thebibliography}{75}

%%% the Web address.

\ifx \showCODEN    \undefined \def \showCODEN     #1{\unskip}     \fi
\ifx \showDOI      \undefined \def \showDOI       #1{#1}\fi
\ifx \showISBNx    \undefined \def \showISBNx     #1{\unskip}     \fi
\ifx \showISBNxiii \undefined \def \showISBNxiii  #1{\unskip}     \fi
\ifx \showISSN     \undefined \def \showISSN      #1{\unskip}     \fi
\ifx \showLCCN     \undefined \def \showLCCN      #1{\unskip}     \fi
\ifx \shownote     \undefined \def \shownote      #1{#1}          \fi
\ifx \showarticletitle \undefined \def \showarticletitle #1{#1}   \fi
\ifx \showURL      \undefined \def \showURL       {\relax}        \fi
% invisible to TeX
\providecommand\bibfield[2]{#2}
\providecommand\bibinfo[2]{#2}
\providecommand\natexlab[1]{#1}
\providecommand\showeprint[2][]{arXiv:#2}

\bibitem[\protect\citeauthoryear{??}{ten}{2022}]%
        {tensorpipe}
 \bibinfo{year}{2022}\natexlab{}.
\newblock \bibinfo{title}{{A Tensor-aware Point-to-point Communication
  Primitive for Machine Learning}}.
\newblock
\newblock
\newblock
\shownote{\url{https://github.com/pytorch/tensorpipe}.}


\bibitem[\protect\citeauthoryear{??}{eul}{2022}]%
        {euler}
 \bibinfo{year}{2022}\natexlab{}.
\newblock \bibinfo{title}{{Euler 2.0: A Distributed Graph Deep Learning
  Framework}}.
\newblock
\newblock
\newblock
\shownote{\url{https://github.com/alibaba/euler}.}


\bibitem[\protect\citeauthoryear{??}{pgl}{2022}]%
        {pgl}
 \bibinfo{year}{2022}\natexlab{}.
\newblock \bibinfo{title}{{PGL: An Efficient and Flexible Graph Learning
  Framework based on PaddlePaddle}}.
\newblock
\newblock
\newblock
\shownote{\url{https://github.com/PaddlePaddle/PGL}.}


\bibitem[\protect\citeauthoryear{??}{qui}{2022}]%
        {quiver}
 \bibinfo{year}{2022}\natexlab{}.
\newblock \bibinfo{title}{{PyTorch Library for Fast and Easy Distributed Graph
  Learning}}.
\newblock
\newblock
\newblock
\shownote{\url{https://github.com/quiver-team/torch-quiver}.}


\bibitem[\protect\citeauthoryear{Abadi, Barham, Chen, Chen, Davis, Dean, Devin,
  Ghemawat, Irving, Isard, Kudlur, Levenberg, Monga, Moore, Murray, Steiner,
  Tucker, Vasudevan, Warden, Wicke, Yu, and Zheng}{Abadi et~al\mbox{.}}{2016}]%
        {abadi2016tensorflow}
\bibfield{author}{\bibinfo{person}{Mart{\'\i}n Abadi}, \bibinfo{person}{Paul
  Barham}, \bibinfo{person}{Jianmin Chen}, \bibinfo{person}{Zhifeng Chen},
  \bibinfo{person}{Andy Davis}, \bibinfo{person}{Jeffrey Dean},
  \bibinfo{person}{Matthieu Devin}, \bibinfo{person}{Sanjay Ghemawat},
  \bibinfo{person}{Geoffrey Irving}, \bibinfo{person}{Michael Isard},
  \bibinfo{person}{Manjunath Kudlur}, \bibinfo{person}{Josh Levenberg},
  \bibinfo{person}{Rajat Monga}, \bibinfo{person}{Sherry Moore},
  \bibinfo{person}{Derek~G. Murray}, \bibinfo{person}{Benoit Steiner},
  \bibinfo{person}{Paul Tucker}, \bibinfo{person}{Vijay Vasudevan},
  \bibinfo{person}{Pete Warden}, \bibinfo{person}{Martin Wicke},
  \bibinfo{person}{Yuan Yu}, {and} \bibinfo{person}{Xiaoqiang Zheng}.}
  \bibinfo{year}{2016}\natexlab{}.
\newblock \showarticletitle{{TensorFlow: A System for Large-scale Machine
  Learning}}. In \bibinfo{booktitle}{\emph{Proceedings of the 12th USENIX
  Symposium on Operating Systems Design and Implementation}}.
\newblock


\bibitem[\protect\citeauthoryear{Ahrabian, Xu, Zhang, Wu, Wang, and
  Coates}{Ahrabian et~al\mbox{.}}{2021}]%
        {ahrabian2021structure}
\bibfield{author}{\bibinfo{person}{Kian Ahrabian}, \bibinfo{person}{Yishi Xu},
  \bibinfo{person}{Yingxue Zhang}, \bibinfo{person}{Jiapeng Wu},
  \bibinfo{person}{Yuening Wang}, {and} \bibinfo{person}{Mark Coates}.}
  \bibinfo{year}{2021}\natexlab{}.
\newblock \showarticletitle{{Structure Aware Experience Replay for Incremental
  Learning in Graph-based Recommender Systems}}. In
  \bibinfo{booktitle}{\emph{Proceedings of the 30th ACM International
  Conference on Information \& Knowledge Management}}.
\newblock


\bibitem[\protect\citeauthoryear{Angles, Antal, Averbuch, Birler, Boncz,
  B{\'u}r, Erling, Gubichev, Haprian, Kaufmann, et~al\mbox{.}}{Angles
  et~al\mbox{.}}{2020}]%
        {angles2020ldbc}
\bibfield{author}{\bibinfo{person}{Renzo Angles},
  \bibinfo{person}{J{\'a}nos~Benjamin Antal}, \bibinfo{person}{Alex Averbuch},
  \bibinfo{person}{Altan Birler}, \bibinfo{person}{Peter Boncz},
  \bibinfo{person}{M{\'a}rton B{\'u}r}, \bibinfo{person}{Orri Erling},
  \bibinfo{person}{Andrey Gubichev}, \bibinfo{person}{Vlad Haprian},
  \bibinfo{person}{Moritz Kaufmann}, {et~al\mbox{.}}}
  \bibinfo{year}{2020}\natexlab{}.
\newblock \showarticletitle{{The LDBC Social Network Benchmark}}.
\newblock \bibinfo{journal}{\emph{arXiv preprint}} (\bibinfo{year}{2020}).
\newblock


\bibitem[\protect\citeauthoryear{Barab{\'a}si and Albert}{Barab{\'a}si and
  Albert}{1999}]%
        {barabasi1999emergence}
\bibfield{author}{\bibinfo{person}{Albert-L{\'a}szl{\'o} Barab{\'a}si} {and}
  \bibinfo{person}{R{\'e}ka Albert}.} \bibinfo{year}{1999}\natexlab{}.
\newblock \showarticletitle{{Emergence of Scaling in Random Networks}}.
\newblock \bibinfo{journal}{\emph{science}} (\bibinfo{year}{1999}).
\newblock


\bibitem[\protect\citeauthoryear{Bennett, Lanning, et~al\mbox{.}}{Bennett
  et~al\mbox{.}}{2007}]%
        {bennett2007netflix}
\bibfield{author}{\bibinfo{person}{James Bennett}, \bibinfo{person}{Stan
  Lanning}, {et~al\mbox{.}}} \bibinfo{year}{2007}\natexlab{}.
\newblock \showarticletitle{{The Netflix Prize}}. In
  \bibinfo{booktitle}{\emph{Proceedings of KDD cup and workshop}}.
\newblock


\bibitem[\protect\citeauthoryear{Besta, Fischer, Kalavri, Kapralov, and
  Hoefler}{Besta et~al\mbox{.}}{2020}]%
        {besta2020practice}
\bibfield{author}{\bibinfo{person}{Maciej Besta}, \bibinfo{person}{Marc
  Fischer}, \bibinfo{person}{Vasiliki Kalavri}, \bibinfo{person}{Michael
  Kapralov}, {and} \bibinfo{person}{Torsten Hoefler}.}
  \bibinfo{year}{2020}\natexlab{}.
\newblock \showarticletitle{{Practice of Streaming and Dynamic graphs:
  Concepts, Models, Systems, and Parallelism}}.
\newblock \bibinfo{journal}{\emph{arXiv}} (\bibinfo{year}{2020}).
\newblock


\bibitem[\protect\citeauthoryear{Busato, Green, Bombieri, and Bader}{Busato
  et~al\mbox{.}}{2018}]%
        {busato2018hornet}
\bibfield{author}{\bibinfo{person}{Federico Busato}, \bibinfo{person}{Oded
  Green}, \bibinfo{person}{Nicola Bombieri}, {and} \bibinfo{person}{David~A
  Bader}.} \bibinfo{year}{2018}\natexlab{}.
\newblock \showarticletitle{{Hornet: An Efficient Data Structure for Dynamic
  Sparse Graphs and Matrices on GPUs}}. In \bibinfo{booktitle}{\emph{2018 IEEE
  High Performance extreme Computing Conference (HPEC)}}.
\newblock


\bibitem[\protect\citeauthoryear{da~Xu, chuanwei ruan, evren korpeoglu, sushant
  kumar, and kannan achan}{da~Xu et~al\mbox{.}}{2020}]%
        {Xu2020Inductive}
\bibfield{author}{\bibinfo{person}{da Xu}, \bibinfo{person}{chuanwei ruan},
  \bibinfo{person}{evren korpeoglu}, \bibinfo{person}{sushant kumar}, {and}
  \bibinfo{person}{kannan achan}.} \bibinfo{year}{2020}\natexlab{}.
\newblock \showarticletitle{{Inductive Representation Learning on Temporal
  Graphs}}. In \bibinfo{booktitle}{\emph{Proceedings of International
  Conference on Learning Representations}}.
\newblock


\bibitem[\protect\citeauthoryear{Dhulipala, Blelloch, and Shun}{Dhulipala
  et~al\mbox{.}}{2019}]%
        {dhulipala2019low}
\bibfield{author}{\bibinfo{person}{Laxman Dhulipala}, \bibinfo{person}{Guy~E
  Blelloch}, {and} \bibinfo{person}{Julian Shun}.}
  \bibinfo{year}{2019}\natexlab{}.
\newblock \showarticletitle{{Low-latency Graph Streaming using Compressed
  Purely-functional Trees}}. In \bibinfo{booktitle}{\emph{Proceedings of the
  40th ACM SIGPLAN conference on programming language design and
  implementation}}.
\newblock


\bibitem[\protect\citeauthoryear{Ding, Feng, He, Liao, Shi, and Zhang}{Ding
  et~al\mbox{.}}{2022}]%
        {ding2022causal}
\bibfield{author}{\bibinfo{person}{Sihao Ding}, \bibinfo{person}{Fuli Feng},
  \bibinfo{person}{Xiangnan He}, \bibinfo{person}{Yong Liao},
  \bibinfo{person}{Jun Shi}, {and} \bibinfo{person}{Yongdong Zhang}.}
  \bibinfo{year}{2022}\natexlab{}.
\newblock \showarticletitle{{Causal Incremental Graph Convolution for
  Recommender System Retraining}}.
\newblock \bibinfo{journal}{\emph{IEEE Transactions on Neural Networks and
  Learning Systems}} (\bibinfo{year}{2022}).
\newblock


\bibitem[\protect\citeauthoryear{Ediger, McColl, Riedy, and Bader}{Ediger
  et~al\mbox{.}}{2012}]%
        {ediger2012stinger}
\bibfield{author}{\bibinfo{person}{David Ediger}, \bibinfo{person}{Rob McColl},
  \bibinfo{person}{Jason Riedy}, {and} \bibinfo{person}{David~A Bader}.}
  \bibinfo{year}{2012}\natexlab{}.
\newblock \showarticletitle{{STINGER: High performance Data Structure for
  Streaming Graphs}}. In \bibinfo{booktitle}{\emph{Proceedings of 2012 IEEE
  Conference on High Performance Extreme Computing}}.
\newblock


\bibitem[\protect\citeauthoryear{Faloutsos, Faloutsos, and Faloutsos}{Faloutsos
  et~al\mbox{.}}{1999}]%
        {faloutsos1999power}
\bibfield{author}{\bibinfo{person}{Michalis Faloutsos}, \bibinfo{person}{Petros
  Faloutsos}, {and} \bibinfo{person}{Christos Faloutsos}.}
  \bibinfo{year}{1999}\natexlab{}.
\newblock \showarticletitle{{On Power-law Relationships of the Internet
  Topology}}.
\newblock \bibinfo{journal}{\emph{ACM SIGCOMM computer communication review}}
  (\bibinfo{year}{1999}).
\newblock


\bibitem[\protect\citeauthoryear{Fan, Andersen, and Kaminsky}{Fan
  et~al\mbox{.}}{2013}]%
        {fan2013memc3}
\bibfield{author}{\bibinfo{person}{Bin Fan}, \bibinfo{person}{David~G
  Andersen}, {and} \bibinfo{person}{Michael Kaminsky}.}
  \bibinfo{year}{2013}\natexlab{}.
\newblock \showarticletitle{{Memc3: Compact and Concurrent Memcache with Dumber
  Caching and Smarter Hashing}}. In \bibinfo{booktitle}{\emph{Presented as part
  of the 10th USENIX Symposium on Networked Systems Design and
  Implementation}}.
\newblock


\bibitem[\protect\citeauthoryear{Fan, Ma, Li, He, Zhao, Tang, and Yin}{Fan
  et~al\mbox{.}}{2019}]%
        {fan2019graph}
\bibfield{author}{\bibinfo{person}{Wenqi Fan}, \bibinfo{person}{Yao Ma},
  \bibinfo{person}{Qing Li}, \bibinfo{person}{Yuan He}, \bibinfo{person}{Eric
  Zhao}, \bibinfo{person}{Jiliang Tang}, {and} \bibinfo{person}{Dawei Yin}.}
  \bibinfo{year}{2019}\natexlab{}.
\newblock \showarticletitle{{Graph Neural Networks for Social Recommendation}}.
  In \bibinfo{booktitle}{\emph{Proceedings of the World Wide Web Conference}}.
\newblock


\bibitem[\protect\citeauthoryear{Feng, Meng, and Ammar}{Feng
  et~al\mbox{.}}{2015}]%
        {feng2015distinger}
\bibfield{author}{\bibinfo{person}{Guoyao Feng}, \bibinfo{person}{Xiao Meng},
  {and} \bibinfo{person}{Khaled Ammar}.} \bibinfo{year}{2015}\natexlab{}.
\newblock \showarticletitle{{DISTINGER: A Distributed Graph Data Structure for
  Massive Dynamic Graph Processing}}. In \bibinfo{booktitle}{\emph{Proceedings
  of 2015 IEEE International Conference on Big Data (Big Data)}}.
\newblock


\bibitem[\protect\citeauthoryear{Fey and Lenssen}{Fey and Lenssen}{2019}]%
        {fey2019fast}
\bibfield{author}{\bibinfo{person}{Matthias Fey} {and}
  \bibinfo{person}{Jan~Eric Lenssen}.} \bibinfo{year}{2019}\natexlab{}.
\newblock \showarticletitle{{Fast Graph Representation Learning with PyTorch
  Geometric}}.
\newblock \bibinfo{journal}{\emph{ICLR workshop on Representation Learning on
  Graphs and Manifolds}} (\bibinfo{year}{2019}).
\newblock


\bibitem[\protect\citeauthoryear{Fout, Byrd, Shariat, and Ben-Hur}{Fout
  et~al\mbox{.}}{2017}]%
        {fout2017protein}
\bibfield{author}{\bibinfo{person}{Alex Fout}, \bibinfo{person}{Jonathon Byrd},
  \bibinfo{person}{Basir Shariat}, {and} \bibinfo{person}{Asa Ben-Hur}.}
  \bibinfo{year}{2017}\natexlab{}.
\newblock \showarticletitle{{Protein Interface Prediction using Graph
  Convolutional Networks}}. In \bibinfo{booktitle}{\emph{Proceedings of
  Advances in Neural Information Processing Systems}}.
\newblock


\bibitem[\protect\citeauthoryear{Fu, Chen, Yang, Shi, Li, and Guo}{Fu
  et~al\mbox{.}}{2023}]%
        {fu2023blad}
\bibfield{author}{\bibinfo{person}{Kaihua Fu}, \bibinfo{person}{Quan Chen},
  \bibinfo{person}{Yuzhuo Yang}, \bibinfo{person}{Jiuchen Shi},
  \bibinfo{person}{Chao Li}, {and} \bibinfo{person}{Minyi Guo}.}
  \bibinfo{year}{2023}\natexlab{}.
\newblock \showarticletitle{{BLAD: Adaptive Load Balanced Scheduling and
  Operator Overlap Pipeline For Accelerating The Dynamic GNN Training}}. In
  \bibinfo{booktitle}{\emph{Proceedings of the International Conference for
  High Performance Computing, Networking, Storage and Analysis}}.
\newblock


\bibitem[\protect\citeauthoryear{Gama, \v{Z}liobaitundefined, Bifet,
  Pechenizkiy, and Bouchachia}{Gama et~al\mbox{.}}{2014}]%
        {concept-drift}
\bibfield{author}{\bibinfo{person}{Jo\~{a}o Gama},
  \bibinfo{person}{Indrundefined \v{Z}liobaitundefined},
  \bibinfo{person}{Albert Bifet}, \bibinfo{person}{Mykola Pechenizkiy}, {and}
  \bibinfo{person}{Abdelhamid Bouchachia}.} \bibinfo{year}{2014}\natexlab{}.
\newblock \showarticletitle{{A Survey on Concept Drift Adaptation}}.
\newblock \bibinfo{journal}{\emph{Comput. Surveys}} (\bibinfo{year}{2014}).
\newblock


\bibitem[\protect\citeauthoryear{Gandhi and Iyer}{Gandhi and Iyer}{2021}]%
        {gandhi2021p3}
\bibfield{author}{\bibinfo{person}{Swapnil Gandhi} {and}
  \bibinfo{person}{Anand~Padmanabha Iyer}.} \bibinfo{year}{2021}\natexlab{}.
\newblock \showarticletitle{{P3: Distributed deep graph learning at scale}}. In
  \bibinfo{booktitle}{\emph{Proceedings of the 15th USENIX Symposium on
  Operating Systems Design and Implementation}}.
\newblock


\bibitem[\protect\citeauthoryear{Gilmer, Schoenholz, Riley, Vinyals, and
  Dahl}{Gilmer et~al\mbox{.}}{2017}]%
        {gilmer2017neural}
\bibfield{author}{\bibinfo{person}{Justin Gilmer}, \bibinfo{person}{Samuel~S
  Schoenholz}, \bibinfo{person}{Patrick~F Riley}, \bibinfo{person}{Oriol
  Vinyals}, {and} \bibinfo{person}{George~E Dahl}.}
  \bibinfo{year}{2017}\natexlab{}.
\newblock \showarticletitle{{Neural Message Passing for Quantum Chemistry}}. In
  \bibinfo{booktitle}{\emph{Proceedings of International Conference on Machine
  Learning}}.
\newblock


\bibitem[\protect\citeauthoryear{Green and Bader}{Green and Bader}{2016}]%
        {green2016custinger}
\bibfield{author}{\bibinfo{person}{Oded Green} {and} \bibinfo{person}{David~A
  Bader}.} \bibinfo{year}{2016}\natexlab{}.
\newblock \showarticletitle{{cuSTINGER: Supporting Dynamic Graph Algorithms for
  GPUs}}. In \bibinfo{booktitle}{\emph{Proceedings of 2016 IEEE High
  Performance Extreme Computing Conference}}.
\newblock


\bibitem[\protect\citeauthoryear{Hamilton, Ying, and Leskovec}{Hamilton
  et~al\mbox{.}}{2017}]%
        {hamilton2017inductive}
\bibfield{author}{\bibinfo{person}{Will Hamilton}, \bibinfo{person}{Zhitao
  Ying}, {and} \bibinfo{person}{Jure Leskovec}.}
  \bibinfo{year}{2017}\natexlab{}.
\newblock \showarticletitle{{Inductive Representation Learning on Large
  Graphs}}. In \bibinfo{booktitle}{\emph{Proceedings of Advances in Neural
  Information Processing Systems}}.
\newblock


\bibitem[\protect\citeauthoryear{Iyer, Pu, Patel, Gonzalez, and Stoica}{Iyer
  et~al\mbox{.}}{2021}]%
        {iyer2021tegra}
\bibfield{author}{\bibinfo{person}{Anand~Padmanabha Iyer},
  \bibinfo{person}{Qifan Pu}, \bibinfo{person}{Kishan Patel},
  \bibinfo{person}{Joseph~E Gonzalez}, {and} \bibinfo{person}{Ion Stoica}.}
  \bibinfo{year}{2021}\natexlab{}.
\newblock \showarticletitle{{TEGRA: Efficient Ad-Hoc Analytics on Evolving
  Graphs}}. In \bibinfo{booktitle}{\emph{Proceedings of the 18th USENIX
  Symposium on Networked Systems Design and Implementation}}.
\newblock


\bibitem[\protect\citeauthoryear{Jaccard}{Jaccard}{1912}]%
        {jaccard1912distribution}
\bibfield{author}{\bibinfo{person}{Paul Jaccard}.}
  \bibinfo{year}{1912}\natexlab{}.
\newblock \showarticletitle{{The Distribution of the Llora in the Alpine
  Zone.}}
\newblock \bibinfo{journal}{\emph{New phytologist}} (\bibinfo{year}{1912}),
  \bibinfo{pages}{37--50}.
\newblock


\bibitem[\protect\citeauthoryear{Jangda, Polisetty, Guha, and Serafini}{Jangda
  et~al\mbox{.}}{2021}]%
        {jangda2021accelerating}
\bibfield{author}{\bibinfo{person}{Abhinav Jangda}, \bibinfo{person}{Sandeep
  Polisetty}, \bibinfo{person}{Arjun Guha}, {and} \bibinfo{person}{Marco
  Serafini}.} \bibinfo{year}{2021}\natexlab{}.
\newblock \showarticletitle{{Accelerating Graph Sampling for Graph Machine
  Learning using GPUs}}. In \bibinfo{booktitle}{\emph{Proceedings of the
  Sixteenth European Conference on Computer Systems}}.
\newblock


\bibitem[\protect\citeauthoryear{Karypis and Kumar}{Karypis and Kumar}{1998}]%
        {metis1998}
\bibfield{author}{\bibinfo{person}{George Karypis} {and} \bibinfo{person}{Vipin
  Kumar}.} \bibinfo{year}{1998}\natexlab{}.
\newblock \showarticletitle{{A Fast and High Quality Multilevel Scheme for
  Partitioning Irregular Graphs}}.
\newblock \bibinfo{journal}{\emph{SIAM Journal on scientific Computing}}
  (\bibinfo{year}{1998}).
\newblock


\bibitem[\protect\citeauthoryear{Kipf and Welling}{Kipf and Welling}{2017}]%
        {kipf2017semisupervised}
\bibfield{author}{\bibinfo{person}{Thomas~N. Kipf} {and} \bibinfo{person}{Max
  Welling}.} \bibinfo{year}{2017}\natexlab{}.
\newblock \showarticletitle{{Semi-Supervised Classification with Graph
  Convolutional Networks}}. In \bibinfo{booktitle}{\emph{Proceedings of
  International Conference on Learning Representations}}.
\newblock


\bibitem[\protect\citeauthoryear{Kumar and Huang}{Kumar and Huang}{2019}]%
        {kumar2019graphone}
\bibfield{author}{\bibinfo{person}{Pradeep Kumar} {and}
  \bibinfo{person}{H~Howie Huang}.} \bibinfo{year}{2019}\natexlab{}.
\newblock \showarticletitle{{GraphOne: A Data Store for Real-time Analytics on
  Evolving Graphs}}. In \bibinfo{booktitle}{\emph{Proceedings of 17th USENIX
  Conference on File and Storage Technologies}}.
\newblock


\bibitem[\protect\citeauthoryear{Kumar, Zhang, and Leskovec}{Kumar
  et~al\mbox{.}}{2019}]%
        {kumar2019predicting}
\bibfield{author}{\bibinfo{person}{Srijan Kumar}, \bibinfo{person}{Xikun
  Zhang}, {and} \bibinfo{person}{Jure Leskovec}.}
  \bibinfo{year}{2019}\natexlab{}.
\newblock \showarticletitle{Predicting dynamic embedding trajectory in temporal
  interaction networks}. In \bibinfo{booktitle}{\emph{Proceedings of the 25th
  ACM SIGKDD International Conference on Knowledge Discovery \& Data Dining}}.
\newblock


\bibitem[\protect\citeauthoryear{Li, Zhao, Varma, Salpekar, Noordhuis, Li,
  Paszke, Smith, Vaughan, Damania, et~al\mbox{.}}{Li et~al\mbox{.}}{2020}]%
        {li13pytorch}
\bibfield{author}{\bibinfo{person}{Shen Li}, \bibinfo{person}{Yanli Zhao},
  \bibinfo{person}{Rohan Varma}, \bibinfo{person}{Omkar Salpekar},
  \bibinfo{person}{Pieter Noordhuis}, \bibinfo{person}{Teng Li},
  \bibinfo{person}{Adam Paszke}, \bibinfo{person}{Jeff Smith},
  \bibinfo{person}{Brian Vaughan}, \bibinfo{person}{Pritam Damania},
  {et~al\mbox{.}}} \bibinfo{year}{2020}\natexlab{}.
\newblock \showarticletitle{{PyTorch Distributed: Experiences on Accelerating
  Data Parallel Training}}.
\newblock \bibinfo{journal}{\emph{Proceedings of the VLDB Endowment}}
  (\bibinfo{year}{2020}).
\newblock


\bibitem[\protect\citeauthoryear{Lin, Sun, Ding, Ke, Gu, Huang, Song, Zhang,
  Yi, Wen, and Chen}{Lin et~al\mbox{.}}{2022}]%
        {platogl}
\bibfield{author}{\bibinfo{person}{Dandan Lin}, \bibinfo{person}{Shijie Sun},
  \bibinfo{person}{Jingtao Ding}, \bibinfo{person}{Xuehan Ke},
  \bibinfo{person}{Hao Gu}, \bibinfo{person}{Xing Huang},
  \bibinfo{person}{Chonggang Song}, \bibinfo{person}{Xuri Zhang},
  \bibinfo{person}{Lingling Yi}, \bibinfo{person}{Jie Wen}, {and}
  \bibinfo{person}{Chuan Chen}.} \bibinfo{year}{2022}\natexlab{}.
\newblock \showarticletitle{{PlatoGL: Effective and Scalable Deep Graph
  Learning System for Graph-Enhanced Real-Time Recommendation}}. In
  \bibinfo{booktitle}{\emph{Proceedings of the 31st ACM International
  Conference on Information \& Knowledge Management}}.
\newblock


\bibitem[\protect\citeauthoryear{Lin, Li, Miao, Liu, and Xu}{Lin
  et~al\mbox{.}}{2020}]%
        {lin2020pagraph}
\bibfield{author}{\bibinfo{person}{Zhiqi Lin}, \bibinfo{person}{Cheng Li},
  \bibinfo{person}{Youshan Miao}, \bibinfo{person}{Yunxin Liu}, {and}
  \bibinfo{person}{Yinlong Xu}.} \bibinfo{year}{2020}\natexlab{}.
\newblock \showarticletitle{{PaGraph: Scaling GNN Training on Large Graphs via
  Computation-aware Caching}}. In \bibinfo{booktitle}{\emph{Proceedings of the
  11th ACM Symposium on Cloud Computing}}.
\newblock


\bibitem[\protect\citeauthoryear{Liu, Chen, Li, Wu, Zhu, He, Peng, Chen, Chen,
  and Guo}{Liu et~al\mbox{.}}{2023}]%
        {liu2021bgl}
\bibfield{author}{\bibinfo{person}{Tianfeng Liu}, \bibinfo{person}{Yangrui
  Chen}, \bibinfo{person}{Dan Li}, \bibinfo{person}{Chuan Wu},
  \bibinfo{person}{Yibo Zhu}, \bibinfo{person}{Jun He},
  \bibinfo{person}{Yanghua Peng}, \bibinfo{person}{Hongzheng Chen},
  \bibinfo{person}{Hongzhi Chen}, {and} \bibinfo{person}{Chuanxiong Guo}.}
  \bibinfo{year}{2023}\natexlab{}.
\newblock \showarticletitle{{BGL: GPU-efficient GNN Training by Optimizing
  Graph Data I/O and Preprocessing}}. In \bibinfo{booktitle}{\emph{Proceedings
  of the 20th USENIX Symposium on Networked Systems Design and
  Implementation}}.
\newblock


\bibitem[\protect\citeauthoryear{Low, Gonzalez, Kyrola, Bickson, Guestrin, and
  Hellerstein}{Low et~al\mbox{.}}{2014}]%
        {low2014graphlab}
\bibfield{author}{\bibinfo{person}{Yucheng Low}, \bibinfo{person}{Joseph~E
  Gonzalez}, \bibinfo{person}{Aapo Kyrola}, \bibinfo{person}{Danny Bickson},
  \bibinfo{person}{Carlos~E Guestrin}, {and} \bibinfo{person}{Joseph
  Hellerstein}.} \bibinfo{year}{2014}\natexlab{}.
\newblock \showarticletitle{Graphlab: A new framework for parallel machine
  learning}.
\newblock \bibinfo{journal}{\emph{arXiv preprint arXiv:1408.2041}}
  (\bibinfo{year}{2014}).
\newblock


\bibitem[\protect\citeauthoryear{Ma, Guo, Ren, Tang, and Yin}{Ma
  et~al\mbox{.}}{2020}]%
        {ma2020streaming}
\bibfield{author}{\bibinfo{person}{Yao Ma}, \bibinfo{person}{Ziyi Guo},
  \bibinfo{person}{Zhaocun Ren}, \bibinfo{person}{Jiliang Tang}, {and}
  \bibinfo{person}{Dawei Yin}.} \bibinfo{year}{2020}\natexlab{}.
\newblock \showarticletitle{{Streaming Graph Neural Networks}}. In
  \bibinfo{booktitle}{\emph{Proceedings of the 43rd International ACM SIGIR
  Conference on Research and Development in Information Retrieval}}.
\newblock


\bibitem[\protect\citeauthoryear{Macko, Marathe, Margo, and Seltzer}{Macko
  et~al\mbox{.}}{2015}]%
        {macko2015llama}
\bibfield{author}{\bibinfo{person}{Peter Macko}, \bibinfo{person}{Virendra~J
  Marathe}, \bibinfo{person}{Daniel~W Margo}, {and} \bibinfo{person}{Margo~I
  Seltzer}.} \bibinfo{year}{2015}\natexlab{}.
\newblock \showarticletitle{{LLAMA: Efficient Graph Analytics using Large
  Multiversioned Arrays}}. In \bibinfo{booktitle}{\emph{Proceedings of 2015
  IEEE 31st International Conference on Data Engineering}}.
\newblock


\bibitem[\protect\citeauthoryear{Malewicz, Austern, Bik, Dehnert, Horn, Leiser,
  and Czajkowski}{Malewicz et~al\mbox{.}}{2010}]%
        {malewicz2010pregel}
\bibfield{author}{\bibinfo{person}{Grzegorz Malewicz},
  \bibinfo{person}{Matthew~H Austern}, \bibinfo{person}{Aart~JC Bik},
  \bibinfo{person}{James~C Dehnert}, \bibinfo{person}{Ilan Horn},
  \bibinfo{person}{Naty Leiser}, {and} \bibinfo{person}{Grzegorz Czajkowski}.}
  \bibinfo{year}{2010}\natexlab{}.
\newblock \showarticletitle{Pregel: a system for large-scale graph processing}.
  In \bibinfo{booktitle}{\emph{Proceedings of the 2010 ACM SIGMOD International
  Conference on Management of data}}.
\newblock


\bibitem[\protect\citeauthoryear{Matani, Shah, and Mitra}{Matani
  et~al\mbox{.}}{2021}]%
        {matani20211}
\bibfield{author}{\bibinfo{person}{Dhruv Matani}, \bibinfo{person}{Ketan Shah},
  {and} \bibinfo{person}{Anirban Mitra}.} \bibinfo{year}{2021}\natexlab{}.
\newblock \showarticletitle{{An O(1) Algorithm for Implementing the LFU Cache
  Eviction Scheme}}.
\newblock \bibinfo{journal}{\emph{arXiv preprint}} (\bibinfo{year}{2021}).
\newblock


\bibitem[\protect\citeauthoryear{Nguyen, Lee, Rossi, Ahmed, Koh, and
  Kim}{Nguyen et~al\mbox{.}}{2018}]%
        {nguyen2018continuous}
\bibfield{author}{\bibinfo{person}{Giang~Hoang Nguyen},
  \bibinfo{person}{John~Boaz Lee}, \bibinfo{person}{Ryan~A Rossi},
  \bibinfo{person}{Nesreen~K Ahmed}, \bibinfo{person}{Eunyee Koh}, {and}
  \bibinfo{person}{Sungchul Kim}.} \bibinfo{year}{2018}\natexlab{}.
\newblock \showarticletitle{{Continuous-time Dynamic Network Embeddings}}. In
  \bibinfo{booktitle}{\emph{Companion Proceedings of the Web Conference}}.
\newblock


\bibitem[\protect\citeauthoryear{O'neil, O'neil, and Weikum}{O'neil
  et~al\mbox{.}}{1993}]%
        {o1993lru}
\bibfield{author}{\bibinfo{person}{Elizabeth~J O'neil},
  \bibinfo{person}{Patrick~E O'neil}, {and} \bibinfo{person}{Gerhard Weikum}.}
  \bibinfo{year}{1993}\natexlab{}.
\newblock \showarticletitle{{The LRU-K Page Replacement Algorithm for Database
  Disk Buffering}}.
\newblock \bibinfo{journal}{\emph{Acm Sigmod Record}} (\bibinfo{year}{1993}).
\newblock


\bibitem[\protect\citeauthoryear{Pandey, Li, Hoisie, Li, and Liu}{Pandey
  et~al\mbox{.}}{2020}]%
        {pandey2020c}
\bibfield{author}{\bibinfo{person}{Santosh Pandey}, \bibinfo{person}{Lingda
  Li}, \bibinfo{person}{Adolfy Hoisie}, \bibinfo{person}{Xiaoye~S Li}, {and}
  \bibinfo{person}{Hang Liu}.} \bibinfo{year}{2020}\natexlab{}.
\newblock \showarticletitle{{C-SAW: A Framework for Graph Sampling and Random
  Walk on GPUs}}. In \bibinfo{booktitle}{\emph{Proceedings of SC20:
  International Conference for High Performance Computing, Networking, Storage
  and Analysis}}.
\newblock


\bibitem[\protect\citeauthoryear{Pareja, Domeniconi, Chen, Ma, Suzumura,
  Kanezashi, Kaler, Schardl, and Leiserson}{Pareja et~al\mbox{.}}{2020}]%
        {pareja2020evolvegcn}
\bibfield{author}{\bibinfo{person}{Aldo Pareja}, \bibinfo{person}{Giacomo
  Domeniconi}, \bibinfo{person}{Jie Chen}, \bibinfo{person}{Tengfei Ma},
  \bibinfo{person}{Toyotaro Suzumura}, \bibinfo{person}{Hiroki Kanezashi},
  \bibinfo{person}{Tim Kaler}, \bibinfo{person}{Tao Schardl}, {and}
  \bibinfo{person}{Charles Leiserson}.} \bibinfo{year}{2020}\natexlab{}.
\newblock \showarticletitle{{EvolveGCN: Evolving Graph Convolutional Networks
  for Dynamic Graphs}}. In \bibinfo{booktitle}{\emph{Proceedings of the AAAI
  Conference on Artificial Intelligence}}.
\newblock


\bibitem[\protect\citeauthoryear{Paszke, Gross, Massa, Lerer, Bradbury, Chanan,
  Killeen, Lin, Gimelshein, Antiga, et~al\mbox{.}}{Paszke
  et~al\mbox{.}}{2019}]%
        {paszke2019pytorch}
\bibfield{author}{\bibinfo{person}{Adam Paszke}, \bibinfo{person}{Sam Gross},
  \bibinfo{person}{Francisco Massa}, \bibinfo{person}{Adam Lerer},
  \bibinfo{person}{James Bradbury}, \bibinfo{person}{Gregory Chanan},
  \bibinfo{person}{Trevor Killeen}, \bibinfo{person}{Zeming Lin},
  \bibinfo{person}{Natalia Gimelshein}, \bibinfo{person}{Luca Antiga},
  {et~al\mbox{.}}} \bibinfo{year}{2019}\natexlab{}.
\newblock \showarticletitle{{PyTorch: An Imperative Style, High-performance
  Deep Learning Library}}. In \bibinfo{booktitle}{\emph{Proceedings of Advances
  in Neural Information Processing Systems}}.
\newblock


\bibitem[\protect\citeauthoryear{Perini, Ramponi, Carbone, and Kalavri}{Perini
  et~al\mbox{.}}{2022}]%
        {perini2022learning}
\bibfield{author}{\bibinfo{person}{Massimo Perini}, \bibinfo{person}{Giorgia
  Ramponi}, \bibinfo{person}{Paris Carbone}, {and} \bibinfo{person}{Vasiliki
  Kalavri}.} \bibinfo{year}{2022}\natexlab{}.
\newblock \showarticletitle{{Learning on Streaming Graphs with Experience
  Replay}}. In \bibinfo{booktitle}{\emph{Proceedings of the 37th ACM/SIGAPP
  Symposium on Applied Computing}}.
\newblock


\bibitem[\protect\citeauthoryear{Petroni, Querzoni, Daudjee, Kamali, and
  Iacoboni}{Petroni et~al\mbox{.}}{2015}]%
        {petroni2015hdrf}
\bibfield{author}{\bibinfo{person}{Fabio Petroni}, \bibinfo{person}{Leonardo
  Querzoni}, \bibinfo{person}{Khuzaima Daudjee}, \bibinfo{person}{Shahin
  Kamali}, {and} \bibinfo{person}{Giorgio Iacoboni}.}
  \bibinfo{year}{2015}\natexlab{}.
\newblock \showarticletitle{{Hdrf: Stream-based Partitioning for Power-law
  Graphs}}. In \bibinfo{booktitle}{\emph{Proceedings of the 24th ACM
  international on conference on information and knowledge management}}.
\newblock


\bibitem[\protect\citeauthoryear{Rossi, Chamberlain, Frasca, Eynard, Monti, and
  Bronstein}{Rossi et~al\mbox{.}}{2021}]%
        {rossi2021temporal}
\bibfield{author}{\bibinfo{person}{Emanuele Rossi}, \bibinfo{person}{Ben
  Chamberlain}, \bibinfo{person}{Fabrizio Frasca}, \bibinfo{person}{Davide
  Eynard}, \bibinfo{person}{Federico Monti}, {and} \bibinfo{person}{Michael
  Bronstein}.} \bibinfo{year}{2021}\natexlab{}.
\newblock \showarticletitle{{Temporal Graph Networks for Deep Learning on
  Dynamic Graphs}}. In \bibinfo{booktitle}{\emph{Proceedings of International
  Conference on Learning Representations}}.
\newblock


\bibitem[\protect\citeauthoryear{Sankar, Wu, Gou, Zhang, and Yang}{Sankar
  et~al\mbox{.}}{2020}]%
        {sankar2020dysat}
\bibfield{author}{\bibinfo{person}{Aravind Sankar}, \bibinfo{person}{Yanhong
  Wu}, \bibinfo{person}{Liang Gou}, \bibinfo{person}{Wei Zhang}, {and}
  \bibinfo{person}{Hao Yang}.} \bibinfo{year}{2020}\natexlab{}.
\newblock \showarticletitle{{DySAT: Deep Neural Representation Learning on
  Dynamic Graphs via Self-attention Networks}}. In
  \bibinfo{booktitle}{\emph{Proceedings of the 13th International Conference on
  Web Search and Data Mining}}.
\newblock


\bibitem[\protect\citeauthoryear{Sun, Su, Shi, Shen, Wang, Wang, Zhang, Li, Yu,
  Zhou, and Wu}{Sun et~al\mbox{.}}{2023}]%
        {sun2023legion}
\bibfield{author}{\bibinfo{person}{Jie Sun}, \bibinfo{person}{Li Su},
  \bibinfo{person}{Zuocheng Shi}, \bibinfo{person}{Wenting Shen},
  \bibinfo{person}{Zeke Wang}, \bibinfo{person}{Lei Wang}, \bibinfo{person}{Jie
  Zhang}, \bibinfo{person}{Yong Li}, \bibinfo{person}{Wenyuan Yu},
  \bibinfo{person}{Jingren Zhou}, {and} \bibinfo{person}{Fei Wu}.}
  \bibinfo{year}{2023}\natexlab{}.
\newblock \showarticletitle{{Legion: Automatically Pushing the Envelope of
  {Multi-GPU} System for {Billion-Scale} {GNN} Training}}. In
  \bibinfo{booktitle}{\emph{Proceedings of USENIX Annual Technical
  Conference}}.
\newblock


\bibitem[\protect\citeauthoryear{Trivedi, Dai, Wang, and Song}{Trivedi
  et~al\mbox{.}}{2017}]%
        {trivedi2017know}
\bibfield{author}{\bibinfo{person}{Rakshit Trivedi}, \bibinfo{person}{Hanjun
  Dai}, \bibinfo{person}{Yichen Wang}, {and} \bibinfo{person}{Le Song}.}
  \bibinfo{year}{2017}\natexlab{}.
\newblock \showarticletitle{{Know-Evolve: Deep Temporal Reasoning for Dynamic
  Lnowledge Graphs}}. In \bibinfo{booktitle}{\emph{Proceedings of International
  Conference on Machine Learning}}.
\newblock


\bibitem[\protect\citeauthoryear{Trivedi, Farajtabar, Biswal, and Zha}{Trivedi
  et~al\mbox{.}}{2019}]%
        {trivedi2019dyrep}
\bibfield{author}{\bibinfo{person}{Rakshit Trivedi}, \bibinfo{person}{Mehrdad
  Farajtabar}, \bibinfo{person}{Prasenjeet Biswal}, {and}
  \bibinfo{person}{Hongyuan Zha}.} \bibinfo{year}{2019}\natexlab{}.
\newblock \showarticletitle{{DyRep: Learning Representations over Dynamic
  Graphs}}. In \bibinfo{booktitle}{\emph{Proceedings of International
  Conference on Learning Representations}}.
\newblock


\bibitem[\protect\citeauthoryear{Tsourakakis, Gkantsidis, Radunovic, and
  Vojnovic}{Tsourakakis et~al\mbox{.}}{2014}]%
        {Streaming:fennel14wsdm}
\bibfield{author}{\bibinfo{person}{Charalampos~E. Tsourakakis},
  \bibinfo{person}{Christos Gkantsidis}, \bibinfo{person}{Bozidar Radunovic},
  {and} \bibinfo{person}{Milan Vojnovic}.} \bibinfo{year}{2014}\natexlab{}.
\newblock \showarticletitle{{FENNEL:} Streaming Graph Partitioning for Massive
  Scale Graphs}. In \bibinfo{booktitle}{\emph{Proceedings of the Seventh {ACM}
  International Conference on Web Search and Data Mining}}.
\newblock


\bibitem[\protect\citeauthoryear{Vaswani, Shazeer, Parmar, Uszkoreit, Jones,
  Gomez, Kaiser, and Polosukhin}{Vaswani et~al\mbox{.}}{2017}]%
        {vaswani2017attention}
\bibfield{author}{\bibinfo{person}{Ashish Vaswani}, \bibinfo{person}{Noam
  Shazeer}, \bibinfo{person}{Niki Parmar}, \bibinfo{person}{Jakob Uszkoreit},
  \bibinfo{person}{Llion Jones}, \bibinfo{person}{Aidan~N Gomez},
  \bibinfo{person}{{\L}ukasz Kaiser}, {and} \bibinfo{person}{Illia
  Polosukhin}.} \bibinfo{year}{2017}\natexlab{}.
\newblock \showarticletitle{{Attention is All You Need}}.
\newblock \bibinfo{journal}{\emph{Proceedings of Advances in Neural Information
  Processing Systems}}.
\newblock


\bibitem[\protect\citeauthoryear{Veličković, Cucurull, Casanova, Romero,
  Liò, and Bengio}{Veličković et~al\mbox{.}}{2018}]%
        {velickovic2018graph}
\bibfield{author}{\bibinfo{person}{Petar Veličković},
  \bibinfo{person}{Guillem Cucurull}, \bibinfo{person}{Arantxa Casanova},
  \bibinfo{person}{Adriana Romero}, \bibinfo{person}{Pietro Liò}, {and}
  \bibinfo{person}{Yoshua Bengio}.} \bibinfo{year}{2018}\natexlab{}.
\newblock \showarticletitle{{Graph Attention Networks}}. In
  \bibinfo{booktitle}{\emph{Proceedings of International Conference on Learning
  Representations}}.
\newblock


\bibitem[\protect\citeauthoryear{Wang, Sun, and Bai}{Wang
  et~al\mbox{.}}{2023}]%
        {wangpipad2023}
\bibfield{author}{\bibinfo{person}{Chunyang Wang}, \bibinfo{person}{Desen Sun},
  {and} \bibinfo{person}{Yuebin Bai}.} \bibinfo{year}{2023}\natexlab{}.
\newblock \showarticletitle{{PiPAD: Pipelined and Parallel Dynamic GNN Training
  on GPUs}}. In \bibinfo{booktitle}{\emph{Proceedings of the 28th ACM SIGPLAN
  Annual Symposium on Principles and Practice of Parallel Programming}}.
\newblock


\bibitem[\protect\citeauthoryear{Wang, Song, Wu, and Wang}{Wang
  et~al\mbox{.}}{2020}]%
        {wang2020streaming}
\bibfield{author}{\bibinfo{person}{Junshan Wang}, \bibinfo{person}{Guojie
  Song}, \bibinfo{person}{Yi Wu}, {and} \bibinfo{person}{Liang Wang}.}
  \bibinfo{year}{2020}\natexlab{}.
\newblock \showarticletitle{{Streaming Graph Neural Networks via Continual
  Learning}}. In \bibinfo{booktitle}{\emph{Proceedings of the 29th ACM
  International Conference on Information \& Knowledge Management}}.
\newblock


\bibitem[\protect\citeauthoryear{Wang, Zheng, Ye, Gan, Li, Song, Zhou, Ma, Yu,
  Gai, et~al\mbox{.}}{Wang et~al\mbox{.}}{2019}]%
        {wang2019deep}
\bibfield{author}{\bibinfo{person}{Minjie Wang}, \bibinfo{person}{Da Zheng},
  \bibinfo{person}{Zihao Ye}, \bibinfo{person}{Quan Gan},
  \bibinfo{person}{Mufei Li}, \bibinfo{person}{Xiang Song},
  \bibinfo{person}{Jinjing Zhou}, \bibinfo{person}{Chao Ma},
  \bibinfo{person}{Lingfan Yu}, \bibinfo{person}{Yu Gai}, {et~al\mbox{.}}}
  \bibinfo{year}{2019}\natexlab{}.
\newblock \showarticletitle{{Deep Graph Library: A Graph-centric,
  Highly-performant Package for Graph Neural Networks}}.
\newblock \bibinfo{journal}{\emph{arXiv preprint}} (\bibinfo{year}{2019}).
\newblock


\bibitem[\protect\citeauthoryear{Wang, Zhang, and Coates}{Wang
  et~al\mbox{.}}{2021}]%
        {wang2021graph}
\bibfield{author}{\bibinfo{person}{Yuening Wang}, \bibinfo{person}{Yingxue
  Zhang}, {and} \bibinfo{person}{Mark Coates}.}
  \bibinfo{year}{2021}\natexlab{}.
\newblock \showarticletitle{Graph Structure Aware Contrastive Knowledge
  Distillation for Incremental Learning in Recommender Systems}. In
  \bibinfo{booktitle}{\emph{Proceedings of the 30th ACM International
  Conference on Information \& Knowledge Management}}.
\newblock


\bibitem[\protect\citeauthoryear{Winter, Mlakar, Zayer, Seidel, and
  Steinberger}{Winter et~al\mbox{.}}{2018}]%
        {winter2018faimgraph}
\bibfield{author}{\bibinfo{person}{Martin Winter}, \bibinfo{person}{Daniel
  Mlakar}, \bibinfo{person}{Rhaleb Zayer}, \bibinfo{person}{Hans-Peter Seidel},
  {and} \bibinfo{person}{Markus Steinberger}.} \bibinfo{year}{2018}\natexlab{}.
\newblock \showarticletitle{{faimGraph: High Performance Management of
  Fully-dynamic Graphs under Tight Memory Constraints on the GPU}}. In
  \bibinfo{booktitle}{\emph{Proceedings of International Conference for High
  Performance Computing, Networking, Storage and Analysis}}.
\newblock


\bibitem[\protect\citeauthoryear{Wu, Tang, Zhu, Wang, Xie, and Tan}{Wu
  et~al\mbox{.}}{2019}]%
        {wu2019session}
\bibfield{author}{\bibinfo{person}{Shu Wu}, \bibinfo{person}{Yuyuan Tang},
  \bibinfo{person}{Yanqiao Zhu}, \bibinfo{person}{Liang Wang},
  \bibinfo{person}{Xing Xie}, {and} \bibinfo{person}{Tieniu Tan}.}
  \bibinfo{year}{2019}\natexlab{}.
\newblock \showarticletitle{{Session-based Recommendation with Graph Neural
  Networks}}. In \bibinfo{booktitle}{\emph{Proceedings of the AAAI conference
  on Artificial Intelligence}}.
\newblock


\bibitem[\protect\citeauthoryear{Xie, Yan, Li, and Zhang}{Xie
  et~al\mbox{.}}{2014}]%
        {dbh}
\bibfield{author}{\bibinfo{person}{Cong Xie}, \bibinfo{person}{Ling Yan},
  \bibinfo{person}{Wu-Jun Li}, {and} \bibinfo{person}{Zhihua Zhang}.}
  \bibinfo{year}{2014}\natexlab{}.
\newblock \showarticletitle{{Distributed Power-law Graph Computing: Theoretical
  and Empirical Analysis}}.
\newblock \bibinfo{journal}{\emph{Advances in neural information processing
  systems}} (\bibinfo{year}{2014}).
\newblock


\bibitem[\protect\citeauthoryear{Xu, Zhang, Guo, Guo, Tang, and Coates}{Xu
  et~al\mbox{.}}{2020}]%
        {xu2020graphsail}
\bibfield{author}{\bibinfo{person}{Yishi Xu}, \bibinfo{person}{Yingxue Zhang},
  \bibinfo{person}{Wei Guo}, \bibinfo{person}{Huifeng Guo},
  \bibinfo{person}{Ruiming Tang}, {and} \bibinfo{person}{Mark Coates}.}
  \bibinfo{year}{2020}\natexlab{}.
\newblock \showarticletitle{{GraphSAIL: Graph Structure Aware Incremental
  Learning for Recommender Systems}}. In \bibinfo{booktitle}{\emph{Proceedings
  of the 29th ACM International Conference on Information \& Knowledge
  Management}}.
\newblock


\bibitem[\protect\citeauthoryear{Yang, Liu, Qi, and Lai}{Yang
  et~al\mbox{.}}{2022a}]%
        {yang2022wholegraph}
\bibfield{author}{\bibinfo{person}{Dongxu Yang}, \bibinfo{person}{Junhong Liu},
  \bibinfo{person}{Jiaxing Qi}, {and} \bibinfo{person}{Junjie Lai}.}
  \bibinfo{year}{2022}\natexlab{a}.
\newblock \showarticletitle{{WholeGraph: A Fast Graph Neural Network Training
  Framework with Multi-GPU Distributed Shared Memory Architecture}}. In
  \bibinfo{booktitle}{\emph{Proceedings of SC22: International Conference for
  High Performance Computing, Networking, Storage and Analysis}}.
\newblock


\bibitem[\protect\citeauthoryear{Yang, Tang, Song, Wang, Yin, Chen, Yu, and
  Zhou}{Yang et~al\mbox{.}}{2022b}]%
        {yang2022gnnlab}
\bibfield{author}{\bibinfo{person}{Jianbang Yang}, \bibinfo{person}{Dahai
  Tang}, \bibinfo{person}{Xiaoniu Song}, \bibinfo{person}{Lei Wang},
  \bibinfo{person}{Qiang Yin}, \bibinfo{person}{Rong Chen},
  \bibinfo{person}{Wenyuan Yu}, {and} \bibinfo{person}{Jingren Zhou}.}
  \bibinfo{year}{2022}\natexlab{b}.
\newblock \showarticletitle{{GNNLab: A Factored System for Sample-based GNN
  Training over GPUs}}. In \bibinfo{booktitle}{\emph{Proceedings of the 17th
  European Conference on Computer Systems}}.
\newblock


\bibitem[\protect\citeauthoryear{Ying, He, Chen, Eksombatchai, Hamilton, and
  Leskovec}{Ying et~al\mbox{.}}{2018}]%
        {ying2018graph}
\bibfield{author}{\bibinfo{person}{Rex Ying}, \bibinfo{person}{Ruining He},
  \bibinfo{person}{Kaifeng Chen}, \bibinfo{person}{Pong Eksombatchai},
  \bibinfo{person}{William~L Hamilton}, {and} \bibinfo{person}{Jure Leskovec}.}
  \bibinfo{year}{2018}\natexlab{}.
\newblock \showarticletitle{{Graph Convolutional Neural Networks for Web-scale
  Recommender Systems}}. In \bibinfo{booktitle}{\emph{Proceedings of the 24th
  ACM SIGKDD International Conference on Knowledge Discovery \& Data Mining}}.
\newblock


\bibitem[\protect\citeauthoryear{Yu, Childers, Huang, Qian, and Wang}{Yu
  et~al\mbox{.}}{2020}]%
        {yu2020quantitative}
\bibfield{author}{\bibinfo{person}{Qi Yu}, \bibinfo{person}{Bruce Childers},
  \bibinfo{person}{Libo Huang}, \bibinfo{person}{Cheng Qian}, {and}
  \bibinfo{person}{Zhiying Wang}.} \bibinfo{year}{2020}\natexlab{}.
\newblock \showarticletitle{A Quantitative Evaluation of Unified Memory in
  GPUs}.
\newblock \bibinfo{journal}{\emph{The Journal of Supercomputing}}
  (\bibinfo{year}{2020}).
\newblock


\bibitem[\protect\citeauthoryear{Zhang, Huang, Liu, Zhou, Hu, Song, Ge, Wang,
  Zhang, and Qi}{Zhang et~al\mbox{.}}{2020}]%
        {DBLP:journals/pvldb/ZhangHL0HSGWZ020}
\bibfield{author}{\bibinfo{person}{Dalong Zhang}, \bibinfo{person}{Xin Huang},
  \bibinfo{person}{Ziqi Liu}, \bibinfo{person}{Jun Zhou},
  \bibinfo{person}{Zhiyang Hu}, \bibinfo{person}{Xianzheng Song},
  \bibinfo{person}{Zhibang Ge}, \bibinfo{person}{Lin Wang},
  \bibinfo{person}{Zhiqiang Zhang}, {and} \bibinfo{person}{Yuan Qi}.}
  \bibinfo{year}{2020}\natexlab{}.
\newblock \showarticletitle{{AGL: {A} Scalable System for Industrial-purpose
  Graph Machine Learning}}.
\newblock \bibinfo{journal}{\emph{{VLDB} Journal}} (\bibinfo{year}{2020}).
\newblock


\bibitem[\protect\citeauthoryear{Zheng, Ma, Wang, Zhou, Su, Song, Gan, Zhang,
  and Karypis}{Zheng et~al\mbox{.}}{2020}]%
        {zheng2020distdgl}
\bibfield{author}{\bibinfo{person}{Da Zheng}, \bibinfo{person}{Chao Ma},
  \bibinfo{person}{Minjie Wang}, \bibinfo{person}{Jinjing Zhou},
  \bibinfo{person}{Qidong Su}, \bibinfo{person}{Xiang Song},
  \bibinfo{person}{Quan Gan}, \bibinfo{person}{Zheng Zhang}, {and}
  \bibinfo{person}{George Karypis}.} \bibinfo{year}{2020}\natexlab{}.
\newblock \showarticletitle{{DistDGL: Distributed Graph Neural Network Training
  for Billion-scale Graphs}}. In \bibinfo{booktitle}{\emph{Proceedings of 2020
  IEEE/ACM 10th Workshop on Irregular Applications: Architectures and
  Algorithms}}.
\newblock


\bibitem[\protect\citeauthoryear{Zhou, Zheng, Nisa, Ioannidis, Song, and
  Karypis}{Zhou et~al\mbox{.}}{2022}]%
        {zhou2022tgl}
\bibfield{author}{\bibinfo{person}{Hongkuan Zhou}, \bibinfo{person}{Da Zheng},
  \bibinfo{person}{Israt Nisa}, \bibinfo{person}{Vasileios Ioannidis},
  \bibinfo{person}{Xiang Song}, {and} \bibinfo{person}{George Karypis}.}
  \bibinfo{year}{2022}\natexlab{}.
\newblock \showarticletitle{{TGL: A General Framework for Temporal GNN Training
  on Billion-Scale Graphs}}. In \bibinfo{booktitle}{\emph{Proceedings of the
  VLDB Endowment}}.
\newblock


\bibitem[\protect\citeauthoryear{Zhou, Zheng, Song, Karypis, and Prasanna}{Zhou
  et~al\mbox{.}}{2023}]%
        {zhou2023disttgl}
\bibfield{author}{\bibinfo{person}{Hongkuan Zhou}, \bibinfo{person}{Da Zheng},
  \bibinfo{person}{Xiang Song}, \bibinfo{person}{George Karypis}, {and}
  \bibinfo{person}{Viktor Prasanna}.} \bibinfo{year}{2023}\natexlab{}.
\newblock \showarticletitle{{DistTGL: Distributed Memory-Based Temporal Graph
  Neural Network Training}}. In \bibinfo{booktitle}{\emph{Proceedings of the
  International Conference for High Performance Computing, Networking, Storage
  and Analysis}}.
\newblock


\bibitem[\protect\citeauthoryear{Zhu, Zhao, Yang, Lin, Zhou, Ai, Li, and
  Zhou}{Zhu et~al\mbox{.}}{2019}]%
        {zhu2019aligraph}
\bibfield{author}{\bibinfo{person}{Rong Zhu}, \bibinfo{person}{Kun Zhao},
  \bibinfo{person}{Hongxia Yang}, \bibinfo{person}{Wei Lin},
  \bibinfo{person}{Chang Zhou}, \bibinfo{person}{Baole Ai},
  \bibinfo{person}{Yong Li}, {and} \bibinfo{person}{Jingren Zhou}.}
  \bibinfo{year}{2019}\natexlab{}.
\newblock \showarticletitle{{AliGraph: A Comprehensive Graph Neural Network
  Platform}}. In \bibinfo{booktitle}{\emph{Proceedings of the VLDB Endowment}}.
\newblock


\end{thebibliography}
